\documentclass[11pt,letterpaper]{article}

\usepackage[T1]{fontenc}
\usepackage[margin=1in]{geometry}
\usepackage{amsmath,amssymb,amsthm,mathtools}
\allowdisplaybreaks
\usepackage{booktabs,tabularx,array}
\usepackage{enumitem}
\usepackage{lmodern}
\usepackage{xcolor}
\usepackage{tikz}
\usetikzlibrary{arrows.meta,calc,fit,positioning,shapes.geometric}
\usepackage[numbers,sort&compress]{natbib}
\usepackage{aliascnt}
\usepackage[unicode,colorlinks=true,linkcolor=blue!55!black,citecolor=green!40!black,urlcolor=blue!65!black]{hyperref}
\usepackage[nameinlink,noabbrev,capitalise]{cleveref}
\hypersetup{
  pdftitle={Single-Exponential FPT Algorithms for Minimum Bisection and Matching Cut via a New Unbreakable Decomposition},
  pdfauthor={Soh Kumabe}
}

\setlist[itemize]{leftmargin=2em,itemsep=.25em,topsep=.35em}
\setlist[enumerate]{leftmargin=2.2em,itemsep=.25em,topsep=.35em}
\renewcommand{\arraystretch}{1.14}

\newaliascnt{lemma}{theorem}
\newtheorem{lemma}[lemma]{Lemma}
\aliascntresetthe{lemma}
\newaliascnt{proposition}{theorem}

\aliascntresetthe{proposition}
\newaliascnt{corollary}{theorem}

\aliascntresetthe{corollary}
\newaliascnt{claim}{theorem}
\newtheorem{claim}[claim]{Claim}
\aliascntresetthe{claim}
\theoremstyle{definition}
\newaliascnt{definition}{theorem}

\aliascntresetthe{definition}
\newaliascnt{example}{theorem}

\aliascntresetthe{example}
\theoremstyle{remark}
\newaliascnt{remark}{theorem}

\aliascntresetthe{remark}
\usepackage{thmtools}

\crefname{lemma}{Lemma}{Lemmas}
\crefname{proposition}{Proposition}{Propositions}
\crefname{corollary}{Corollary}{Corollaries}
\crefname{claim}{Claim}{Claims}
\crefname{definition}{Definition}{Definitions}
\crefname{example}{Example}{Examples}
\crefname{remark}{Remark}{Remarks}

\newcommand{\Q}{\mathbb Q}
\newcommand{\cC}{\mathcal C}
\newcommand{\cD}{\mathcal D}
\newcommand{\bigO}{\mathcal O}
\newcommand{\Ostar}{\mathcal O^*}
\newcommand{\boxproblem}[4]{%
  \begin{center}
  \fbox{\begin{minipage}{0.94\linewidth}
    \vspace{2pt}\noindent\normalsize\textsc{#1}\par\vspace{2pt}
    \setlength{\tabcolsep}{3pt}\renewcommand{\arraystretch}{1.0}
    \begin{tabularx}{\linewidth}{@{}lX@{}}
      \textbf{Input:}& #2\\
      \textbf{Question:}& #3\\
      \textbf{Parameter:}& #4
    \end{tabularx}
  \end{minipage}}
  \end{center}}
\newcommand{\boxproblemwithoutparameter}[3]{%
  \begin{center}
  \fbox{\begin{minipage}{0.94\linewidth}
    \vspace{2pt}\noindent\normalsize\textsc{#1}\par\vspace{2pt}
    \setlength{\tabcolsep}{3pt}\renewcommand{\arraystretch}{1.0}
    \begin{tabularx}{\linewidth}{@{}lX@{}}
      \textbf{Input:}& #2\\
      \textbf{Question:}& #3
    \end{tabularx}
  \end{minipage}}
  \end{center}}

\title{Single-Exponential FPT Algorithms for Minimum Bisection and Matching Cut via a New Unbreakable Decomposition}
\author{Soh Kumabe\thanks{CyberAgent, Inc., Tokyo, Japan. Email: \texttt{kumabe\_soh@cyberagent.co.jp}.}}
\date{October 2026}

\begin{document}
\maketitle

\begin{abstract}
\textsc{Minimum Bisection} asks whether the vertices of a graph
can be partitioned into two parts whose sizes differ by at most one and with
at most $k$ edges between the two parts.  Cygan, Lokshtanov, Marcin
Pilipczuk, Micha{\l} Pilipczuk, and Saurabh (SICOMP 2019) gave the first
fixed-parameter algorithm, which runs in $\Ostar(2^{\bigO(k^3)})$ time, by
introducing an unbreakable decomposition.  Cygan, Komosa, Lokshtanov,
Marcin Pilipczuk, Micha{\l} Pilipczuk, Saurabh, and Wahlstr{\"o}m (TALG 2021)
strengthened the decomposition and improved the running time to
$\Ostar(2^{\bigO(k\log k)})$.  However, whether the running time could be
improved to single-exponential remained open.

In this paper, we give a deterministic $\Ostar(2^{\bigO(k)})$-time algorithm
for \textsc{Minimum Bisection}.  Our algorithm also counts all feasible
bisections.  To obtain an $\Ostar(2^{\bigO(k)})$-time algorithm, we introduce
two new techniques.
First, we give a new algorithm that computes a $(2k,k)$-edge-unbreakable
decomposition with adhesion at most $2k$ in $\Ostar(2^{\bigO(k)})$ time by
relaxing the unbreakability and adhesion guarantees of the decomposition in
(TALG 2021).  Second, we improve the running time of the dynamic program over
the decomposition from
$\Ostar(2^{\bigO(k\log k)})$ to $\Ostar(2^{\bigO(k)})$ using important cuts,
inclusion--exclusion, and formal logarithms and exponentials.

Together with color coding, the same techniques give a
deterministic $\Ostar(2^{\bigO(k)})$-time algorithm for \textsc{Matching Cut},
which asks for a cut whose crossing edges form a matching.  This
improves the $\Ostar(2^{\bigO(k\log k)})$-time algorithm of Aravind and Saxena
(IWOCA 2021).
\end{abstract}

\clearpage
\tableofcontents
\clearpage

\section{Introduction}\label{sec:introduction}

\textsc{Minimum Bisection} asks whether the vertices of a graph
can be divided into two parts of equal size, up to one vertex, with at most
$k$ edges between the two parts.  Throughout the paper, $G$ is a simple
undirected graph and $n=|V(G)|$.  For $A\subseteq V(G)$, let
$\delta_G(A):=\{uv\in E(G):|\{u,v\}\cap A|=1\}$.

\boxproblem{Minimum Bisection}
  {A simple undirected graph $G$ and an integer $k\ge0$.}
  {Is there a set $A\subseteq V(G)$ with
   $|A|=\lfloor n/2\rfloor$ and $|\delta_G(A)|\le k$?}
  {$k$.}

Kernighan and Lin~\cite{KernighanLin1970} formulated the problem of
partitioning the vertices of an edge-weighted graph into parts of prescribed
sizes so as to minimize the weight of the crossing edges.  The unweighted
two-part case with equal part sizes is \textsc{Minimum Bisection}.  Garey,
Johnson, and Stockmeyer~\cite{GareyJohnsonStockmeyer1976} proved that
\textsc{Minimum Bisection} is NP-complete.

Cygan, Lokshtanov, Marcin Pilipczuk, Micha{\l} Pilipczuk, and
Saurabh~\cite{CyganEtAl2019} gave the first fixed-parameter algorithm for
\textsc{Minimum Bisection}, with running time
$\Ostar(2^{\bigO(k^3)})$, and asked whether \textsc{Minimum Bisection} admits
a single-exponential fixed-parameter algorithm.\footnote{The $\Ostar$ notation
  suppresses factors polynomial in the input size.}
Cygan, Komosa, Lokshtanov, Marcin Pilipczuk, Micha{\l} Pilipczuk, Saurabh,
and Wahlstr{\"o}m \cite{CyganEtAl2021} improved the running time to
$\Ostar(2^{\bigO(k\log k)})$, making substantial progress toward this goal
while leaving the question open.

In this paper, we answer the question posed in~\cite{CyganEtAl2019}
affirmatively by giving a deterministic $\Ostar(2^{\bigO(k)})$-time algorithm
for \textsc{Minimum Bisection}.

\paragraph{Existing framework for Minimum Bisection by
\cite{CyganEtAl2019} and \cite{CyganEtAl2021}.}
Before describing our algorithm, we review the framework common to the two
previous algorithms.  Both algorithms first compute a tree decomposition with
a certain property, called an \emph{unbreakable decomposition}, and then
perform bottom-up dynamic programming over the decomposition.  Unbreakable
decomposition itself was introduced in \cite{CyganEtAl2019} as an
algorithmically tractable relaxation of \emph{lean tree
decompositions}~\cite{Thomas1990,BellenbaumDiestel2002}.

For a graph $H$, a set $X\subseteq V(H)$ is
\emph{$(q,k)$-vertex-unbreakable} if every separation
$(L,R)$\footnote{A pair $(L,R)$ is a \emph{separation} when
$L\cup R=V(H)$ and there is no edge between $L\setminus R$ and
$R\setminus L$.} of $H$ with $|L\cap R|\le k$ satisfies
$\min\{|L\cap X|,|R\cap X|\}\le q$.  A tree decomposition is
\emph{$(q,k)$-vertex-unbreakable} if every bag is
$(q,k)$-vertex-unbreakable.  \cite{CyganEtAl2019} constructs in
$\Ostar(2^{\bigO(k^2)})$ time a
$(2^{\bigO(k)},k)$-vertex-unbreakable decomposition with adhesion at most
$2^{\bigO(k)}$, where the \emph{adhesion} of a tree decomposition is the
maximum size of the intersection of two adjacent bags.
\cite{CyganEtAl2021} substantially strengthens the decomposition guarantee
and constructs in $\Ostar(2^{\bigO(k\log k)})$ time a tree decomposition with
adhesion at most $k$ that is
$(i,i)$-vertex-unbreakable for every $1\le i\le k$.

Both algorithms then perform bottom-up dynamic programming over their tree
decompositions.  At a node $t$ with parent $t'$, a state records which
vertices of the adhesion $B_t\cap B_{t'}$ lie on one side of the cut, the
number of vertices on that side in the subgraph below $t$, and the number of
cut edges in that subgraph.

The difficult step is the transition at a large bag $B_t$.  The
unbreakability guarantee in \cite{CyganEtAl2021} ensures that every set
$A\subseteq V(G)$ with $|\delta_G(A)|\le k$ satisfies
$|A\cap B_t|\le k$ or $|B_t\setminus A|\le k$, but enumerating all such
intersections takes $n^{\bigO(k)}$ time.  Their algorithm avoids this
enumeration by color coding.  In particular, at most $k$ child adhesions are
split by the cut, and each has size at most $k$.  Hence, the coloring must
separate the smaller of $A\cap B_t$ and $B_t\setminus A$ from
$\bigO(k^2)$ vertices on the other side of these adhesions.  If each vertex
is colored red with probability $\Theta(1/k)$, the probability that every
vertex of the smaller set is red and all $\bigO(k^2)$ opposite-side
vertices are blue is $2^{-\bigO(k\log k)}$.  The color-coding step therefore
contributes a factor of $2^{\bigO(k\log k)}$ to the transition time.

\paragraph{Our techniques.}
In \cite{CyganEtAl2021}, both the construction of the unbreakable
decomposition and the transition at a large bag take
$\Ostar(2^{\bigO(k\log k)})$ time.  We perform both steps in
$\Ostar(2^{\bigO(k)})$ time and thereby obtain a single-exponential algorithm.

We first consider the construction of the unbreakable decomposition.
We do not know whether a vertex-unbreakable decomposition with the bounds
of \cite{CyganEtAl2021}, that is, adhesions of size at most $k$ and
$(i,i)$-vertex-unbreakable bags for every $1\le i\le k$, can be constructed in
$\Ostar(2^{\bigO(k)})$ time.
However, as observed in \cite{CyganEtAl2021}, all applications considered
there, including \textsc{Minimum Bisection}, require only the weaker notion of
\emph{edge-unbreakability}.  For a graph $H$, a
set $X\subseteq V(H)$ is \emph{$(q,k)$-edge-unbreakable} if
$\min\{|A\cap X|,|X\setminus A|\}\le q$ for every $A\subseteq V(H)$ with
$|\delta_H(A)|\le k$.  A tree decomposition of $H$ is
\emph{$(q,k)$-edge-unbreakable} if every bag is $(q,k)$-edge-unbreakable.
We construct an edge-unbreakable decomposition in single-exponential
time.

\begin{restatable}{theorem}{decompositionthm}\label{thm:decomp}
Given a graph $G$ and $k$, one can construct in
$\Ostar(2^{\bigO(k)})$ time a $(2k,k)$-edge-unbreakable decomposition
with adhesion at most $2k$.
\end{restatable}

We next consider the transition at a large bag.  Unlike the previous
algorithms, we do not use color coding.  We instead apply algebraic
operations to formal power series that represent the dynamic-programming
tables, including inclusion--exclusion and formal logarithms and exponentials,
all of which can be performed in $\Ostar(2^{\bigO(k)})$ time.  These operations
cannot be performed on Boolean existence values alone, so each state stores
the number of partial solutions that it represents.  Consequently, our
algorithm counts all feasible solutions to \textsc{Minimum Bisection} rather
than merely deciding feasibility.

\begin{restatable}{theorem}{bisectionthm}\label{thm:bisection}
The feasible solutions to \textnormal{\textsc{Minimum Bisection}} can be
counted deterministically in $\Ostar(2^{\bigO(k)})$ time.  If a feasible
solution exists, one can recover one within the same running time.
\end{restatable}

\paragraph{Application.}
Our approach also applies to \textsc{Matching Cut}, parameterized by the
number of crossing edges.

\boxproblem{Matching Cut}
  {A simple undirected graph $G$ and an integer $k\ge0$.}
  {Is there a set $\emptyset\neq A\subsetneq V(G)$ such that
   $|\delta_G(A)|\le k$, and
   $|\delta_G(A)\cap\delta_G(\{v\})|\le1$ for every $v\in V(G)$?}
  {$k$.}

Gomes and Sau~\cite{GomesSau2019} observed that \textsc{Matching Cut}
parameterized by $k$ reduces to \textsc{Stable Cutset}, for which Marx,
O'Sullivan, and Razgon~\cite{MarxOSullivanRazgon2010} gave a fixed-parameter
algorithm using Courcelle's theorem.  Aravind and
Saxena~\cite{AravindSaxena2021} later made the dependence on $k$ explicit with
an $\Ostar(2^{\bigO(k\log k)})$-time algorithm by applying the framework of
\cite{CyganEtAl2021}.

Our algorithm for \textsc{Matching Cut} combines the \textsc{Minimum
Bisection} algorithm with color coding that assigns distinct colors to the
endpoints of the cut.  In particular, we obtain the following result.

\begin{restatable}{theorem}{matchingcutthm}\label{thm:matching-cut}
\textnormal{\textsc{Matching Cut}} can be solved deterministically, and a
witness can be recovered, in
$\Ostar(2^{\bigO(k)})$ time.
\end{restatable}

Unlike our algorithm for \textsc{Minimum Bisection}, our algorithm for
\textsc{Matching Cut} does not count feasible solutions, because the initial
color-coding step preserves only the existence of a solution.

\paragraph{Remark.}
It is natural to ask whether our techniques also yield single-exponential
algorithms for other applications of unbreakable decompositions.
\cite{CyganEtAl2021} applied the unbreakable-decomposition framework to
\textsc{Steiner Cut}, a generalization of \textsc{$p$-Way Cut}, and to
\textsc{Steiner Multicut}.
As noted above, these applications use only edge-unbreakability, so our
decomposition is sufficient for the same framework, up to constant changes
in the parameters.
This replacement alone, however, does not yield single-exponential
algorithms, because the bottleneck is already present in the number of
dynamic-programming states.
For \textsc{$p$-Way Cut}, a solution may have up to $p\le k+1$ parts, and
the direct dynamic program assigns each adhesion vertex to one of these
parts, already resulting in
$p^{\bigO(k)}=2^{\bigO(k\log k)}$ dynamic-programming states, whereas the
corresponding number for \textsc{Minimum Bisection} is only
$2^{\bigO(k)}$.
We do not know whether these states can be compressed.
Nevertheless, we believe that our construction provides a useful first step
toward speeding up other applications of unbreakable decompositions, such as
\textsc{$p$-Way Cut}, \textsc{Steiner Cut}, and
\textsc{Steiner Multicut}.

\paragraph{Lower bound.}
Aravind and Saxena~\cite{AravindSaxena2021} gave a lower-bound result for
\textsc{Matching Cut} parameterized by the number $k$ of crossing edges.
More quantitatively, let
$g\colon\mathbb N\to\mathbb R_{\geq 0}$ be unbounded and nondecreasing,
satisfy $g(x)=\bigO(x)$, and satisfy $g(\lambda x)=\bigO(g(x))$ for every fixed
constant $\lambda>0$.  Then \textsc{Matching Cut} cannot be solved in
$\Ostar(2^{o(g(k))})$ time unless \textsc{3-SAT} can be solved in
$\Ostar(2^{o(g(s))})$ time on instances of size $s$, where $s$ is the total
number of variables and clauses in the input 3-CNF formula.\footnote{Their
stated result was that, under the Exponential Time Hypothesis,
\textsc{Matching Cut} admits no $\Ostar(2^{o(k)})$-time algorithm.
However, very recently, OpenAI and Anthropic independently released draft
manuscripts claiming to refute ETH, each accompanied by a Lean
formalization~\cite{OpenAI2026ETH,ClaudeAnthropic2026ETH}.
Although neither claim has yet been independently validated by the research
community and we have not verified either claim ourselves, in view of these
developments, we think that it is better to state the quantitative consequences
of ETH-based reductions rather than only their consequences under ETH.}

For \textsc{Minimum Bisection}, we prove
\cref{thm:bisection-lower-bound}.

\begin{restatable}{theorem}{bisectionlowerboundthm}\label{thm:bisection-lower-bound}
Let $g\colon\mathbb N\to\mathbb R_{\geq 0}$ be unbounded and nondecreasing,
satisfy $g(x)=\bigO(x)$, and satisfy $g(\lambda x)=\bigO(g(x))$ for every fixed
constant $\lambda>0$.  Suppose that \textnormal{\textsc{Minimum Bisection}}
can be solved in $\Ostar(2^{o(g(k))})$ time.  Then \textnormal{\textsc{3-SAT}}
can be solved in $2^{o(g(qs))+\bigO(s\log q/q+\log s)}$ time for every integer
$q$ with $2\leq q\leq s$, where $s$ is the total number of variables and
clauses in the input 3-CNF formula.\footnote{The usual convention was to state
ETH in terms of the number $n$ of variables and invoke the sparsification
lemma~\cite{ImpagliazzoPaturiZane2001} to assume $\bigO(n)$ clauses.  For every
fixed $\varepsilon>0$, however, sparsification incurs a $2^{\varepsilon n}$
overhead.  This overhead is harmless for ETH itself, but it cannot be absorbed
into the finer $2^{o(s^\alpha)}$ and $2^{o(s/(\log s)^c)}$ bounds.  We therefore
measure $s$ by the total number of variables and clauses.}

In particular, the theorem yields three consequences.
\begin{itemize}
  \item For every fixed $0<\alpha<1$, if \textnormal{\textsc{3-SAT}}
  has no $\Ostar(2^{o(s^\alpha)})$-time algorithm, then
  \textnormal{\textsc{Minimum Bisection}} has no
  $\Ostar(2^{o((k/\log k)^{\alpha/(2-\alpha)})})$-time algorithm.

  \item For every fixed $c>0$, if \textnormal{\textsc{3-SAT}} has no
  $\Ostar(2^{o(s/(\log s)^c)})$-time algorithm, then
  \textnormal{\textsc{Minimum Bisection}} has no
  $\Ostar(2^{o(k/((\log k)^{2c}\log\log k))})$-time algorithm.

  \item If \textnormal{\textsc{3-SAT}} has no
  $\Ostar(2^{o(s)})$-time algorithm, equivalently, if ETH holds, then
  \textnormal{\textsc{Minimum Bisection}} has no
  $\Ostar(2^{o(k)})$-time algorithm.
\end{itemize}
\end{restatable}

We remark that \cite{CyganEtAl2021} asked whether an
$\Ostar(2^{o(k)})$-time algorithm for \textsc{Minimum Bisection} could be
ruled out under ETH.  The ETH consequence in
\cref{thm:bisection-lower-bound} answers this question affirmatively.
However, OpenAI and Anthropic recently released independent manuscripts
claiming that ETH is false~\cite{OpenAI2026ETH,ClaudeAnthropic2026ETH}.
If either claim is correct, ETH no longer provides an algorithmic barrier.
We nevertheless believe that \cref{thm:bisection-lower-bound} retains the
spirit of the question in~\cite{CyganEtAl2021}, because the first and second
items transfer near-exponential lower bounds for \textsc{3-SAT} to
near-single-exponential lower bounds for \textsc{Minimum Bisection}, with the
exponent remaining arbitrarily close to linear.

\subsection{Related work}

\paragraph{Lean and unbreakable decompositions.}
A tree decomposition
$(\mathcal T,\{B_t\}_{t\in V(\mathcal T)})$ of a graph $G$ is \emph{lean} if, for all
$s,t\in V(\mathcal T)$ and all $Z_s\subseteq B_s$ and $Z_t\subseteq B_t$ with
$|Z_s|=|Z_t|=\ell$, the graph $G$ contains $\ell$ pairwise vertex-disjoint
$Z_s$--$Z_t$ paths unless there is an adhesion of size less than $\ell$ along
the $s$--$t$ path in $\mathcal T$.  Thomas~\cite{Thomas1990} introduced the
notion of \emph{lean tree decompositions} and proved that every graph $G$ admits
a lean tree decomposition of width $\operatorname{tw}(G)$.\footnote{We write
$\operatorname{tw}(G)$ for the treewidth of $G$.}  Bellenbaum and
Diestel~\cite{BellenbaumDiestel2002} gave a shorter proof of Thomas's theorem.
Carmesin, Diestel, Hamann, and Hundertmark~\cite{CarmesinEtAl2014} introduced
\emph{$k$-lean tree decompositions}, a variant of lean tree decompositions
that has adhesion at most $k$ and restricts the quantification over $\ell$ to
$\ell\le k+1$, and proved that every graph admits a $k$-lean tree decomposition
for every $k$.  Erde~\cite{Erde2018} developed a
common framework for lean-decomposition existence theorems in abstract
separation systems and proved that every matroid $M$ admits a lean tree
decomposition of width $\operatorname{tw}(M)$.

Cygan, Lokshtanov, Marcin Pilipczuk, Micha{\l} Pilipczuk, and
Saurabh~\cite{CyganEtAl2019} introduced the algorithmic notion of an
unbreakable decomposition as part of the first fixed-parameter algorithm for
\textsc{Minimum Bisection}.
Cygan, Komosa, Lokshtanov, Marcin Pilipczuk, Micha{\l} Pilipczuk, Saurabh, and
Wahlstr{\"o}m~\cite{CyganEtAl2021} subsequently gave an algorithm that computes
a tree decomposition with adhesion at most $k$ whose bags are
$(i,i)$-vertex-unbreakable for every $1\le i\le k$, and applied the
decomposition to \textsc{Minimum Bisection}, \textsc{Steiner Cut}, and
\textsc{Steiner Multicut}.

Lokshtanov, Saurabh, and Surianarayanan~\cite{LokshtanovEtAl2024} gave a
polynomial-time algorithm that computes a
$((k+1)^5,k)$-edge-unbreakable decomposition with adhesion at most $k$, which
they used in a parameterized approximation scheme for
\textsc{Minimum $p$-Way Cut}.  Inamdar, Lokshtanov, Saurabh, and
Surianarayanan~\cite{InamdarEtAl2023} obtained a
$(9k,k)$-edge-unbreakable decomposition with adhesion at most $8k$ and
logarithmic depth, and used it to obtain an FPT-approximation algorithm for
\textsc{Fair Bisection}.
For every $0<\epsilon\le1$, Anand, Lee, Li, Long, and
Saranurak~\cite{AnandEtAl2025SODA} give a randomized algorithm that computes with
high probability a $(\bigO(k/\epsilon),k)$-vertex-unbreakable decomposition
with adhesion $\bigO(k/\epsilon)$ in
$2^{\bigO((k/\epsilon)\log(k/\epsilon))}m^{1+\epsilon}$ time.  They use the
decomposition to obtain the first close-to-linear-time FPT algorithm for
\textsc{Minimum $p$-Way Cut}.
Korhonen~\cite{Korhonen2025} gave the first algorithm for constructing a
$k$-lean tree decomposition, which runs in $k^{\bigO(k^2)}(n+m)$ time.  The
decomposition is $(i,i)$-vertex-unbreakable for every $1\le i\le k$ and has
adhesion at most $k$.  The construction yields, for every fixed $k$, a
linear-time algorithm for computing the $k$-edge-connected components, as
well as algorithms for computing $k$-Gomory--Hu trees and testing
$k$-vertex-connectivity.  Li and
Vaughn~\cite{LiVaughn2026} gave a randomized near-linear-time construction of
an $(\bigO(k\log^2 n\log\log n),k)$-edge-unbreakable decomposition with
adhesion $\bigO(k\log^2 n\log\log n)$.  They use the decomposition to obtain
an improved FPT algorithm and a near-linear-time approximation algorithm for
\textsc{Minimum $p$-Way Cut}.
Most recently, Anand, Lee, Long, and Saranurak~\cite{AnandEtAl2027SODA} gave a
randomized algorithm that computes, with high probability, a
$(5k,k)$-vertex-unbreakable decomposition with adhesion $4k$ in
$\Ostar(2^{\bigO(k\log k)})$ time, with near-linear dependence on the input
size.
\Cref{tab:decomposition-comparison} summarizes existing results on lean,
vertex-unbreakable, and edge-unbreakable decompositions together with our
result in \cref{thm:decomp}.

\begin{table}[t]
\centering
\small
\setlength{\tabcolsep}{4pt}
\begin{tabularx}{\linewidth}{@{}
  >{\centering\arraybackslash}p{23mm}
  >{\centering\arraybackslash}p{30mm}
  >{\centering\arraybackslash}p{28mm}
  >{\centering\arraybackslash}X
  >{\centering\arraybackslash}p{23mm}@{}}
\toprule
Cut model
  & Small side in bag
  & Adhesion
  & Time
  & Reference \\
\midrule
\mbox{vertex (lean)}
  & $i\;(i\ge 1)$
  & $\operatorname{tw}(G)+1$
  & --
  & \mbox{\cite{Thomas1990,BellenbaumDiestel2002}} \\
\mbox{vertex ($k$-lean)}
  & $i\;(1\le i\le k)$
  & $k$
  & --
  & \cite{CarmesinEtAl2014} \\
\addlinespace
vertex
  & $2^{\bigO(k)}$
  & $2^{\bigO(k)}$
  & $\Ostar(2^{\bigO(k^2)})$
  & \cite{CyganEtAl2019} \\
vertex
  & $i\;(1\le i\le k)$
  & $k$
  & $\Ostar(2^{\bigO(k\log k)})$
  & \cite{CyganEtAl2021} \\
edge
  & $9k$
  & $8k$
  & $\Ostar(2^{\bigO(k\log k)})$
  & \cite{InamdarEtAl2023} \\
edge
  & $(k+1)^5$
  & $k$
  & $n^{\bigO(1)}$
  & \cite{LokshtanovEtAl2024} \\
vertex
  & $\bigO(k/\epsilon)$
  & $\bigO(k/\epsilon)$
  & $\Ostar(2^{\bigO((k/\epsilon)\log(k/\epsilon))})$
  & \cite{AnandEtAl2025SODA} \\
\mbox{vertex ($k$-lean)}
  & $i\;(1\le i\le k)$
  & $k$
  & $\Ostar(k^{\bigO(k^2)})$
  & \cite{Korhonen2025} \\
edge
  & {\scriptsize $\bigO(k\log^2 n\log\log n)$}
  & {\scriptsize $\bigO(k\log^2 n\log\log n)$}
  & $\widetilde{\bigO}(m+kn)$
  & \cite{LiVaughn2026} \\
vertex
  & $5k$
  & $4k$
  & $\Ostar(2^{\bigO(k\log k)})$
  & \cite{AnandEtAl2027SODA} \\
edge
  & $2k$
  & $2k$
  & $\Ostar(2^{\bigO(k)})$
  & \cref{thm:decomp} \\
\bottomrule
\end{tabularx}
\caption{Comparison of lean and unbreakable decomposition guarantees.
The results in the first two rows prove only existence and give no running-time
bounds.}
\label{tab:decomposition-comparison}
\end{table}

\paragraph{Algorithms for Minimum Bisection.}
Early algorithmic work on \textsc{Minimum Bisection} concerned
heuristics~\cite{KernighanLin1970,KarypisKumar1998} and average-case
guarantees~\cite{BuiEtAl1987}.
For approximation with worst-case guarantees, Krauthgamer and
Feige~\cite{KrauthgamerFeige2006} obtained approximation ratios
$\bigO(\log^{3/2} n)$ for \textsc{Minimum Bisection} on general graphs and
$\bigO(\log n)$ for every fixed-minor-free graph class.
R{\"a}cke~\cite{Racke2008} improved the approximation ratio for general
graphs to $\bigO(\log n)$.  Arora, Karger, and
Karpinski~\cite{AroraKargerKarpinski1999} gave a polynomial-time approximation
scheme (PTAS) for graphs with minimum degree linear in $n$.  Fox, Klein, and
Mozes~\cite{FoxKleinMozes2015} gave a bicriteria PTAS for weighted planar
instances.  Khot~\cite{Khot2006} ruled out a PTAS for \textsc{Minimum
Bisection} on general graphs under the assumption
$\mathsf{NP}\not\subseteq\bigcap_{\varepsilon>0}
\mathsf{BPTIME}(2^{n^\varepsilon})$.

For exact computation, Scott and Sorkin~\cite{ScottSorkin2009} formulated
\textsc{Minimum Bisection} as a polynomial-valued constraint-satisfaction
problem and gave a polynomial-space $\Ostar(2^{19m/100})$-time algorithm that
also supports counting.
Hanaka, Kobayashi, and Sone~\cite{HanakaKobayashiSone2021} gave an
$\bigO(2^t(tn)^2)$-time algorithm on a given tree decomposition of width $t$.
Cygan, Lokshtanov, Marcin Pilipczuk, Micha{\l} Pilipczuk, and
Saurabh~\cite{CyganEtAl2019} gave the first fixed-parameter algorithm for
\textsc{Minimum Bisection} parameterized by the number of cut edges, with
running time
$\Ostar(2^{\bigO(k^3)})$.  Cygan, Komosa, Lokshtanov, Marcin
Pilipczuk, Micha{\l} Pilipczuk, Saurabh, and
Wahlstr{\"o}m~\cite{CyganEtAl2021} improved the running time to
$\Ostar(2^{\bigO(k\log k)})$.

\paragraph{Algorithms for Matching Cut.}
Graham~\cite{Graham1970} introduced \emph{decomposable graphs}, namely graphs
admitting matching cuts, to solve a number-theoretic problem on cube numbering.
Subsequent work studied the complexity of \textsc{Matching Cut} on graph
classes such as bounded-degree graphs~\cite{Chvatal1984}, line
graphs~\cite{Moshi1989}, series-parallel graphs~\cite{PatrignaniPizzonia2001},
planar graphs, claw-free graphs, cographs, and graphs of bounded treewidth or
clique-width~\cite{Bonsma2009}, bounded-diameter graphs, both general and
bipartite~\cite{LeLe2019}, bounded-radius graphs and $P_6$-free
graphs~\cite{LuckePaulusmaRies2022}, and $P_{15}$-free
graphs~\cite{LuckePaulusmaRies2023}.

Exact and parameterized algorithms for general graphs appeared more recently.
Kratsch and Le~\cite{KratschLe2016} gave the first branching algorithm, which
runs in $\Ostar(2^{n/2})$ time, and a single-exponential algorithm
parameterized by the vertex-cover number.  Komusiewicz, Kratsch, and
Le~\cite{KomusiewiczKratschLe2020} gave a randomized exact algorithm running
in $\Ostar(1.3071^n)$ time, a quadratic kernel parameterized by distance to a
cluster graph, a linear kernel parameterized by distance to a clique, and
single-exponential algorithms parameterized by distance to a cluster graph
and distance to a co-cluster graph.  Gomes and Sau~\cite{GomesSau2019} proved
that \textsc{Matching Cut} is fixed-parameter tractable with respect to the
number $k$ of crossing edges by combining the line-graph reduction to
\textsc{Stable Cutset} with the fixed-parameter algorithm of Marx,
O'Sullivan, and Razgon~\cite{MarxOSullivanRazgon2010} for
\textsc{Stable Cutset} parameterized by the cutset size.  Aravind and
Saxena~\cite{AravindSaxena2021} gave an
$\Ostar(2^{\bigO(k\log k)})$-time algorithm.

\subsection{Technical overview}\label{sec:technical-overview}

\subsubsection{A single-exponential algorithm for
Minimum Bisection}\label{sec:technical-overview-bisection}

\paragraph{Dynamic programming.}
Recall that \cref{thm:decomp} constructs a rooted tree decomposition whose
adhesions have size at most $2k$ and whose bags are
$(2k,k)$-edge-unbreakable.  Fix such a decomposition $\mathcal T$.  For a node
$t$ of $\mathcal T$, let $\partial G_t$ be the adhesion between the bags
corresponding to $t$ and the parent of $t$, with
$\partial G_t:=\varnothing$ at the root.

We perform bottom-up dynamic programming on $\mathcal T$.  We use the standard
representation of a cut on an adhesion: for every
$P\subseteq\partial G_t$, we introduce a state indexed by $P$, which stores
information about cuts $A\subseteq V(G)$ satisfying
$A\cap\partial G_t=P$.  Since $\lvert\partial G_t\rvert\le 2k$, there are at
most $2^{2k}=2^{\bigO(k)}$ choices for $P$.  More precisely, the dynamic
programming keeps two additional scalar indices, which we suppress in this
overview: the number of vertices on the chosen side, needed for the bisection
constraint, and the number of cut edges.  They have only polynomially many
possible values and therefore increase the number of states only by a
polynomial factor.

In the usual dynamic programming on a tree decomposition of bounded width, a
single-exponential bound on the number of states typically yields a
single-exponential running time.  This is not the case here, because
$\mathcal T$ bounds only the adhesion size and its bags can be arbitrarily
large.  Hence we need to design the transitions carefully to obtain a
single-exponential running time.

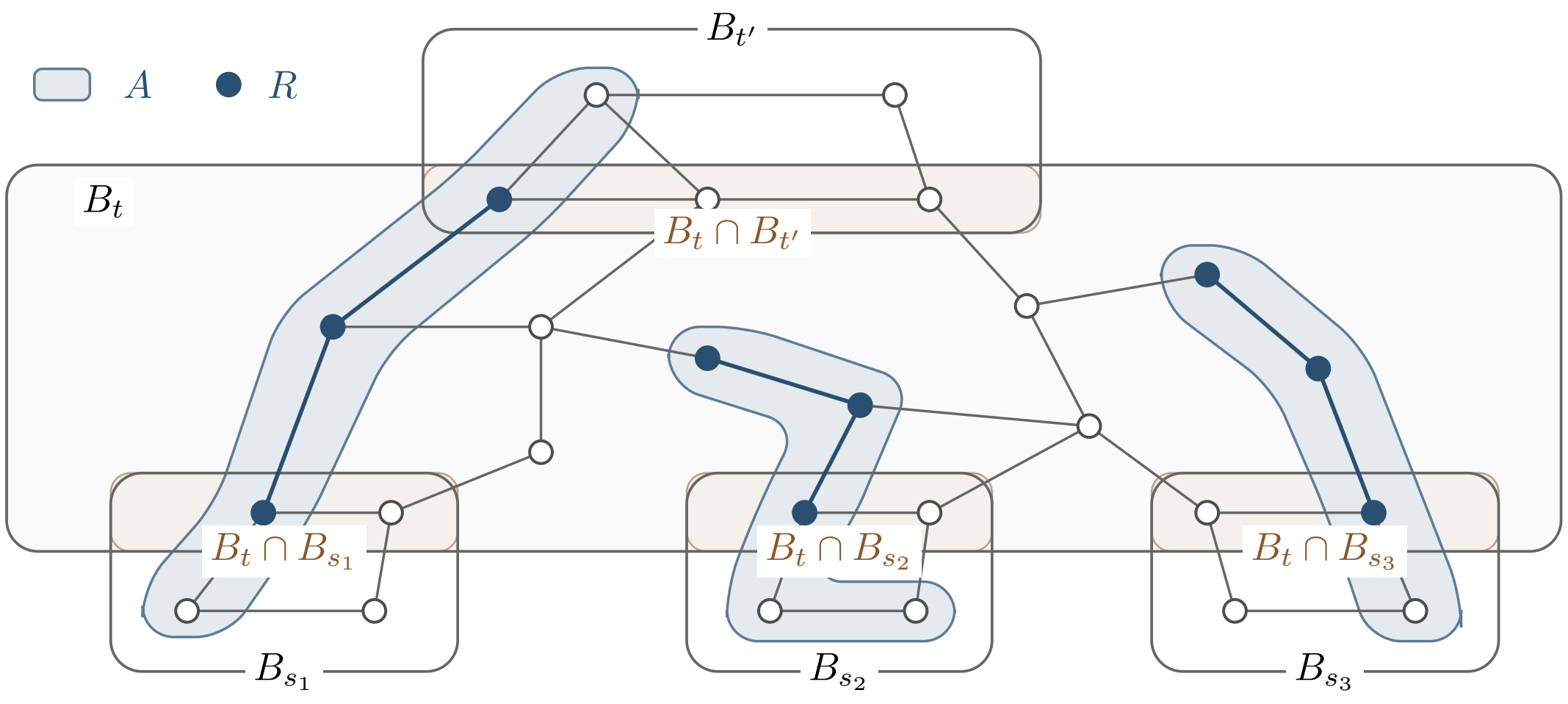
\begin{figure}[tb]
\centering
\begingroup
\definecolor{traceblue}{RGB}{42,79,113}
\definecolor{traceadh}{RGB}{137,87,43}
\begin{tikzpicture}[
  x=1.25cm,y=.94cm,font=\small,
  bag/.style={draw=black!60,line width=.7pt,rounded corners=8pt},
  adhesion/.style={fill=traceadh!9,draw=traceadh!55,
    line width=.5pt,rounded corners=5pt},
  component/.style={fill=traceblue!12,draw=traceblue!75,
    line width=.65pt,rounded corners=8pt},
  vertex/.style={circle,fill=white,draw=black!70,line width=.75pt,
    minimum size=5.8pt,inner sep=0pt},
  rvertex/.style={vertex,fill=traceblue,draw=traceblue},
  edge/.style={draw=black!60,line width=.65pt},
  set label/.style={fill=white,inner sep=2pt}]

  \path[fill=black!2,rounded corners=8pt]
    (.35,1.35) rectangle (11.55,5.05);
  \path[adhesion] (3.35,4.40) rectangle (7.80,5.05);
  \path[adhesion] (1.10,1.35) rectangle (3.60,2.10);
  \path[adhesion] (5.25,1.35) rectangle (7.45,2.10);
  \path[adhesion] (8.60,1.35) rectangle (11.10,2.10);

  \path[component]
    (4.32,5.98)--(4.90,5.98)--(4.90,5.48)--(4.22,4.49)
    --(3.10,3.27)--(2.53,1.65)--(1.94,.53)--(1.33,.53)
    --(1.33,1.02)--(1.86,1.82)--(2.32,3.63)--(3.60,4.98)--cycle;
  \path[component]
    (5.12,3.50)--(5.68,3.50)--(6.87,2.97)
    --(6.44,1.62)--(6.15,1.06)--(7.18,1.06)--(7.18,.49)
    --(5.54,.49)--(5.54,1.02)--(5.77,1.79)--(6.05,2.55)
    --(5.12,2.94)--cycle;
  \path[component]
    (8.67,4.28)--(9.25,4.28)--(10.13,3.30)
    --(10.83,.92)--(10.83,.49)--(10.17,.49)
    --(9.86,1.72)--(9.47,2.92)
    --(8.67,3.72)--cycle;

  \draw[bag] (.35,1.35) rectangle (11.55,5.05);
  \draw[bag] (3.35,4.40) rectangle (7.80,6.35);
  \draw[bag] (1.10,.20) rectangle (3.60,2.10);
  \draw[bag] (5.25,.20) rectangle (7.45,2.10);
  \draw[bag] (8.60,.20) rectangle (11.10,2.10);

  \coordinate (r1) at (3.90,4.72);
  \coordinate (r2) at (2.70,3.50);
  \coordinate (r3) at (2.20,1.72);
  \coordinate (r4) at (5.40,3.20);
  \coordinate (r5) at (6.50,2.75);
  \coordinate (r6) at (6.10,1.72);
  \coordinate (r7) at (9.00,4.00);
  \coordinate (r8) at (9.80,3.10);
  \coordinate (r9) at (10.20,1.72);
  \coordinate (v1) at (5.40,4.72);
  \coordinate (v2) at (7.00,4.72);
  \coordinate (v3) at (4.20,3.50);
  \coordinate (v4) at (7.70,3.70);
  \coordinate (v5) at (4.20,2.30);
  \coordinate (v6) at (8.15,2.55);
  \coordinate (v7) at (3.12,1.72);
  \coordinate (v8) at (7.00,1.72);
  \coordinate (v9) at (9.00,1.72);
  \coordinate (p1) at (4.60,5.72);
  \coordinate (p2) at (6.75,5.72);
  \coordinate (a1) at (1.65,.78);
  \coordinate (a2) at (3.00,.78);
  \coordinate (b1) at (5.85,.78);
  \coordinate (b2) at (6.90,.78);
  \coordinate (c1) at (9.20,.78);
  \coordinate (c2) at (10.50,.78);

  \draw[edge]
    (r1)--(v1)--(v2) (v1)--(v3)--(v5)--(v7)
    (v2)--(v4)--(v6)--(v8) (v6)--(v9)
    (r2)--(v3) (r4)--(v3) (r5)--(v6) (r7)--(v4)
    (r3)--(v7) (r6)--(v8) (r9)--(v9)
    (p1)--(r1) (p1)--(p2)--(v2) (p1)--(v1)
    (r3)--(a1)--(a2)--(v7)
    (r6)--(b1)--(b2)--(v8)
    (r9)--(c2)--(c1)--(v9);
  \draw[traceblue,line width=1.05pt]
    (r1)--(r2)--(r3) (r4)--(r5)--(r6) (r7)--(r8)--(r9);

  \foreach \v in {v1,v2,v3,v4,v5,v6,v7,v8,v9,p1,p2,a1,a2,b1,b2,c1,c2}
    \node[vertex] at (\v) {};
  \foreach \v in {r1,r2,r3,r4,r5,r6,r7,r8,r9}
    \node[rvertex] at (\v) {};

  \node[set label] at (1.05,4.70) {$B_t$};
  \node[set label] at (5.58,6.35) {$B_{t'}$};
  \node[set label,text=traceadh] at (5.58,4.40) {$B_t\cap B_{t'}$};
  \node[set label,text=traceadh] at (2.35,1.35) {$B_t\cap B_{s_1}$};
  \node[set label,text=traceadh] at (6.35,1.35) {$B_t\cap B_{s_2}$};
  \node[set label,text=traceadh] at (9.85,1.35) {$B_t\cap B_{s_3}$};
  \node[set label] at (2.35,.20) {$B_{s_1}$};
  \node[set label] at (6.35,.20) {$B_{s_2}$};
  \node[set label] at (9.85,.20) {$B_{s_3}$};

  \path[component,rounded corners=2pt] (.55,5.67) rectangle (.95,5.97);
  \node[anchor=west,text=traceblue] at (1.08,5.82) {$A$};
  \node[rvertex] at (1.95,5.82) {};
  \node[anchor=west,text=traceblue] at (2.12,5.82) {$R$};
\end{tikzpicture}
\endgroup
\caption{A bag $B_t$, its parent bag $B_{t'}$, and three child bags
  $B_{s_1},B_{s_2},B_{s_3}$.  The shaded regions enclose the three connected
  components of $G[A]$.  Blue vertices form $R=A\cap B_t$.}
\label{fig:bag-cut-trace}
\end{figure}

Consider computing the table at a node $t$ from the already computed tables of
its children, and let $B_t$ denote the bag corresponding to $t$.  A naive way
to compute the transition is to enumerate the possible sets $R:=A\cap B_t$ for
a hypothetical cut $A\subseteq V(G)$.  Once $R$ is fixed, every child $s$ must
use the state
indexed by $R\cap\partial G_s$.  Hence the contribution of cuts satisfying
$A\cap B_t=R$ can be computed from the child table entries indexed by
$R\cap\partial G_s$ and added to the entry at $t$ indexed by
$R\cap\partial G_t$.
\Cref{fig:bag-cut-trace} illustrates $A$, the set $R=A\cap B_t$, and the
intersections of $B_t$ with the adjacent bags.

We now analyze the number of possible sets $R$.  A straightforward approach
enumerates all $R\subseteq B_t$, which gives $2^{|B_t|}$ possibilities and can
be as large as $2^n$, which is far too large.  Instead of enumerating every
$R\subseteq B_t$, we observe that only cuts with at most $k$ cut edges need to
be considered, since a cut with more than $k$ cut edges cannot be a feasible
solution.  For every such cut $A$, the $(2k,k)$-edge-unbreakability of $B_t$ gives
$\min\{|A\cap B_t|,|B_t\setminus A|\}\le 2k$.  Hence the number of possible
sets $R$ can be reduced to
$2\sum_{i=0}^{2k}\binom{|B_t|}{i}=n^{\bigO(k)}$.  However, this is still too
large.

Our key observation is that, we can break this $n^{\bigO(k)}$ bound when we
only consider cuts $A$ for which $G[A]$ is connected.  More
precisely, in $\Ostar(2^{\bigO(k)})$ time we can compute a family
$\mathcal C_t\subseteq2^{B_t}$ of size $\Ostar(2^{\bigO(k)})$ such that
for every cut $A$ for which $G[A]$ is connected, $|\delta_G(A)|\le k$, and
$|A\cap B_t|\le2k$, the set $R=A\cap B_t$ belongs to $\mathcal C_t$.  To see
this, we use important cuts, introduced by Marx~\cite{Marx2006}.  For distinct
vertices $x,y$ in a graph $H$, a connected $x$--$y$ cut $C$ is important if
no connected $x$--$y$ cut $C'\supsetneq C$ satisfies
$|\delta_H(C')|\le|\delta_H(C)|$.\footnote{A cut $C$ of $H$ is a connected
$x$--$y$ cut if $x\in C$, $y\notin C$, and $H[C]$ is connected.}  It is known
that every connected
$x$--$y$ cut $C$ is contained in an important cut $D$ with
$|\delta_H(D)|\le|\delta_H(C)|$, and that there are at most $4^\ell$
important $x$--$y$ cuts with at most $\ell$ cut edges
(\Cref{lem:important-cuts}); see \cite[Section~8.2]{CyganEtAl2015} for further
details.

To construct $\mathcal C_t$, form $H_t$ from $G$ by adding a vertex $r_t$
adjacent to every vertex of $B_t$.  For each $v\in B_t$, enumerate the
important connected $v$--$r_t$ cuts $D$ with at most $3k$ cut edges, and add
every subset of $D\cap B_t$ to $\mathcal C_t$.  There are at most
$|B_t|4^{3k}$ choices for $D$.  Since $r_t\notin D$ and every vertex of
$D\cap B_t$ is adjacent to $r_t$, we have
$|D\cap B_t|\le|\delta_{H_t}(D)|\le3k$.  Thus $\mathcal C_t$ has size
$\Ostar(2^{\bigO(k)})$ and can be computed within the same time bound.  Now let
$A$ satisfy $G[A]$ connected, $|\delta_G(A)|\le k$, and
$0<|A\cap B_t|\le2k$, and choose $v\in A\cap B_t$.  In $H_t$, the set $A$ is
a connected $v$--$r_t$ cut with
$|\delta_{H_t}(A)|=|\delta_G(A)|+|A\cap B_t|\le3k$.  Since every connected
$v$--$r_t$ cut is contained in an important $v$--$r_t$ cut with no more cut
edges, there is an important $v$--$r_t$ cut $D$ with $A\subseteq D$ and
$|\delta_{H_t}(D)|\le|\delta_{H_t}(A)|\le3k$.  Hence $D$ is enumerated in the
construction of $\mathcal C_t$, and $A\cap B_t\subseteq D\cap B_t$ gives
$A\cap B_t\in\mathcal C_t$.

\paragraph{Formal logarithm.}
Before describing our use of the formal logarithm, we give a standard example
of how the formal logarithm recovers a generating function for connected
objects from one that counts arbitrary objects.  The example is not used in
our algorithm for \textsc{Minimum Bisection}: our algorithm uses a different
generating polynomial and proves connected support by another argument.  The
example only provides the underlying intuition.

Let $f_n=2^{\binom{n}{2}}$ be the number of graphs on the labeled vertex set
$[n]$, and let $g_n$ be the number of connected such graphs.  Every graph is
uniquely determined by the partition of its vertex set into connected
components and a connected graph on each part.  Therefore, for every $n\ge1$,
\[
  f_n
  =\sum_{\{B_1,\ldots,B_r\}}\prod_{j=1}^r g_{|B_j|}
  =n!\sum_{r=1}^n\left(
    \frac{1}{r!}
      \sum_{\substack{n_1+\cdots+n_r=n\\n_1,\ldots,n_r\ge1}}
      \prod_{j=1}^r\frac{g_{n_j}}{n_j!}
  \right),
\]
where the sum $\sum_{\{B_1,\ldots,B_r\}}$ ranges over all partitions
$\{B_1,\ldots,B_r\}$ of $[n]$.  The factor
$n!/(n_1!\cdots n_r!)$ assigns the labeled vertices to components of the
specified sizes, while $1/r!$ removes the ordering of the components.

Define the exponential generating functions
\[
  F(x):=\sum_{n\ge0}f_n\frac{x^n}{n!}
  ,\qquad
  G(x):=\sum_{n\ge1}g_n\frac{x^n}{n!}.
\]
Substituting the expression for $f_n$ and reordering the sums gives
\[
\begin{aligned}
  F(x)
  &=1+\sum_{n\ge1}\sum_{r=1}^n\left(
    \frac{1}{r!}
      \sum_{\substack{n_1+\cdots+n_r=n\\n_1,\ldots,n_r\ge1}}
      \prod_{j=1}^r\frac{g_{n_j}}{n_j!}
  \right)x^n \\
  &=1+\sum_{r\ge1}\frac{1}{r!}
    \sum_{n_1,\ldots,n_r\ge1}
      \prod_{j=1}^r g_{n_j}\frac{x^{n_j}}{n_j!}
   =\sum_{r\ge0}\frac{G(x)^r}{r!}.
\end{aligned}
\]
The series $\sum_{r\ge0}G(x)^r/r!$ is the Taylor series of the exponential
function evaluated formally at $G(x)$.  We write $\exp H:=\sum_{r\ge0}H^r/r!$ and
$\log H:=\sum_{r\ge1}(-1)^{r+1}(H-1)^r/r$ for the formal exponential and
logarithm, respectively.  Then the equality
$F(x)=\sum_{r\ge0}G(x)^r/r!$ gives $F(x)=\exp G(x)$.  Since the formal
exponential and logarithm are inverse to each other, it follows that
$G(x)=\log F(x)$.  Thus, the formal logarithm recovers the generating function
for connected labeled graphs from the generating function for all labeled
graphs.

Now we return to \textsc{Minimum Bisection}.  The construction of
$\mathcal C_t$ computes a family in $2^{B_t}$ of size
$\Ostar(2^{\bigO(k)})$ containing
$A\cap B_t$ for every cut $A$ such that $G[A]$ is connected,
$|\delta_G(A)|\le k$, and $|A\cap B_t|\le2k$.  To handle an arbitrary cut
$A$, we need to convolve the contributions of the connected components of
$G[A]$.  A direct convolution again considers the $n^{\bigO(k)}$ possible
unions $R=A\cap B_t$, so it does not yield a single-exponential running time.
We instead use a technique combining the formal logarithm and degree
truncation of the generating polynomial.

To focus on the algebraic intuition, we temporarily forget about dynamic
programming and consider the special case in which the entire vertex set
$V(G)$ forms a single $(2k,k)$-edge-unbreakable bag.  We also drop the
bisection constraint and instead study a generating polynomial that represents
every cut $(A,V(G)\setminus A)$.  Indeed, if $A\subseteq V(G)$ satisfies
$|\delta_G(A)|\le k$ and $|A|=|V(G)\setminus A|$, edge-unbreakability gives
$|A|\le2k$ and hence $|V(G)|\le4k$, but in this case, all
$2^{|V(G)|}\le2^{4k}$ choices of $A$ can be enumerated directly, so there is
nothing to do.

Introduce a variable $y_v$ for every vertex $v\in V(G)$, and write
$\boldsymbol y=(y_v)_{v\in V(G)}$.  For every $S\subseteq V(G)$, define
\[
  Z_S(\boldsymbol y,w)
  :=\sum_{A\subseteq S}
    \left(w^{|\delta_G(A)|}\cdot\prod_{v\in A}y_v\right).
\]
Every term of $Z_S(\boldsymbol y,w)$ represents a set $A\subseteq S$.
The vertex variables record $A$, and the exponent of $w$ records
$|\delta_G(A)|$.  The coefficient of $w^cy^a$ in the polynomial obtained from
$Z_{V(G)}(\boldsymbol y,w)$ by the substitution $y_v\mapsto y$ for
every $v\in V(G)$ equals the number of sets $A\subseteq V(G)$ with
$|\delta_G(A)|=c$ and $|A|=a$.  The goal here is to compute all such
coefficients.

We consider the formal logarithm of the polynomial
$Z_{V(G)}(\boldsymbol y,w)$, defined by
\begin{equation}\label{eq:overview-formal-logarithm}
  \log Z_{V(G)}(\boldsymbol y,w)
  :=\sum_{j\ge1}\frac{(-1)^{j+1}}{j}
    \bigl(Z_{V(G)}(\boldsymbol y,w)-1\bigr)^j.
\end{equation}

The key observation is that every monomial occurring in
$\log Z_{V(G)}(\boldsymbol y,w)$ has connected support, where
the \emph{support} of $w^c\prod_{v\in V(G)}y_v^{a_v}$ is defined as
$\{v\in V(G):a_v>0\}$.  To prove this, fix $S\subseteq V(G)$.  The
coefficient of a monomial with support exactly $S$
is the same in $\log Z_{V(G)}(\boldsymbol y,w)$ and
$\log Z_S(\boldsymbol y,w)$, since a product contributing to such a
monomial cannot use a variable $y_v$ with $v\notin S$.

Let $S_1,\ldots,S_r$ be the connected components of $G[S]$.  For
$A\subseteq S$, put $A_i:=A\cap S_i$.  No edge joins distinct sets $S_i$, so
$A$ is the disjoint union of $A_1,\ldots,A_r$ and $\delta_G(A)$ is the
disjoint union of $\delta_G(A_1),\ldots,\delta_G(A_r)$.  Consequently,
\[
  w^{|\delta_G(A)|}\prod_{v\in A}y_v
  =\prod_{i=1}^r
    \left(w^{|\delta_G(A_i)|}\prod_{v\in A_i}y_v\right).
\]
The sum of the monomial factorization over all choices $A_i\subseteq S_i$
gives
\[
  Z_S(\boldsymbol y,w)
  =\prod_{i=1}^r Z_{S_i}(\boldsymbol y,w).
\]
Taking formal logarithms turns the product into a sum:
\[
  \log Z_S(\boldsymbol y,w)
  =\sum_{i=1}^r\log Z_{S_i}(\boldsymbol y,w).
\]
Suppose that $G[S]$ is disconnected.  Then $r\ge2$ and
each $S_i$ is a proper subset of $S$.  Every monomial in
$\log Z_{S_i}(\boldsymbol y,w)$ has support contained in $S_i$, so the
coefficient of any monomial with support exactly $S$ is zero in
$\log Z_S(\boldsymbol y,w)$.  Since such coefficients are equal in
$\log Z_S(\boldsymbol y,w)$ and $\log Z_{V(G)}(\boldsymbol y,w)$, they are
also zero in $\log Z_{V(G)}(\boldsymbol y,w)$.  Hence every monomial with a nonzero
coefficient in $\log Z_{V(G)}(\boldsymbol y,w)$ has connected support.

Although every monomial in
$\log Z_{V(G)}(\boldsymbol y,w)$ has connected support, the series may
still have many distinct supports and, even worse, infinitely many terms with
the same support, so we cannot compute it directly.  To resolve this, we
truncate the series by discarding every
monomial whose $w$-degree exceeds $k$ or whose total $\boldsymbol y$-degree
exceeds $2k$.  The goal is to show that only
$\Ostar(2^{\bigO(k)})$ monomial supports remain after truncation.  To this end,
we use the family $\mathcal C_t$ constructed for the bag $B_t$.  In the
situation considered here, $B_t=V(G)$, so we denote $\mathcal C_t$ simply by
$\mathcal C$.  As described earlier, the family $\mathcal C$ can be computed in
$\Ostar(2^{\bigO(k)})$ time, has size
$\Ostar(2^{\bigO(k)})$, and contains every $A\subseteq V(G)$ such that $G[A]$
is connected, $|\delta_G(A)|\le k$, and $|A|\le2k$.

Consider a retained monomial $m$, and let $S$ be its support.  We show that
$S\in\mathcal C$.  We already know that $G[S]$ is connected, so it remains to
show that $|S|\le2k$ and $|\delta_G(S)|\le k$.  The truncation at total
$\boldsymbol y$-degree $2k$ gives
$|S|\le\sum_{v\in V(G)}\deg_{y_v}(m)\le2k$.
Moreover, every product that occurs in the expansion of a summand of
\cref{eq:overview-formal-logarithm} and contributes to $m$ counts every edge of
$\delta_G(S)$ at least once in its $w$-degree.  Hence
$|\delta_G(S)|\le\deg_w(m)\le k$.
Thus $S\in\mathcal C$.

Consequently, we can compute the truncated logarithm as
\[
  \left\langle\log Z_{V(G)}(\boldsymbol y,w)\right\rangle
  =\sum_{T\in\mathcal C}L_T(\boldsymbol y,w).
\]
Here $\langle F\rangle$ discards every monomial of $F$ whose $w$-degree
exceeds $k$ or whose total $\boldsymbol y$-degree exceeds $2k$, and
$L_T(\boldsymbol y,w)$ is the sum of the monomials in
$\left\langle\log Z_{V(G)}(\boldsymbol y,w)\right\rangle$ with support
exactly $T$.  It is not hard to see that we can compute
$L_T(\boldsymbol y,w)$ in
$\Ostar(2^{\bigO(k)})$ time using subset M\"obius inversion.  Since
$|\mathcal C|=\Ostar(2^{\bigO(k)})$, the truncated
logarithm is computable in $\Ostar(2^{\bigO(k)})$ time.

Recall that the goal here is to compute the coefficients of $w^cy^a$ for
$c\le k$ and $a\le2k$ in the polynomial obtained from
$Z_{V(G)}(\boldsymbol y,w)$ by substituting $y_v\mapsto y$ for every
$v\in V(G)$.  To this end, we first make this substitution in the truncated
logarithm.  We then take the formal exponential, defined by
\[
  \exp\left(\left\langle\log Z_{V(G)}(\boldsymbol y,w)\right\rangle\right)
  :=\sum_{j\ge0}\frac{1}{j!}
  \left\langle\log Z_{V(G)}(\boldsymbol y,w)\right\rangle^j.
\]
Using the truncated logarithm--exponential identity
\[
  \left\langle
    \exp\left(
      \left\langle\log Z_{V(G)}(\boldsymbol y,w)\right\rangle
    \right)
  \right\rangle
  =\left\langle Z_{V(G)}(\boldsymbol y,w)\right\rangle,
\]
we can compute all desired coefficients.
The substitution $y_v\mapsto y$ sends total $\boldsymbol y$-degree to
$y$-degree, so truncation does not discard any coefficient of $w^cy^a$ with
$c\le k$ and $a\le2k$.

We now return to the dynamic-programming transition at a large bag $B_t$.  In
addition to $B_t$, the transition must account for the vertices contained in
the child subtrees, so some additional care is needed.  Nevertheless, the
same technique as described above computes the transition in
$\Ostar(2^{\bigO(k)})$ time (\cref{sec:bisection}).  Hence the bottom-up
dynamic programming solves \textsc{Minimum Bisection} in
$\Ostar(2^{\bigO(k)})$ time.

\subsubsection{Unbreakable decomposition}
\label{sec:technical-overview-decomposition}

The remaining task is to construct a rooted $(2k,k)$-edge-unbreakable
decomposition with adhesions of size at most $2k$ in
$\Ostar(2^{\bigO(k)})$ time.  Our strategy is as follows.  If $V(G)$ is
$(2k,k)$-edge-unbreakable in $G$, we use $V(G)$ as the only bag.  Otherwise,
we find a set $B\subsetneq V(G)$ such that $B$ is
$(2k,k)$-edge-unbreakable in $G$ and has adhesion at most $2k$, use $B$ as the
root bag, and recurse on every component of $G-B$.

For this overview, we focus on constructing the root bag.  Choose an arbitrary vertex
$v\in V(G)$ that must remain in the root bag, and start with $B:=V(G)$.  If
$B$ is $(2k,k)$-edge-unbreakable, we stop.  Otherwise, we find a set $B'$ such
that $v\in B'$, $|B'|<|B|$, and $B'$ has adhesion at most $2k$, and replace
$B$ by $B'$.  Each replacement decreases the bag size, so the process
eventually produces the root bag.

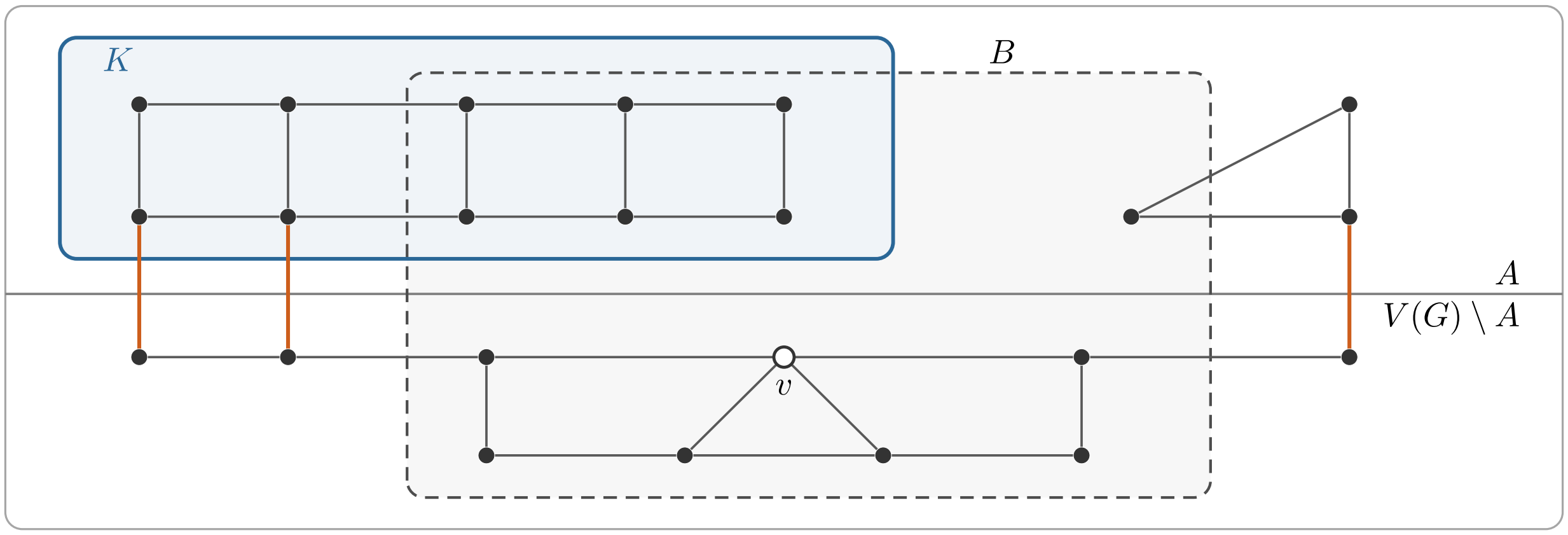
\begin{figure}[tb]
\centering
\begingroup
\definecolor{overviewblue}{RGB}{42,103,151}
\definecolor{overvieworange}{RGB}{205,94,28}
\pgfmathsetlengthmacro{\overviewunit}{\linewidth/16}
\begin{tikzpicture}[
  x=\overviewunit,y=.73cm,font=\small,
  edge/.style={draw=black!65,line width=.7pt},
  cut edge/.style={draw=overvieworange,line width=1.15pt},
  vertex/.style={circle,fill=black!80,draw=white,line width=.2pt,
    minimum size=4.9pt,inner sep=0pt}]
  \path[use as bounding box] (0,0) rectangle (16,7.75);

  \path[fill=black!3,rounded corners=5pt]
    (4.20,.60) rectangle (12.30,6.65);
  \path[fill=overviewblue!7,rounded corners=5pt]
    (.70,4.00) rectangle (9.10,7.15);
  \draw[draw=black!35,line width=.6pt,rounded corners=5pt]
    (.15,.15) rectangle (15.85,7.60);
  \draw[draw=black!50,line width=.7pt] (.15,3.50)--(15.85,3.50);
  \draw[draw=black!70,line width=.8pt,dash pattern=on 4pt off 2.6pt,rounded corners=5pt]
    (4.20,.60) rectangle (12.30,6.65);
  \draw[draw=overviewblue,line width=1.1pt,rounded corners=5pt]
    (.70,4.00) rectangle (9.10,7.15);

  \coordinate (a1) at (1.50,4.60);
  \coordinate (a2) at (1.50,6.20);
  \coordinate (a3) at (3.00,4.60);
  \coordinate (a4) at (3.00,6.20);
  \coordinate (b1) at (4.80,4.60);
  \coordinate (b2) at (4.80,6.20);
  \coordinate (b3) at (6.40,4.60);
  \coordinate (b4) at (6.40,6.20);
  \coordinate (b5) at (8.00,4.60);
  \coordinate (b6) at (8.00,6.20);
  \coordinate (d1) at (11.50,4.60);
  \coordinate (d2) at (13.70,4.60);
  \coordinate (d3) at (13.70,6.20);
  \coordinate (w1) at (1.50,2.60);
  \coordinate (w2) at (3.00,2.60);
  \coordinate (w3) at (13.70,2.60);
  \coordinate (s1) at (5.00,1.20);
  \coordinate (s2) at (7.00,1.20);
  \coordinate (s3) at (9.00,1.20);
  \coordinate (s4) at (11.00,1.20);
  \coordinate (s5) at (5.00,2.60);
  \coordinate (v) at (8.00,2.60);
  \coordinate (s7) at (11.00,2.60);

  \foreach \s/\v in {a1/a2,a1/a3,a2/a4,a3/a4,a3/b1,a4/b2,b1/b2,b1/b3,b2/b4,b3/b4,b3/b5,b4/b6,b5/b6,d1/d2,d2/d3,d3/d1,w1/w2,w2/s5,w3/s7,s1/s2,s2/s3,s3/s4,s5/v,v/s7,s1/s5,s4/s7,s2/v,s3/v}
    \draw[edge] (\s)--(\v);
  \foreach \s/\v in {a1/w1,a3/w2,d2/w3}
    \draw[cut edge] (\s)--(\v);

  \foreach \v in {a1,a2,a3,a4,b1,b2,b3,b4,b5,b6,d1,d2,d3,w1,w2,w3,s1,s2,s3,s4,s5,s7}
    \node[vertex] at (\v) {};
  \node[vertex,fill=white,draw=black!80,line width=.9pt,minimum size=6pt] at (v) {};

  \node[anchor=west,text=overviewblue] at (1.00,6.84) {$K$};
  \node at (10.20,6.95) {$B$};
  \node[anchor=east] at (15.55,3.80) {$A$};
  \node[anchor=east] at (15.55,3.16) {$V(G)\setminus A$};
  \node at (8.00,2.17) {$v$};
\end{tikzpicture}
\endgroup
\caption{A cut $A$ and a component $K$ of $G[A]$, for $k=3$.
The horizontal line separates $A$ from $V(G)\setminus A$, the dashed line
encloses the vertices of $B$, and the blue region marks $K$.
The three orange edges form $\delta_G(A)$.}
\label{fig:decomposition-overview}
\end{figure}

Suppose that $B$ is not $(2k,k)$-edge-unbreakable.  Then, there is a cut with at
most $k$ cut edges and more than $2k$ vertices of $B$ on each side.  We choose
the side $A$ that does not contain $v$.  The components of $G[A]$ together
contain more than $2k$ vertices of $B$, while their cut-edge sets contain at
most $k$ edges in total; hence some component $K$ satisfies
$|K\cap B|>|\delta_G(K)|$.  We can prove that adding at most
$|\delta_G(K)|$ vertices to $B\setminus K$ gives a set $B'$ that contains $v$
and has adhesion at most $2k$.  Since $|\delta_G(K)|<|K\cap B|$, we also have
$|B'|<|B|$.  \Cref{fig:decomposition-overview} illustrates $A$, $B$, and $K$.

The remaining task is to find such a component $K$.  Note that the witness
cut $A$ is not given, and finding it directly is already nontrivial.  We
instead use important cuts~\cite{Marx2006}.  We can prove that, for some choice
of $r\in B$, one of the important $r$--$v$ cuts with at
most $k$ cut edges yields a suitable $K$.  There are at most $4^k$ such cuts
for each $r$.  Thus we enumerate them for every $r\in B$ and test the bag
obtained from each cut.  Whenever $B$ is not $(2k,k)$-edge-unbreakable, at
least one enumerated cut yields a valid smaller bag.

For a nonroot recursive call, the new root bag must contain the adhesion to
its parent; otherwise, the bags containing an adhesion vertex may fail to form
a connected subtree.  To this end, instead of preserving the single vertex $v$, we
preserve a set $X$ consisting of the parent adhesion together with one vertex
of the current component.  Since the parent adhesion has size at most $2k$,
we have $|X|\le2k+1$.  The same shrinking procedure works with $X$ in place
of $\{v\}$, and recursion gives the complete decomposition.
\Cref{sec:structure} gives the details.  The running time is dominated by the
number of important cuts, which is $\Ostar(2^{\bigO(k)})$, and this proves
\cref{thm:decomp}.

\subsection{Organization}

The rest of this paper is organized as follows.
\Cref{sec:preliminaries} collects the preliminaries.
\Cref{sec:bisection} proves \cref{thm:bisection} by giving a
single-exponential algorithm for \textsc{Minimum Bisection}, assuming
\cref{thm:decomp}.
\Cref{sec:matching-cut} proves \cref{thm:matching-cut} by applying an
algorithm similar to that in \cref{sec:bisection} to \textsc{Matching Cut}.
\Cref{sec:structure} proves \cref{thm:decomp}.
\Cref{sec:lower-bound} proves \cref{thm:bisection-lower-bound}.

\section{Preliminaries}\label{sec:preliminaries}

\paragraph{Graphs.}

All graphs in this paper are finite, simple, and undirected.  A \emph{cut} of a graph $H$ is a set
$A\subseteq V(H)$.  Its \emph{cut-edge set} $\delta_H(A)$ consists of the
edges with exactly one endpoint in $A$.  The function
$A\mapsto|\delta_H(A)|$ is submodular: $|\delta_H(A)|+|\delta_H(B)|\ge
|\delta_H(A\cap B)|+|\delta_H(A\cup B)|$ for all $A,B\subseteq V(H)$.
A \emph{vertex separation} of $H$ is a pair $(L,R)$ with
$L\cup R=V(H)$ and no edge from $L\setminus R$ to $R\setminus L$.  Its
\emph{separator} is $L\cap R$.  For
vertex sets $S,T\subseteq V(H)$, an \emph{$S$--$T$ vertex separator} is a set
that intersects every path from $S$ to $T$; it may contain vertices of
$S\cup T$.
Menger's theorem states that the minimum size of an $S$--$T$ vertex separator
equals the maximum number of pairwise vertex-disjoint $S$--$T$ paths.  A
minimum $S$--$T$ vertex separator can be computed in polynomial time by vertex
splitting and a maximum-flow computation.

Let $S,T\subseteq V(H)$ be disjoint.  A \emph{connected $S$--$T$ cut} is a
cut $C$ of $H$ such that $S\subseteq C\subseteq V(H)\setminus T$ and $H[C]$
is connected.  It is \emph{important} if there is no connected $S$--$T$ cut
$D\supsetneq C$ with $|\delta_H(D)|\le |\delta_H(C)|$.
Every connected $S$--$T$ cut $C$ is contained in an
important connected $S$--$T$ cut $D$ with
$|\delta_H(D)|\le |\delta_H(C)|$.  The following lemma bounds the number of
important cuts.

\begin{lemma}[{\cite[Theorems~8.11 and~8.13]{CyganEtAl2015}}]
\label{lem:important-cuts}
Let $H$ be an unweighted undirected graph, let $S,T\subseteq V(H)$ be
disjoint, and suppose that $S$ is nonempty and connected.  For every
$\ell\ge0$, there are at most $4^\ell$ important connected $S$--$T$ cuts $D$ with
$|\delta_H(D)|\le\ell$.  All such cuts can be enumerated in
$\Ostar(4^\ell)$ time.
\end{lemma}

A \emph{tree decomposition} of $H$ is a tree $\mathcal T$ together with bags
$B_t\subseteq V(H)$ for $t\in V(\mathcal T)$ such that
$\bigcup_{t\in V(\mathcal T)}B_t=V(H)$, every edge of $H$ has both endpoints
in some bag, and, for every $v\in V(H)$, the nodes $t$ with $v\in B_t$ induce
a connected subtree of $\mathcal T$.  For $tt'\in E(\mathcal T)$, the set
$B_t\cap B_{t'}$ is the \emph{adhesion} of $tt'$; the \emph{adhesion} of the
decomposition is the maximum size of an adhesion, or zero if $\mathcal T$ has
no edge.  A \emph{rooted tree decomposition} additionally has a designated
root.

\paragraph{Perfect hash families.}

For $1\le r\le n$, an \emph{$(n,r)$-perfect hash family} is a family
$\mathcal H$ of functions $h:[n]\to[r]$ such that, for every
$X\subseteq[n]$ with $|X|\le r$, some $h\in\mathcal H$ is injective on $X$.
An $(n,r)$-perfect hash family of size $\Ostar(2^{\bigO(r)})$ can be
constructed in $\Ostar(2^{\bigO(r)})$ time~\cite{NaorSchulmanSrinivasan1995}.

\paragraph{Formal power series.}

Let $\mathcal X$ be a finite set of variables.  A formal power series over
$\mathcal X$ is a possibly infinite $\Q$-linear combination of monomials in
$\mathcal X$.  We write $\Q[[\mathcal X]]$ for the ring of such series.  For a
monomial $m$ and a variable $z$, let
$\deg_z(m)$ denote the exponent of $z$ in $m$.  For
$\boldsymbol u\subseteq\mathcal X$, every $F\in\Q[[\mathcal X]]$ has a unique
expansion $F=\sum_m F_m m$, where $m$ ranges over the monomials in
$\boldsymbol u$ and $F_m\in\Q[[\mathcal X\setminus\boldsymbol u]]$; define
$[m]_{\boldsymbol u}F:=F_m$.

For a series $F$ whose constant term is one, define
\begin{equation}\label{eq:fps-log-exp}
  \log F:=\sum_{j\ge1}\frac{(-1)^{j+1}}j(F-1)^j.
\end{equation}
For a series $F$ whose constant term is zero, define
\begin{equation}\label{eq:fps-exp}
  \exp F:=\sum_{j\ge0}\frac1{j!}F^j.
\end{equation}
The formal logarithm and exponential are inverse to each other.

For a series $F$ and a nonnegative degree bound $D$, let
$\left\langle F\right\rangle_{\deg\le D}$ denote the
series obtained by discarding the monomials of degree greater than $D$.

\begin{lemma}\label{lem:truncated-log-exp}
If the formal logarithm of $Z$ is defined, then
\[
  \left\langle
    \exp\left(\left\langle\log Z\right\rangle_{\deg\le D}\right)
  \right\rangle_{\deg\le D}
  =\left\langle Z\right\rangle_{\deg\le D}.
\]
\end{lemma}

\begin{proof}
Every discarded monomial has degree greater than $D$, and multiplying it by
any monomial cannot decrease its degree.  Hence it cannot affect any retained
coefficient in the exponential.  The claim follows from
$\exp(\log Z)=Z$.
\end{proof}

\paragraph{Zeta and M\"obius transforms.}
For a finite set $Y$ and a function
$f:2^Y\to\mathbb Q$, its \emph{subset zeta transform} is the
function $g:2^Y\to\mathbb Q$ defined by
$g(T):=\sum_{J\subseteq T}f(J)$.  The inverse operation is the
\emph{subset M\"obius transform}, which recovers $f$ by
$f(S)=\sum_{T\subseteq S}(-1)^{|S|-|T|}g(T)$.

\section{Minimum Bisection}\label{sec:bisection}

In this section, we prove \cref{thm:bisection}.  We use the guarantee of
\cref{thm:decomp}, leaving the decomposition construction to
\cref{sec:structure}.

\subsection{Polynomial representation}\label{sec:bisection-state}

Fix a rooted tree decomposition $\mathcal T$ with bags $B_t$ supplied by
\cref{thm:decomp}, and denote its root by $t_0$.  For a nonroot node $t$ with
parent $t'$, let $G_t$ be the graph obtained from the subgraph of $G$ induced
by the union of the bags in the subtree rooted at $t$ by deleting
$E(G[B_t\cap B_{t'}])$; at the root, set $G_{t_0}:=G$.  We call
$\partial G_t:=B_t\cap B_{t'}$ the \emph{parent boundary} of $G_t$, and set
$\partial G_{t_0}:=\varnothing$.  For a nonroot node $t$, deleting
$E(G[B_t\cap B_{t'}])$ prevents an edge with both endpoints in $\partial G_t$
from being charged at both $t$ and $t'$.

For $A_t\subseteq V(G_t)$, define the minimum completion cost of $A_t$ by
\[
  \mu_t(A_t):=\min\{|\delta_G(A)|:A\cap V(G_t)=A_t\}.
\]
In other words, $\mu_t(A_t)$ is the smallest number of cut edges obtainable
when $A_t$ is fixed inside $V(G_t)$ and the vertices outside $V(G_t)$ may be
chosen freely.
Call $A_t$ \emph{$k$-extendable} if $\mu_t(A_t)\le k$.

Our strategy is dynamic programming on $\mathcal T$.  A natural way is to
keep, for every node $t$, states indexed by the following data:
\begin{itemize}
  \item a set $P\subseteq\partial G_t$, which specifies
        $A_t\cap\partial G_t$;
  \item an integer $b$, which specifies $|A_t\setminus\partial G_t|$;
  \item an integer $c\in\{0,\ldots,k\}$, which specifies $\mu_t(A_t)$.
\end{itemize}
The state $(P,b,c)$ corresponds to the sets $A_t\subseteq V(G_t)$ satisfying
$A_t\cap\partial G_t=P$, $|A_t\setminus\partial G_t|=b$, and
$\mu_t(A_t)=c$.  The answer to the \textsc{Minimum Bisection} instance can be
recovered from the state values at the root $t_0$.

We next decide what each state should store.  The most natural way is to store
a Boolean value indicating whether any such set $A_t$ exists.  However,
Boolean state values are insufficient even if the goal is only to decide
whether a solution exists: our algorithm applies algebraic operations, such
as M\"obius inversion and the formal logarithm, to the state values, which is
impossible with existence information alone.
We therefore store the number of sets $A_t$ represented by each state.  As our
algorithm applies algebraic operations to these counts, it is convenient to
encode them in a generating polynomial as follows.

Let $\boldsymbol y=(y_v)_{v\in V(G_t)}$.  Define
\[
  Z_t(\boldsymbol y,w)
  :=\sum_{\substack{A_t\subseteq V(G_t)\\\mu_t(A_t)\le k}}
    \left(w^{\mu_t(A_t)}\cdot\prod_{v\in A_t}y_v\right)
\]
The polynomial $Z_t$ lists all $k$-extendable subsets of $V(G_t)$.  For the
monomial indexed by $A_t$, the exponent of $w$ records $\mu_t(A_t)$.

In our dynamic-programming states, the vertices in
$A_t\setminus\partial G_t$ are not distinguished; each state records only
$|A_t\setminus\partial G_t|=b$.  In contrast, the monomial indexed by $A_t$ in
$Z_t$ contains $\prod_{v\in A_t}y_v$, which distinguishes every vertex of $A_t$.
To use $Z_t$ for dynamic programming, we must forget the identities of the
vertices in $A_t\setminus\partial G_t$.  Throughout the algorithm, we implement
such a forgetting operation by variable substitution.  For $X\subseteq V(G_t)$,
define the substitution
$\boldsymbol\omega_t^X(\boldsymbol y,x):=(\omega_{t,v}^X)_{v\in V(G_t)}$,
where, for every $v\in V(G_t)$,
\[
  \omega_{t,v}^X:=
  \begin{cases}
    y_v,&v\in X,\\
    x,&v\in V(G_t)\setminus X.
  \end{cases}
\]
We write $Z_t(\boldsymbol\omega_t^X(\boldsymbol y,x),w)$ for the polynomial
obtained by applying $\boldsymbol\omega_t^X(\boldsymbol y,x)$ to
$Z_t(\boldsymbol y,w)$.  Precisely, we have
\[
  Z_t\bigl(\boldsymbol\omega_t^{\partial G_t}(\boldsymbol y,x),w\bigr)
  =\sum_{\substack{A_t\subseteq V(G_t)\\ \mu_t(A_t)\le k}}
    \left(
      x^{|A_t\setminus\partial G_t|}
      \cdot\left(\prod_{v\in A_t\cap\partial G_t}y_v\right)
      \cdot w^{\mu_t(A_t)}
    \right).
\]
Grouping the terms of
$Z_t\bigl(\boldsymbol\omega_t^{\partial G_t}(\boldsymbol y,x),w\bigr)$ by
$P=A_t\cap\partial G_t$ gives
\[
  Z_t\bigl(\boldsymbol\omega_t^{\partial G_t}(\boldsymbol y,x),w\bigr)
  =\sum_{P\subseteq\partial G_t}
    \left(
      \left(\prod_{v\in P}y_v\right)
      \cdot
      \sum_{\substack{A_t\subseteq V(G_t)\\
          \mu_t(A_t)\le k,\ A_t\cap\partial G_t=P}}
        \left(x^{|A_t\setminus\partial G_t|}
          \cdot w^{\mu_t(A_t)}\right)
    \right).
\]
For $P\subseteq\partial G_t$,
$0\le b\le|V(G_t)\setminus\partial G_t|$, and $0\le c\le k$, taking the
coefficient of $x^bw^c\prod_{v\in P}y_v$ gives
\[
\begin{aligned}
  &\left[x^bw^c\prod_{v\in P}y_v\right]
    Z_t\bigl(\boldsymbol\omega_t^{\partial G_t}(\boldsymbol y,x),w\bigr)\\
  &\qquad={}
    \left|\left\{A_t\subseteq V(G_t):
      A_t\cap\partial G_t=P,\quad
      |A_t\setminus\partial G_t|=b,\quad
      \mu_t(A_t)=c
    \right\}\right|,
\end{aligned}
\]
which is exactly the number of sets represented by the dynamic-programming
state $(P,b,c)$.

Our algorithm computes
$Z_t(\boldsymbol\omega_t^{\partial G_t}(\boldsymbol y,x),w)$ for every node
$t$.
Since $0\le b\le n$, $0\le c\le k$, and $|\partial G_t|\le2k$, the polynomial
$Z_t(\boldsymbol\omega_t^{\partial G_t}(\boldsymbol y,x),w)$ contains only
$O^*(2^{|\partial G_t|})=O^*(2^{O(k)})$ monomials.

To recover the answer to the \textsc{Minimum Bisection} instance, consider the
root $t_0$.  Since $\partial G_{t_0}=\varnothing$,
$\boldsymbol\omega_{t_0}^{\partial G_{t_0}}(\boldsymbol y,x)$ replaces every
vertex variable by $x$.  Moreover, $G_{t_0}=G$ and
$\mu_{t_0}(A)=|\delta_G(A)|$, so
\begin{equation}\label{eq:root-generating-polynomial}
  Z_{t_0}\bigl(
    \boldsymbol\omega_{t_0}^{\partial G_{t_0}}(\boldsymbol y,x),w
  \bigr)
  =\sum_{\substack{A\subseteq V(G)\\|\delta_G(A)|\le k}}
    x^{|A|}w^{|\delta_G(A)|}.
\end{equation}
The number of feasible solutions to the \textsc{Minimum Bisection} instance is
\[
  \sum_{c=0}^k
    [x^{\lfloor n/2\rfloor}w^c]
    Z_{t_0}\bigl(
      \boldsymbol\omega_{t_0}^{\partial G_{t_0}}(\boldsymbol y,x),w
    \bigr).
\]

\subsection{Conceptual transition}
\label{sec:bisection-transition}

We now consider how to compute
$Z_t(\boldsymbol\omega_t^{\partial G_t}(\boldsymbol y,x),w)$ from the polynomials
$Z_s(\boldsymbol\omega_s^{\partial G_s}(\boldsymbol y,x),w)$ for the children
$s$ of $t$.  We first consider a conceptually simple strategy.  We enumerate
every $R\subseteq B_t$, compute the contribution to
$Z_t(\boldsymbol\omega_t^{\partial G_t}(\boldsymbol y,x),w)$ of the sets $A_t$
satisfying $A_t\cap B_t=R$, and sum the contributions over all
$R\subseteq B_t$.

Here, we should first remark that the strategy is only conceptual; our
algorithm does not compute
$Z_t(\boldsymbol\omega_t^{\partial G_t}(\boldsymbol y,x),w)$ by enumerating
$R\subseteq B_t$ because the strategy requires enumerating all $2^{|B_t|}$
subsets $R\subseteq B_t$, which cannot be done within the desired
$\Ostar(2^{\bigO(k)})$ time.  Nevertheless, the strategy is useful for
explaining the $\Ostar(2^{\bigO(k)})$-time algorithm, so we describe the
strategy first.

Formally, $Z_t(\boldsymbol\omega_t^{\partial G_t}(\boldsymbol y,x),w)$ is
represented as the following sum:
\begin{align}
  Z_t\bigl(\boldsymbol\omega_t^{\partial G_t}(\boldsymbol y,x),w\bigr)
  &=\sum_{\substack{A_t\subseteq V(G_t)\\ \mu_t(A_t)\le k}}
    \left(
      x^{|A_t\setminus\partial G_t|}
      \cdot\left(\prod_{v\in A_t\cap\partial G_t}y_v\right)
      \cdot w^{\mu_t(A_t)}
    \right)
  \notag\\
  &=\sum_{R\subseteq B_t}
    \sum_{\substack{A_t\subseteq V(G_t)\\
        \mu_t(A_t)\le k,\ A_t\cap B_t=R}}
    \left(
      x^{|R\setminus\partial G_t|}
      \cdot x^{|A_t\setminus B_t|}
      \cdot\left(\prod_{v\in R\cap\partial G_t}y_v\right)
      \cdot w^{\mu_t(A_t)}
    \right)
  \notag\\
  &=\sum_{R\subseteq B_t}
    \left(
      x^{|R\setminus\partial G_t|}
      \cdot\left(\prod_{v\in R\cap\partial G_t}y_v\right)
      \cdot
      \underbrace{
        \sum_{\substack{A_t\subseteq V(G_t)\\
            \mu_t(A_t)\le k,\ A_t\cap B_t=R}}
          \left(x^{|A_t\setminus B_t|}\cdot w^{\mu_t(A_t)}\right)
      }_{=:Q_{t,R}(x,w)}
    \right).
  \label{eq:bag-to-boundary}
\end{align}

Now, the next task is to represent $Q_{t,R}(x,w)$ using the child polynomials
$Z_s(\boldsymbol\omega_s^{\partial G_s}(\boldsymbol y,x),w)$.  Let
$\operatorname{child}(t)$ denote the set of children of $t$ in $\mathcal T$.
Fix $R\subseteq B_t$.  For a child $s$ of $t$, call a set
$A_s\subseteq V(G_s)$ \emph{$R$-consistent} if
$A_s\cap\partial G_s=R\cap\partial G_s$.  By the connectedness axiom of the
tree decomposition $\mathcal T$, every set $A_t\subseteq V(G_t)$ with
$A_t\cap B_t=R$ is obtained uniquely by gluing $R$ with one $R$-consistent
set $A_s$ for each child $s$ of $t$ in the form
\[
  A_t=R\mathbin{\dot\cup}
      \mathop{\dot\bigcup}_{s\in\operatorname{child}(t)}
        (A_s\setminus\partial G_s).
\]
Consequently, $Q_{t,R}(x,w)$ can be written entirely in terms of the sets
$A_s$ as
\begin{equation}\label{eq:child-set-sum}
  Q_{t,R}(x,w)
  =\sum
    \left(
      \left(
        \prod_{s\in\operatorname{child}(t)}
          x^{|A_s\setminus\partial G_s|}
      \right)
      \cdot
      w^{\mu_t\left(
        R\mathbin{\dot\cup}
        \mathop{\dot\bigcup}_{s\in\operatorname{child}(t)}
          (A_s\setminus\partial G_s)
      \right)}
    \right).
\end{equation}
The sum is taken over all choices of an $R$-consistent set
$A_s\subseteq V(G_s)$ for every child $s$ of $t$ that satisfy
\[
  \mu_t\left(
    R\mathbin{\dot\cup}
    \mathop{\dot\bigcup}_{s\in\operatorname{child}(t)}
      (A_s\setminus\partial G_s)
  \right)\le k.
\]
We would like to rewrite the sum in \cref{eq:child-set-sum} as a product, over
the children $s$ of $t$, of sums indexed by $R$-consistent sets $A_s$.  The
$x$-weight already factors over the children, but the exponent of $w$ is not
additive over them.  To do so, we correct the $w$-degree.  For
$P\subseteq\partial G_t$, define
\[
  \eta_t(P):=\min\bigl\{
    |\delta_{G-(V(G_t)\setminus\partial G_t)}(A)|:
    A\subseteq V(G)\setminus(V(G_t)\setminus\partial G_t),
    \ A\cap\partial G_t=P
  \bigr\}.
\]
The quantity $\eta_t(P)$ is the minimum cut contribution outside $G_t$
compatible with the boundary choice $P$.
Thus, if $A_t\cap\partial G_t=P$, then
\begin{equation}\label{eq:completion}
  \mu_t(A_t)=|\delta_{G_t}(A_t)|+\eta_t(P).
\end{equation}
The next lemma corrects the $w$-degrees and gives an exact recurrence for
$Q_{t,R}(x,w)$.

\begin{lemma}\label{lem:projected-series}
For every $R\subseteq B_t$, the polynomial
$Q_{t,R}(x,w)$ satisfies
\begin{equation}\label{eq:fixed-pattern-recurrence}
\begin{aligned}
  Q_{t,R}(x,w)
  ={}&\mathopen{\left\langle
      \vphantom{\displaystyle\prod_{s\in\operatorname{child}(t)}}
      \right.}
      w^{|\delta_{G_t[B_t]}(R)|+\eta_t(R\cap\partial G_t)}\\[-1mm]
    &{}\cdot
      \prod_{s\in\operatorname{child}(t)}
      \left(
        w^{-\eta_s(R\cap\partial G_s)}
        \cdot
        \left[\prod_{v\in R\cap\partial G_s}y_v\right]_{\boldsymbol y}
        Z_s\bigl(\boldsymbol\omega_s^{\partial G_s}(\boldsymbol y,x),w\bigr)
      \right)
      \mathclose{\left.
      \vphantom{\displaystyle\prod_{s\in\operatorname{child}(t)}}
      \right\rangle}_{\deg_w\le k}.
\end{aligned}
\end{equation}
\end{lemma}

\begin{proof}
Fix $R\subseteq B_t$.  For every child $s$ of $t$, the coefficient of the
boundary monomial indexed by $R\cap\partial G_s$ is
\[
  \left[\prod_{v\in R\cap\partial G_s}y_v\right]_{\boldsymbol y}
  Z_s\bigl(\boldsymbol\omega_s^{\partial G_s}(\boldsymbol y,x),w\bigr)
  =\sum_{\substack{A_s\subseteq V(G_s)\\
      \mu_s(A_s)\le k,\ A_s\cap\partial G_s=R\cap\partial G_s}}
    \left(x^{|A_s\setminus\partial G_s|}\cdot w^{\mu_s(A_s)}\right).
\]
Thus, expanding the product in \cref{eq:fixed-pattern-recurrence} amounts to
choosing one $R$-consistent set $A_s$ for every child $s$ of $t$.  For these
choices, let
\[
  A_t:=R\mathbin{\dot\cup}
    \mathop{\dot\bigcup}_{s\in\operatorname{child}(t)}
      (A_s\setminus\partial G_s).
\]
By \cref{eq:child-set-sum}, it suffices to verify that the right-hand side of
\cref{eq:fixed-pattern-recurrence} assigns $A_t$ the monomial
$x^{|A_t\setminus B_t|}w^{\mu_t(A_t)}$ if $\mu_t(A_t)\le k$, and zero
otherwise.  For the $x$-weight,
\[
  \prod_{s\in\operatorname{child}(t)}
    x^{|A_s\setminus\partial G_s|}
  =x^{|A_t\setminus B_t|}.
\]

For the $w$-weight, the edge sets of $G_t[B_t]$ and of the
graphs $G_s$ for the children $s$ of $t$ partition $E(G_t)$.  Hence
\cref{eq:completion}, applied at $t$ and at every child $s$, gives
\begin{align*}
  &\mu_t\left(
    R\mathbin{\dot\cup}
    \mathop{\dot\bigcup}_{s\in\operatorname{child}(t)}
      (A_s\setminus\partial G_s)
  \right)\notag\\
  &\quad={}
    |\delta_{G_t[B_t]}(R)|
    +\sum_{s\in\operatorname{child}(t)}|\delta_{G_s}(A_s)|
    +\eta_t(R\cap\partial G_t)\notag\\
  &\quad={}
    |\delta_{G_t[B_t]}(R)|
    +\eta_t(R\cap\partial G_t)
    +\sum_{s\in\operatorname{child}(t)}
      \bigl(\mu_s(A_s)-\eta_s(R\cap\partial G_s)\bigr).
\end{align*}
Consequently, the selected terms contribute the required monomial
$x^{|A_t\setminus B_t|}w^{\mu_t(A_t)}$.  The truncation in
\cref{eq:fixed-pattern-recurrence} keeps exactly the choices with
$\mu_t(A_t)\le k$.

Conversely, every set $A_t$ with $\mu_t(A_t)\le k$ is obtained from the sets
$A_s=A_t\cap V(G_s)$, which satisfy $\mu_s(A_s)\le k$ for every child $s$
because a minimum completion of $A_t$ is also a completion of every $A_s$.
Thus,
\cref{eq:child-set-sum} shows that the retained sum is $Q_{t,R}(x,w)$.
\end{proof}

If $|B_t|\le4k+1$, we compute $Q_{t,R}(x,w)$ from
\cref{eq:fixed-pattern-recurrence} for every $R\subseteq B_t$ and then compute
$Z_t(\boldsymbol\omega_t^{\partial G_t}(\boldsymbol y,x),w)$ from
\cref{eq:bag-to-boundary}.  The computation takes
$\Ostar(2^{\bigO(k)})$ time.

We now assume that $|B_t|>4k+1$.  Enumerating all $2^{|B_t|}$ subsets
$R\subseteq B_t$ is too expensive, so we need a different algorithm.  By the
edge-unbreakability guarantee for the bags of our tree decomposition, every
$k$-extendable set $A_t$ satisfies
$|A_t\cap B_t|\le2k$ or $|B_t\setminus A_t|\le2k$.  Since
$|B_t|>4k+1$, the two inequalities are mutually exclusive.  We compute the
two contributions separately and add them.  Define
\[
  Z_t^{\le2k}(\boldsymbol y,w)
  :=\sum_{\substack{A_t\subseteq V(G_t),\ \mu_t(A_t)\le k\\
                     |A_t\cap B_t|\le2k}}
    \left(w^{\mu_t(A_t)}\cdot\prod_{v\in A_t}y_v\right).
\]
$Z_t^{\le2k}(\boldsymbol y,w)$ is the restriction of
$Z_t(\boldsymbol y,w)$ to the sets $A_t$ satisfying
$|A_t\cap B_t|\le2k$.
The next lemma represents the restriction of $Z_t(\boldsymbol y,w)$ to the
sets $A_t$ satisfying $|B_t\setminus A_t|\le2k$ in terms of
$Z_t^{\le2k}(\boldsymbol y,w)$, and hence expresses
$Z_t(\boldsymbol y,w)$ as the sum of the two restrictions.

\begin{lemma}\label{lem:large-bag-message}
Let $\boldsymbol y^{-1}:=(y_v^{-1})_{v\in V(G_t)}$.  Then
\[
  Z_t(\boldsymbol y,w)
  =Z_t^{\le2k}(\boldsymbol y,w)
   +\left(\prod_{v\in V(G_t)}y_v\right)
    Z_t^{\le2k}(\boldsymbol y^{-1},w).
\]
\end{lemma}

\begin{proof}
Edge-unbreakability and $|B_t|>4k+1$ imply that every set $A_t$ with
$\mu_t(A_t)\le k$ satisfies exactly one of $|A_t\cap B_t|\le2k$ and
$|B_t\setminus A_t|\le2k$.  The polynomial
$Z_t^{\le2k}(\boldsymbol y,w)$ contains exactly the terms indexed by the sets
satisfying $|A_t\cap B_t|\le2k$.

Fix a set $A_t\subseteq V(G_t)$ satisfying $\mu_t(A_t)\le k$ and
$|B_t\setminus A_t|\le2k$.  Since
$(V(G_t)\setminus A_t)\cap B_t=B_t\setminus A_t$ and
$\mu_t(V(G_t)\setminus A_t)=\mu_t(A_t)$, we have
\[
  \left(\prod_{v\in V(G_t)}y_v\right)
  \left(w^{\mu_t(A_t)}
        \prod_{v\in V(G_t)\setminus A_t}y_v^{-1}\right)
  =w^{\mu_t(A_t)}\prod_{v\in A_t}y_v.
\]
Summing this identity over all such $A_t$ gives
\[
  \left(\prod_{v\in V(G_t)}y_v\right)
  Z_t^{\le2k}(\boldsymbol y^{-1},w)
  =\sum_{\substack{A_t\subseteq V(G_t),\ \mu_t(A_t)\le k\\
                    |B_t\setminus A_t|\le2k}}
     w^{\mu_t(A_t)}\prod_{v\in A_t}y_v.
\]
Together with the terms satisfying $|A_t\cap B_t|\le2k$, this proves the
lemma.
\end{proof}

Applying \cref{lem:large-bag-message} after the substitution
$\boldsymbol\omega_t^{\partial G_t}(\boldsymbol y,x)$ gives
\[
\begin{aligned}
  Z_t\bigl(\boldsymbol\omega_t^{\partial G_t}(\boldsymbol y,x),w\bigr)
  ={}&Z_t^{\le2k}
    \bigl(\boldsymbol\omega_t^{\partial G_t}(\boldsymbol y,x),w\bigr)\\
   &+x^{|V(G_t)\setminus\partial G_t|}
     \left(\prod_{v\in\partial G_t}y_v\right)
     Z_t^{\le2k}
     \bigl(\boldsymbol\omega_t^{\partial G_t}
       (\boldsymbol y^{-1},x^{-1}),w\bigr).
\end{aligned}
\]
The right-hand side can be computed directly from the coefficients of
$Z_t^{\le2k}(\boldsymbol\omega_t^{\partial G_t}(\boldsymbol y,x),w)$.
Thus, we now focus on computing
$Z_t^{\le2k}(\boldsymbol\omega_t^{\partial G_t}(\boldsymbol y,x),w)$.

\subsection{Large-bag transition via formal logarithms}
\label{sec:computing-labelled-series}

By the previous subsection, it remains to compute
$Z_t^{\le2k}(\boldsymbol\omega_t^{\partial G_t}(\boldsymbol y,x),w)$.
Consider the substitution
$\boldsymbol\omega_t^{B_t}(\boldsymbol y,x)$.  By definition,
\[
  \omega_{t,v}^{B_t}=
  \begin{cases}
    y_v,&v\in B_t,\\
    x,&v\in V(G_t)\setminus B_t.
  \end{cases}
  \qquad (v\in V(G_t)).
\]
It gives
\[
  Z_t^{\le2k}
    \bigl(\boldsymbol\omega_t^{B_t}(\boldsymbol y,x),w\bigr)
  =\sum_{\substack{A_t\subseteq V(G_t),\ \mu_t(A_t)\le k\\
                    |A_t\cap B_t|\le2k}}
    x^{|A_t\setminus B_t|}
    \left(\prod_{v\in A_t\cap B_t}y_v\right)
    w^{\mu_t(A_t)}.
\]
If we can compute
$Z_t^{\le2k}(\boldsymbol\omega_t^{B_t}(\boldsymbol y,x),w)$, then
substituting $y_v=x$ for every $v\in B_t\setminus\partial G_t$ yields
$Z_t^{\le2k}(\boldsymbol\omega_t^{\partial G_t}(\boldsymbol y,x),w)$.
However, computing all coefficients of this polynomial appears infeasible:
in the worst case, for every $R\subseteq B_t$ with $|R|\le2k$, it contains a
monomial $x^b w^\ell\prod_{v\in R}y_v$ for some $b\ge0$ and $0\le\ell\le k$.
It can therefore have at
least $\sum_{i=0}^{2k}\binom{|B_t|}{i}=n^{\Theta(k)}$ distinct monomials, so
even writing all its coefficients is too expensive.

The key idea is to consider instead its formal logarithm
$\log Z_t^{\le2k}
  (\boldsymbol\omega_t^{B_t}(\boldsymbol y,x),w)$, viewed as a formal power
series.  The polynomial has constant term $1$, and its formal logarithm is
well-defined; indeed, the set $A_t=\emptyset$ contributes the term $1$,
whereas every other set contributes a monomial of positive degree.

For a monomial $m=x^b w^\ell\prod_{v\in V(G_t)}y_v^{a_v}$, call
$\{v\in B_t:a_v>0\}$ the \emph{$B_t$-support} of $m$.  The next lemma shows
that, after restricting to monomials of $w$-degree at most $k$ and
$B_t$-support size at most $2k$, only $\Ostar(2^{\bigO(k)})$ distinct
supports occur.  It also shows that these supports can be enumerated in
$\Ostar(2^{\bigO(k)})$ time.

\begin{restatable}{lemma}{logsupportfamilylemma}\label{lem:log-support-family}
In $\Ostar(2^{\bigO(k)})$ time, one can enumerate a downward-closed family
$\cC_t\subseteq2^{B_t}$ of size $\Ostar(2^{\bigO(k)})$ such that every member
has size at most $2k$ and
\[
  \left[x^b w^\ell\prod_{v\in B_t}y_v^{a_v}\right]
  \log Z_t^{\le2k}
    (\boldsymbol\omega_t^{B_t}(\boldsymbol y,x),w)\ne0
  \quad\Longrightarrow\quad
  \{v\in B_t:a_v>0\}\in\cC_t
\]
whenever $\ell\le k$ and
$|\{v\in B_t:a_v>0\}|\le2k$.
\end{restatable}

We prove \cref{lem:log-support-family} in \cref{sec:log-support-proof}.
For each $S\subseteq B_t$, let $L_{t,S}(\boldsymbol y,w)$ be the formal power
series consisting of the terms of
$\log Z_t^{\le2k}(\boldsymbol y,w)$ whose monomials have $B_t$-support $S$.
Then
\[
  \log Z_t^{\le2k}
    (\boldsymbol\omega_t^{B_t}(\boldsymbol y,x),w)
  =\sum_{S\subseteq B_t}
    L_{t,S}(\boldsymbol\omega_t^{B_t}(\boldsymbol y,x),w).
\]
We would like to compute the left-hand side from the series on the right-hand
side.  Since these series are generally infinite, we truncate both sides to
obtain:
\[
  \left\langle
    \log Z_t^{\le2k}
      (\boldsymbol\omega_t^{B_t}(\boldsymbol y,x),w)
  \right\rangle_{\le(k,n,2k)}
  =
  \sum_{S\subseteq B_t}
  \left\langle
    L_{t,S}(\boldsymbol\omega_t^{B_t}(\boldsymbol y,x),w)
  \right\rangle_{\le(k,n,2k)},
\]
where, for a formal power series $F$,
\[
  \left\langle F\right\rangle_{\le(k,n,2k)}
  :=
  \left\langle F\right\rangle_{
    \deg_w\le k,\ \deg_x\le n,\ \sum_{v\in B_t}\deg_{y_v}\le2k}.
\]
Applying \cref{lem:log-support-family} to the right-hand side of the preceding
identity, we see that every summand indexed by $S\notin\cC_t$ is zero.
Consequently, we obtain
\begin{equation}\label{eq:truncated-log-support-sum}
  \left\langle
    \log Z_t^{\le2k}
      (\boldsymbol\omega_t^{B_t}(\boldsymbol y,x),w)
  \right\rangle_{\le(k,n,2k)}
  =
  \sum_{S\in\cC_t}
  \left\langle
    L_{t,S}(\boldsymbol\omega_t^{B_t}(\boldsymbol y,x),w)
  \right\rangle_{\le(k,n,2k)}.
\end{equation}
Fix $S\in\cC_t$.  The next lemma computes
$\left\langle
  L_{t,S}(\boldsymbol\omega_t^{B_t}(\boldsymbol y,x),w)
\right\rangle_{\le(k,n,2k)}$
from the polynomials $Q_{t,R}$ with $R\subseteq S$.

\begin{lemma}\label{lem:local-log-support}
All coefficients of
$\left\langle
  L_{t,S}(\boldsymbol\omega_t^{B_t}(\boldsymbol y,x),w)
\right\rangle_{\le(k,n,2k)}$
can be computed in
$\Ostar(2^{\bigO(k)})$ time.
\end{lemma}

\begin{proof}
For $T\subseteq S$, define the substitution
$\boldsymbol\rho_t^T(\boldsymbol y,x):=(\rho_{t,v}^T)_{v\in V(G_t)}$, where
\[
  \rho_{t,v}^T:=
  \begin{cases}
    y_v,&v\in T,\\
    0,&v\in B_t\setminus T,\\
    x,&v\in V(G_t)\setminus B_t.
  \end{cases}
  \qquad (v\in V(G_t)).
\]
The substitution $\boldsymbol\rho_t^T(\boldsymbol y,x)$ is obtained from
$\boldsymbol\omega_t^{B_t}(\boldsymbol y,x)$ by additionally setting $y_v=0$
for every $v\in B_t\setminus T$.  Since $L_{t,J}$ consists of the terms of
$\log Z_t^{\le2k}(\boldsymbol\omega_t^{B_t}(\boldsymbol y,x),w)$ with
$B_t$-support exactly $J$, this substitution leaves
$L_{t,J}(\boldsymbol\omega_t^{B_t}(\boldsymbol y,x),w)$ unchanged when
$J\subseteq T$ and annihilates
$L_{t,J}(\boldsymbol\omega_t^{B_t}(\boldsymbol y,x),w)$ when
$J\not\subseteq T$.  Thus,
\[
  \log Z_t^{\le2k}
    (\boldsymbol\rho_t^T(\boldsymbol y,x),w)
  =\sum_{J\subseteq T}
    L_{t,J}(\boldsymbol\omega_t^{B_t}(\boldsymbol y,x),w).
\]
This is the subset zeta transform of the family
$L_{t,J}(\boldsymbol\omega_t^{B_t}(\boldsymbol y,x),w)$.  Hence M\"obius
inversion gives
\begin{equation}\label{eq:LS}
  L_{t,S}(\boldsymbol\omega_t^{B_t}(\boldsymbol y,x),w)
  =\sum_{T\subseteq S}
    (-1)^{|S|-|T|}
    \log Z_t^{\le2k}
      (\boldsymbol\rho_t^T(\boldsymbol y,x),w).
\end{equation}
Truncating \cref{eq:LS} gives
\begin{equation}\label{eq:truncated-LS}
  \left\langle
    L_{t,S}(\boldsymbol\omega_t^{B_t}(\boldsymbol y,x),w)
  \right\rangle_{\le(k,n,2k)}
  =\sum_{T\subseteq S}
    (-1)^{|S|-|T|}
    \left\langle
      \log Z_t^{\le2k}
        (\boldsymbol\rho_t^T(\boldsymbol y,x),w)
    \right\rangle_{\le(k,n,2k)}.
\end{equation}

For each $T\subseteq S$, we compute the summand in
\cref{eq:truncated-LS} as follows.  As in
\cref{sec:bisection-transition}, we group the terms by $R=A_t\cap B_t$:
\begin{equation}\label{eq:restricted-bag-polynomial}
  Z_t^{\le2k}
    (\boldsymbol\rho_t^T(\boldsymbol y,x),w)
  =\sum_{\substack{R\subseteq B_t\\|R|\le2k}}
    \left(
      \left(\prod_{v\in R}\rho_{t,v}^T\right)\cdot Q_{t,R}(x,w)
    \right)
  =\sum_{R\subseteq T}
    \left(
      \left(\prod_{v\in R}y_v\right)\cdot Q_{t,R}(x,w)
    \right),
\end{equation}
where the second equality follows because $\rho_{t,v}^T=0$ for
$v\in B_t\setminus T$.
By \cref{lem:projected-series}, $Q_{t,R}(x,w)$ can be computed from the polynomials
$Z_s(\boldsymbol\omega_s^{\partial G_s}(\boldsymbol y,x),w)$ for the children
$s$ of $t$, for every $R\subseteq T$.  Hence all coefficients of
$Z_t^{\le2k}(\boldsymbol\rho_t^T(\boldsymbol y,x),w)$ can be computed.  For
every $T\subseteq S$, we take its formal logarithm and truncate it to degrees
$\le(k,n,2k)$.  The sum in \cref{eq:truncated-LS} then computes
$\left\langle
  L_{t,S}(\boldsymbol\omega_t^{B_t}(\boldsymbol y,x),w)
\right\rangle_{\le(k,n,2k)}$.

We analyze the running time.  For each $T\subseteq S$, the sum over
$R\subseteq T$ in \cref{eq:restricted-bag-polynomial} has at most $2^{2k}$
summands because $|T|\le|S|\le2k$.  By \cref{lem:projected-series}, each
$Q_{t,R}(x,w)$ can be computed in polynomial time.  Hence
$Z_t^{\le2k}(\boldsymbol\rho_t^T(\boldsymbol y,x),w)$ can be computed from
\cref{eq:restricted-bag-polynomial} in
$\Ostar(2^{\bigO(k)})$ time.  A term with monomial
$x^bw^\ell\prod_{v\in T}y_v^{a_v}$ in
$\left\langle
  \log Z_t^{\le2k}(\boldsymbol\rho_t^T(\boldsymbol y,x),w)
\right\rangle_{\le(k,n,2k)}$
satisfies
$0\le b\le n$, $0\le\ell\le k$, and $\sum_{v\in T}a_v\le2k$.  Hence $b$ has
$n+1$ choices, $\ell$ has $k+1$ choices, and $(a_v)_{v\in T}$ has
$\binom{|T|+2k}{|T|}$ choices; the binomial coefficient counts the
nonnegative integer vectors whose coordinate sum is at most $2k$.  Therefore,
the number of retained terms is at most
$(n+1)(k+1)\binom{|T|+2k}{|T|}=\Ostar(2^{\bigO(k)})$ because
$|T|\le2k$.  Using \cref{eq:fps-log-exp} with truncation after every
multiplication computes all coefficients of the truncated logarithm in
$\Ostar(2^{\bigO(k)})$ time.  Finally, $|S|\le2k$, so computing the truncated
logarithm for all $2^{|S|}$ sets $T\subseteq S$ and evaluating
\cref{eq:truncated-LS} takes $\Ostar(2^{\bigO(k)})$ time.
\end{proof}

By \cref{eq:truncated-log-support-sum,lem:local-log-support}, we can compute
in $\Ostar(2^{\bigO(k)})$ time the formal logarithm of
$Z_t^{\le2k}(\boldsymbol\omega_t^{B_t}(\boldsymbol y,x),w)$ truncated by the
bounds $(k,n,2k)$.  We want to
recover $Z_t^{\le2k}(\boldsymbol\omega_t^{\partial G_t}(\boldsymbol y,x),w)$
from the truncated logarithm.  Intuitively, this requires two operations:
changing
$\boldsymbol\omega_t^{B_t}(\boldsymbol y,x)$ to
$\boldsymbol\omega_t^{\partial G_t}(\boldsymbol y,x)$ by the substitution
\[
  y_v\longmapsto
  \begin{cases}
    y_v,&v\in\partial G_t,\\
    x,&v\in B_t\setminus\partial G_t.
  \end{cases}
  \qquad (v\in B_t),
\]
and taking the exponential using the relation
\[
  \left\langle
    \exp\left(
      \left\langle
        \log Z_t^{\le2k}
      \right\rangle_{\le(k,n,2k)}
    \right)
  \right\rangle_{\le(k,n,2k)}
  =Z_t^{\le2k},
\]
which follows from \cref{lem:truncated-log-exp}.

A naive approach is to first take the exponential and then replace
$\boldsymbol\omega_t^{B_t}(\boldsymbol y,x)$ with
$\boldsymbol\omega_t^{\partial G_t}(\boldsymbol y,x)$.  However, this approach
does not work because it must first compute
$Z_t^{\le2k}(\boldsymbol\omega_t^{B_t}(\boldsymbol y,x),w)$, which can contain
$\sum_{i=0}^{2k}\binom{|B_t|}{i}=n^{\Theta(k)}$ distinct monomials.

Alternatively, one could substitute $x$ for $y_v$ for every
$v\in B_t\setminus\partial G_t$ to obtain
$\log Z_t^{\le2k}\allowbreak
  (\boldsymbol\omega_t^{\partial G_t}(\boldsymbol y,x),w)$ and then take the
exponential.
However, substituting before exponentiation also fails because the
substitution moves the degrees of $y_v$ for
$v\in B_t\setminus\partial G_t$ into the $x$-degree.  Consequently, a term
that should be discarded because its total degree in the variables $y_v$ with
$v\in B_t$ exceeds $2k$ may satisfy all three degree bounds
$\le(k,n,2k)$ after the substitution and contribute to the exponential.

To avoid these issues, we use an intermediate substitution.  Introduce a
fresh variable $x'$ and define
\[
  \boldsymbol\sigma_t(\boldsymbol y,x,x')
  :=(\sigma_{t,v})_{v\in V(G_t)},
  \qquad
  \sigma_{t,v}:=
  \begin{cases}
    y_v,&v\in\partial G_t,\\
    x',&v\in B_t\setminus\partial G_t,\\
    x,&v\in V(G_t)\setminus B_t.
  \end{cases}
\]
The substitution
$\boldsymbol\omega_t^{B_t}(\boldsymbol y,x)\mapsto
\boldsymbol\sigma_t(\boldsymbol y,x,x')$ fixes $x$, $w$, and every $y_v$ with
$v\in\partial G_t$, and replaces $y_v$ by $x'$ for every
$v\in B_t\setminus\partial G_t$.  The important observation is that if $m'$
is the image of a monomial $m$ under this substitution, then
\[
  \sum_{v\in B_t}\deg_{y_v}(m)
  =\deg_{x'}(m')+\sum_{v\in\partial G_t}\deg_{y_v}(m').
\]
That is, the original total degree in the variables $y_v$ with $v\in B_t$
is preserved as the sum of the $x'$-degree and the total degree in the
remaining variables $y_v$ with $v\in\partial G_t$.
We first substitute $\boldsymbol\sigma_t(\boldsymbol y,x,x')$ for
$\boldsymbol\omega_t^{B_t}(\boldsymbol y,x)$ to obtain
$\left\langle
  \log Z_t^{\le2k}\allowbreak
    (\boldsymbol\sigma_t(\boldsymbol y,x,x'),w)
\right\rangle_{\le(k,n,2k)}$,
where, by a slight abuse of notation, the third coordinate of
$\le(k,n,2k)$ is
$\deg_{x'}+\sum_{v\in\partial G_t}\deg_{y_v}$ instead of
$\sum_{v\in B_t}\deg_{y_v}$.
We next take the exponential and apply
$\left\langle\,\cdot\,\right\rangle_{\le(k,n,2k)}$, and finally substitute
$x$ for $x'$.  The next lemma proves that the polynomial obtained after the
substitution $x'\mapsto x$ is
$Z_t^{\le2k}(\boldsymbol\omega_t^{\partial G_t}(\boldsymbol y,x),w)$ and can
be computed in $\Ostar(2^{\bigO(k)})$ time.

\begin{lemma}\label{lem:reconstruct-small-side}
The polynomial
$Z_t^{\le2k}(\boldsymbol\omega_t^{\partial G_t}(\boldsymbol y,x),w)$ can be
computed in $\Ostar(2^{\bigO(k)})$ time.
\end{lemma}

\begin{proof}
First, apply the substitution $\boldsymbol\sigma_t(\boldsymbol y,x,x')$ to
$\left\langle
  \log Z_t^{\le2k}\allowbreak
    (\boldsymbol\omega_t^{B_t}(\boldsymbol y,x),w)
\right\rangle_{\le(k,n,2k)}$.
We can claim that this substitution yields
$\left\langle
  \log Z_t^{\le2k}\allowbreak
    (\boldsymbol\sigma_t(\boldsymbol y,x,x'),w)
\right\rangle_{\le(k,n,2k)}$.  Indeed, let
\[
  m:=x^bw^\ell\prod_{v\in B_t}y_v^{a_v},
  \qquad
  m':=x^b(x')^{\sum_{v\in B_t\setminus\partial G_t}a_v}w^\ell
    \prod_{v\in\partial G_t}y_v^{a_v}
\]
be corresponding monomials of
$\log Z_t^{\le2k}
  (\boldsymbol\omega_t^{B_t}(\boldsymbol y,x),w)$ and
$\log Z_t^{\le2k}(\boldsymbol\sigma_t(\boldsymbol y,x,x'),w)$, respectively.
Since
\[
  \begin{aligned}
    &\left(\deg_w(m'),\deg_x(m'),
      \deg_{x'}(m')+\sum_{v\in\partial G_t}\deg_{y_v}(m')\right)\\
    &\qquad={}
      \left(\ell,b,\sum_{v\in B_t}a_v\right)
      =\left(\deg_w(m),\deg_x(m),
        \sum_{v\in B_t}\deg_{y_v}(m)\right),
  \end{aligned}
\]
$m$ survives the truncation under
$\boldsymbol\omega_t^{B_t}(\boldsymbol y,x)$ if and only if $m'$ survives the
truncation under $\boldsymbol\sigma_t(\boldsymbol y,x,x')$.

Next, take the exponential and truncate again to $\le(k,n,2k)$.  Since every
term of $Z_t^{\le2k}\allowbreak
  (\boldsymbol\sigma_t(\boldsymbol y,x,x'),w)$ satisfies
these bounds, \cref{lem:truncated-log-exp} gives
\[
  \left\langle
    \exp\left(
      \left\langle
        \log Z_t^{\le2k}
          (\boldsymbol\sigma_t(\boldsymbol y,x,x'),w)
      \right\rangle_{\le(k,n,2k)}
    \right)
  \right\rangle_{\le(k,n,2k)}
  =
  Z_t^{\le2k}
    (\boldsymbol\sigma_t(\boldsymbol y,x,x'),w).
\]

Finally, set $x'=x$.  Since
$\boldsymbol\sigma_t(\boldsymbol y,x,x)
  =\boldsymbol\omega_t^{\partial G_t}(\boldsymbol y,x)$, we obtain
$Z_t^{\le2k}\allowbreak
  (\boldsymbol\sigma_t(\boldsymbol y,x,x),w)
  =Z_t^{\le2k}\allowbreak
  (\boldsymbol\omega_t^{\partial G_t}(\boldsymbol y,x),w)$.

We now analyze the running time.  By
\cref{eq:truncated-log-support-sum,lem:local-log-support}, we can compute
the truncated polynomial
$\bigl\langle\log Z_t^{\le2k}\allowbreak
  (\boldsymbol\omega_t^{B_t}(\boldsymbol y,x),w)
 \bigr\rangle_{\le(k,n,2k)}$
in $\Ostar(2^{\bigO(k)})$ time.  Substituting
$\boldsymbol\sigma_t(\boldsymbol y,x,x')$ only relabels variables and combines
terms, so it does not increase the number of terms and also takes
$\Ostar(2^{\bigO(k)})$ time.

We evaluate the exponential using \cref{eq:fps-exp} and truncate after
each multiplication.  Every intermediate polynomial retains at most
\[
  (n+1)(k+1)
  \binom{|\partial G_t|+2k+1}{|\partial G_t|+1}
  =\Ostar(2^{\bigO(k)}),
\]
terms.  Hence the truncated exponential can be computed in
$\Ostar(2^{\bigO(k)})$ time.  The substitution $x'=x$ only combines
coefficients and takes the same time.
\end{proof}

The next lemma proves that the transitions compute the required polynomials.

\begin{lemma}\label{lem:bisection-message-correctness}
For every node $t$, given
$Z_s(\boldsymbol\omega_s^{\partial G_s}(\boldsymbol y,x),w)$ for every child
$s$ of $t$, the algorithm computes
$Z_t(\boldsymbol\omega_t^{\partial G_t}(\boldsymbol y,x),w)$ in
$\Ostar(2^{\bigO(k)})$ time.
\end{lemma}

\begin{proof}
If $|B_t|\le4k+1$, then \cref{lem:projected-series} computes
$Q_{t,R}(x,w)$ for every
$R\subseteq B_t$, and substituting these polynomials into
\cref{eq:bag-to-boundary} gives
$Z_t(\boldsymbol\omega_t^{\partial G_t}(\boldsymbol y,x),w)$.  There are
$2^{|B_t|}=2^{\bigO(k)}$ choices of $R$, so this case takes
$\Ostar(2^{\bigO(k)})$ time.

If $|B_t|>4k+1$, then \cref{lem:reconstruct-small-side} computes
$Z_t^{\le2k}(\boldsymbol\omega_t^{\partial G_t}(\boldsymbol y,x),w)$ in
$\Ostar(2^{\bigO(k)})$ time.  Applying \cref{lem:large-bag-message} under
$\boldsymbol\omega_t^{\partial G_t}(\boldsymbol y,x)$ recovers
$Z_t(\boldsymbol\omega_t^{\partial G_t}(\boldsymbol y,x),w)$ by a
coefficientwise transformation, which takes the same time.
\end{proof}

Now we prove \cref{thm:bisection}.
\bisectionthm*

\begin{proof}
Let $G_1,\ldots,G_r$ be the connected components of $G$, and let $t_i$ be
the root of the decomposition of $G_i$.  For every $i\in[r]$, we compute
$Z_{t_i}(\boldsymbol\omega_{t_i}^{\partial G_{t_i}}(\boldsymbol y,x),w)$ using
\cref{thm:decomp,lem:bisection-message-correctness}.  By
\cref{eq:root-generating-polynomial} and additivity over
components, the number of feasible solutions is
\[
  \sum_{c=0}^k
  \left[x^{\lfloor n/2\rfloor}w^c\right]
  \prod_{i=1}^r
  Z_{t_i}(\boldsymbol\omega_{t_i}^{\partial G_{t_i}}(\boldsymbol y,x),w).
\]
The decompositions contain at most $n$ bags in total,
each transition takes $\Ostar(2^{\bigO(k)})$ time, and the root products take
polynomial time.  The intermediate numbers have polynomial bit length, so
the total running time is $\Ostar(2^{\bigO(k)})$.

For witness recovery, first find the minimum feasible cost $c\le k$.  While
$c>0$, delete an edge $e$ for which the resulting instance is feasible with
cost $c-1$, and decrement $c$; an edge of a minimum bisection always has this
property.  At cost zero, subset sum over the sizes of the connected
components recovers a bisection.  This uses only polynomially many calls to
the decision algorithm and preserves the running-time bound.
\end{proof}

\subsection{Proof of Lemma~\ref{lem:log-support-family}}
\label{sec:log-support-proof}

In this subsection, we prove \cref{lem:log-support-family}.  The proof has two
main steps.  The first step considers $\log Z_t(\boldsymbol y,w)$, where $y_v$
is a distinct variable for each $v\in V(G_t)$.  Take any monomial with
nonzero coefficient in $\log Z_t(\boldsymbol y,w)$, write it as
$w^c\prod_{v\in V(G_t)}y_v^{a_v}$, and let
$S:=\{v\in V(G_t):a_v>0\}$.  We show that $S$ is $c$-extendable to a connected
set; that is, there is a connected set $A\subseteq V(G)$
such that $A\cap V(G_t)=S$ and $|\delta_G(A)|\le c$
(\Cref{lem:connected-support}).

For the specialized series
$\log Z_t^{\le2k}(\boldsymbol\omega_t^{B_t}(\boldsymbol y,x),w)$ in
\cref{lem:log-support-family}, we can state that such an $A$ satisfies
$|A\cap B_t|\le2k$.  The second step is to use important cuts to enumerate
in $\Ostar(2^{\bigO(k)})$ time a family $\cC_t$ that contains $A\cap B_t$ for
every connected set $A\subseteq V(G)$ satisfying $|\delta_G(A)|\le k$ and
$|A\cap B_t|\le2k$ (\Cref{lem:important-support-family}).  Hence
$S=A\cap B_t\in\cC_t$.

We now prove the first step.  We would like to choose a minimum-cost
extension of $S$ and factor the part of $Z_t(\boldsymbol y,w)$ supported
on subsets of $S$ as a product over its connected components, so that
taking the formal logarithm turns the product into a sum.  The $y$-part
of each monomial factors automatically over these components.  The
difficulty lies in the $w$-part.  The exponent of $w$ in the term indexed
by $R\subseteq S$ is $\mu_t(R)$, and since $\mu_t(R)$ is defined by
minimizing over extensions outside $G_t$, it is not immediate that these
exponents are additive over the components.

To establish this additivity, for each $S\subseteq V(G_t)$, let
$\Phi_t(S)$ be the
inclusion-wise largest member of
\[
  \left\{A\subseteq V(G):
    A\cap V(G_t)=S,\quad |\delta_G(A)|=\mu_t(S)
  \right\}.
\]
We first prove that $\Phi_t(S)$ is well-defined for every
$S\subseteq V(G_t)$.

\begin{lemma}\label{lem:canonical-minimizer-well-defined}
For every $S\subseteq V(G_t)$, the set $\Phi_t(S)$ is well-defined.
\end{lemma}

\begin{proof}
The family defining $\Phi_t(S)$ is nonempty by the definition of
$\mu_t(S)$.  Let
$A,A'\subseteq V(G)$ satisfy
$A\cap V(G_t)=A'\cap V(G_t)=S$ and
$|\delta_G(A)|=|\delta_G(A')|=\mu_t(S)$.  Both $A\cap A'$ and
$A\cup A'$ intersect $V(G_t)$ in $S$, and thus
$|\delta_G(A\cap A')|\ge\mu_t(S)$ and
$|\delta_G(A\cup A')|\ge\mu_t(S)$.  By submodularity of the cut function,
\[
  2\mu_t(S)
  =|\delta_G(A)|+|\delta_G(A')|
  \ge |\delta_G(A\cap A')|+|\delta_G(A\cup A')|
  \ge 2\mu_t(S).
\]
Thus, both inequalities are equalities.  In particular, $A\cup A'$ also satisfies
$|\delta_G(A\cup A')|=\mu_t(S)$, and hence the family defining
$\Phi_t(S)$ is closed under unions, which implies that its inclusion-wise
largest member $\Phi_t(S)$ is unique.
\end{proof}

We also have the following.

\begin{lemma}\label{lem:canonical-minimizers}
If $S'\subseteq S\subseteq V(G_t)$, then
$\Phi_t(S')\subseteq\Phi_t(S)$.
\end{lemma}

\begin{proof}
Let $S'\subseteq S$.  The sets
$\Phi_t(S')\cap\Phi_t(S)$ and
$\Phi_t(S')\cup\Phi_t(S)$ intersect $V(G_t)$ in $S'$ and $S$,
respectively.  Hence
\begin{align*}
\mu_t(S')+\mu_t(S)
  &=|\delta_G(\Phi_t(S'))|+|\delta_G(\Phi_t(S))|\\
  &\ge |\delta_G(\Phi_t(S')\cap\Phi_t(S))|
     +|\delta_G(\Phi_t(S')\cup\Phi_t(S))|
   \ge\mu_t(S')+\mu_t(S).
\end{align*}
Thus $\Phi_t(S')\cup\Phi_t(S)$ is a minimizer in the definition of
$\mu_t(S)$.
The maximality of $\Phi_t(S)$ gives
$\Phi_t(S')\cup\Phi_t(S)=\Phi_t(S)$, and hence
$\Phi_t(S')\subseteq\Phi_t(S)$.
\end{proof}

For $S\subseteq V(G_t)$, let $\mathcal K_S$ be the family of connected
components of $G[\Phi_t(S)]$.  The next lemma shows that every member of
$\mathcal K_S$ intersects $V(G_t)$.

\begin{lemma}\label{lem:component-intersection}
For every $S\subseteq V(G_t)$ and every $K\in\mathcal K_S$, we have
$K\cap V(G_t)\ne\varnothing$.
\end{lemma}

\begin{proof}
Suppose that $K\cap V(G_t)=\varnothing$ for some
$K\in\mathcal K_S$.  The set $\Phi_t(S)\setminus K$ still intersects
$V(G_t)$ in $S$.
Since $K\cap V(G_t)=\varnothing$ and $V(G_t)\ne\varnothing$, we have
$K\ne V(G)$.  Since $G$ is connected, $|\delta_G(K)|>0$.  Moreover, there
is no edge between $K$ and $\Phi_t(S)\setminus K$, so
\[
  |\delta_G(\Phi_t(S)\setminus K)|
  =|\delta_G(\Phi_t(S))|-|\delta_G(K)|
  <|\delta_G(\Phi_t(S))|
  =\mu_t(S).
\]
This contradicts the definition of $\mu_t(S)$.
\end{proof}

The next lemma is the goal of the first step toward
\cref{lem:log-support-family}.

\begin{lemma}\label{lem:connected-support}
Suppose that $c\le k$ and that the coefficient of
$w^c\prod_{v\in V(G_t)}y_v^{a_v}$ in
$\log Z_t(\boldsymbol y,w)$ is nonzero.  For
$S:=\{v\in V(G_t):a_v>0\}$, the set $\Phi_t(S)$ is connected and
$|\delta_G(\Phi_t(S))|\le c$.
\end{lemma}

\begin{proof}
Define the substitution
$\boldsymbol\pi_t^S(\boldsymbol y)
:=(\pi_{t,v}^S)_{v\in V(G_t)}$, where, for each $v\in V(G_t)$,
\[
  \pi_{t,v}^S:=
  \begin{cases}
    y_v,&v\in S,\\
    0,&v\in V(G_t)\setminus S.
  \end{cases}
\]
Consider $\log Z_t(\boldsymbol\pi_t^S(\boldsymbol y),w)$.
Setting $y_v=0$ for every $v\notin S$ leaves only the monomials of
$\log Z_t(\boldsymbol y,w)$ whose support is contained in $S$.  In particular,
the coefficient of
$w^c\prod_{v\in V(G_t)}y_v^{a_v}$ is unchanged.

Let $\mathcal K_S$ be the family of connected components of $G[\Phi_t(S)]$.
We will prove below that the polynomial
$Z_t(\boldsymbol\pi_t^S(\boldsymbol y),w)$ has the following factorization:
\begin{equation}\label{eq:restricted-factorization}
  Z_t(\boldsymbol\pi_t^S(\boldsymbol y),w)
  =\left\langle
    \prod_{K\in\mathcal K_S}
      \left(\sum_{R\subseteq K\cap V(G_t)}
        \left(w^{\mu_t(R)}\cdot\prod_{v\in R}y_v\right)\right)
  \right\rangle_{\deg_w\le k},
\end{equation}
Taking formal logarithms in \cref{eq:restricted-factorization} and retaining
only terms of $w$-degree at most $k$ gives
\[
  \left\langle
    \log Z_t(\boldsymbol\pi_t^S(\boldsymbol y),w)
  \right\rangle_{\deg_w\le k}
  =\left\langle
    \sum_{K\in\mathcal K_S}
      \log\left(\sum_{R\subseteq K\cap V(G_t)}
        \left(w^{\mu_t(R)}\cdot\prod_{v\in R}y_v\right)\right)
  \right\rangle_{\deg_w\le k}.
\]
Each summand on the right-hand side uses variables from only one member of
$\mathcal K_S$.  If $\mathcal K_S$ has at least two members, then
\cref{lem:component-intersection} implies that every member intersects $S$, so
no summand has support $S$; if $\mathcal K_S$ is empty, the right-hand side is
zero.  Thus, unless $\mathcal K_S$ is a singleton, the coefficient of
$w^c\prod_{v\in V(G_t)}y_v^{a_v}$ is zero, contradicting the hypothesis and
its preservation under $\boldsymbol\pi_t^S$.

It remains to prove \cref{eq:restricted-factorization}.  It suffices to show
that, for every $R\subseteq S$,
\begin{equation}\label{eq:mu-additivity}
  \mu_t(R)=\sum_{K\in\mathcal K_S}\mu_t(R\cap K).
\end{equation}
Indeed, if this is proved, then, since the sets $K\cap V(G_t)$ for
$K\in\mathcal K_S$ partition $S$, we have
\[
\begin{aligned}
  Z_t(\boldsymbol\pi_t^S(\boldsymbol y),w)
  &=\left\langle
      \sum_{R\subseteq S}
        \left(w^{\mu_t(R)}\cdot\prod_{v\in R}y_v\right)
    \right\rangle_{\deg_w\le k}\\
  &=\left\langle
      \prod_{K\in\mathcal K_S}
        \left(\sum_{R\subseteq K\cap V(G_t)}
          \left(w^{\mu_t(R)}\cdot\prod_{v\in R}y_v\right)\right)
    \right\rangle_{\deg_w\le k}.
\end{aligned}
\]
which is \cref{eq:restricted-factorization}.

Now we prove \cref{eq:mu-additivity}.  By \cref{lem:canonical-minimizers},
$\Phi_t(R)\subseteq\Phi_t(S)$.  The sets
$\Phi_t(R)\cap K$ for $K\in\mathcal K_S$ partition $\Phi_t(R)$, and there
are no edges between distinct members of $\mathcal K_S$.  Hence
\[
  \mu_t(R)
  =\sum_{K\in\mathcal K_S}|\delta_G(\Phi_t(R)\cap K)|
  \ge\sum_{K\in\mathcal K_S}\mu_t(R\cap K).
\]
\begin{figure}[t]
\centering
\begingroup
\definecolor{cutblue}{RGB}{42,79,113}
\definecolor{cutgreen}{RGB}{42,120,92}
\begin{tikzpicture}[
  x=.85cm,y=.85cm,font=\small,
  vertex/.style={circle,draw=black!75,fill=white,line width=.8pt,
    minimum size=6pt,inner sep=0pt},
  rvertex/.style={vertex,draw=cutblue,fill=cutblue},
  edge/.style={draw=black!65,line width=.85pt},
  firstcut/.style={draw=cutblue,line width=1.25pt},
  secondcut/.style={draw=cutgreen,line width=1.25pt},
  component/.style={fill=black!4,draw=black!50,
    line width=.7pt,rounded corners=11pt},
  first/.style={fill=cutblue!17,draw=cutblue!85,
    line width=.75pt,rounded corners=13pt},
  second/.style={fill=cutgreen!16,draw=cutgreen!90,
    line width=.75pt,rounded corners=13pt}]

  \path[fill=black!2,rounded corners=5pt]
    (.15,.20) rectangle (17.60,2.00);
  \path[component] (.65,.55) rectangle (9.65,6.10);
  \path[component] (11.35,.55) rectangle (17.10,6.10);

  \path[first] (1.04,.78) rectangle (2.16,4.30);
  \path[first]
    (7.00,.78)--(9.50,.78)--(9.50,1.70)--(8.81,4.30)
    --(7.69,4.30)--(7.00,1.70)--cycle;
  \path[second] (11.64,.78) rectangle (12.76,4.30);
  \path[second] (15.84,.78) rectangle (16.96,4.30);

  \draw[black!45,densely dashed,rounded corners=5pt]
    (.15,.20) rectangle (17.60,2.00);

  \coordinate (c) at (1.60,1.50);
  \coordinate (v) at (1.60,3.65);
  \coordinate (h1) at (3.20,1.50);
  \coordinate (h2) at (4.50,1.50);
  \coordinate (h3) at (5.80,1.50);
  \coordinate (x) at (4.50,4.80);
  \coordinate (b) at (7.50,1.50);
  \coordinate (a) at (9.00,1.50);
  \coordinate (u) at (8.25,3.65);
  \coordinate (d) at (10.50,1.50);
  \coordinate (p) at (12.20,1.50);
  \coordinate (y) at (12.20,3.65);
  \coordinate (h4) at (14.30,1.50);
  \coordinate (q) at (16.40,1.50);
  \coordinate (z) at (16.40,3.65);

  \draw[edge] (c)--(v) (a)--(u)--(b)
    (h1)--(x)--(h2) (x)--(h3) (p)--(y) (q)--(z);
  \draw[firstcut] (v)--(x)--(u) (a)--(d);
  \draw[secondcut] (d)--(p) (y)--(h4)--(z);

  \foreach \name in {a,b,c,p,q} \node[rvertex] at (\name) {};
  \foreach \name in {d,h1,h2,h3,h4,u,v,x,y,z}
    \node[vertex] at (\name) {};

  \node[fill=white,inner sep=2pt] at (4.95,6.10) {$K_1$};
  \node[fill=white,inner sep=2pt] at (14.22,6.10) {$K_2$};

  \node[text=cutblue] at (4.95,5.65) {$\Phi_t(R\cap K_1)$};
  \draw[cutblue!70,line width=.5pt]
    (3.55,5.54)--(1.85,4.33) (6.35,5.54)--(8.00,4.33);
  \node[text=cutgreen] at (14.22,5.65) {$\Phi_t(R\cap K_2)$};
  \draw[cutgreen!75,line width=.5pt]
    (13.10,5.51)--(12.35,4.33) (15.34,5.51)--(16.25,4.33);

  \node[fill=white,inner sep=2pt] at (10.50,.20) {$V(G_t)$};

  \node[rvertex] at (2.00,-.42) {};
  \node[anchor=west,inner sep=0pt] at (2.26,-.42) {$R$};
  \draw[firstcut] (4.00,-.42)--(4.65,-.42);
  \node[anchor=west,inner sep=0pt] at (4.90,-.42)
    {$\delta_G(\Phi_t(R\cap K_1))$};
  \draw[secondcut] (10.80,-.42)--(11.45,-.42);
  \node[anchor=west,inner sep=0pt] at (11.70,-.42)
    {$\delta_G(\Phi_t(R\cap K_2))$};
\end{tikzpicture}
\endgroup
\caption{The gray regions are the components $K_1,K_2$ of
$G[\Phi_t(S)]$.  The darker regions show $\Phi_t(R\cap K_1)$ and
$\Phi_t(R\cap K_2)$, each with two connected components, all meeting $R$.}
\label{fig:component-localization}
\end{figure}
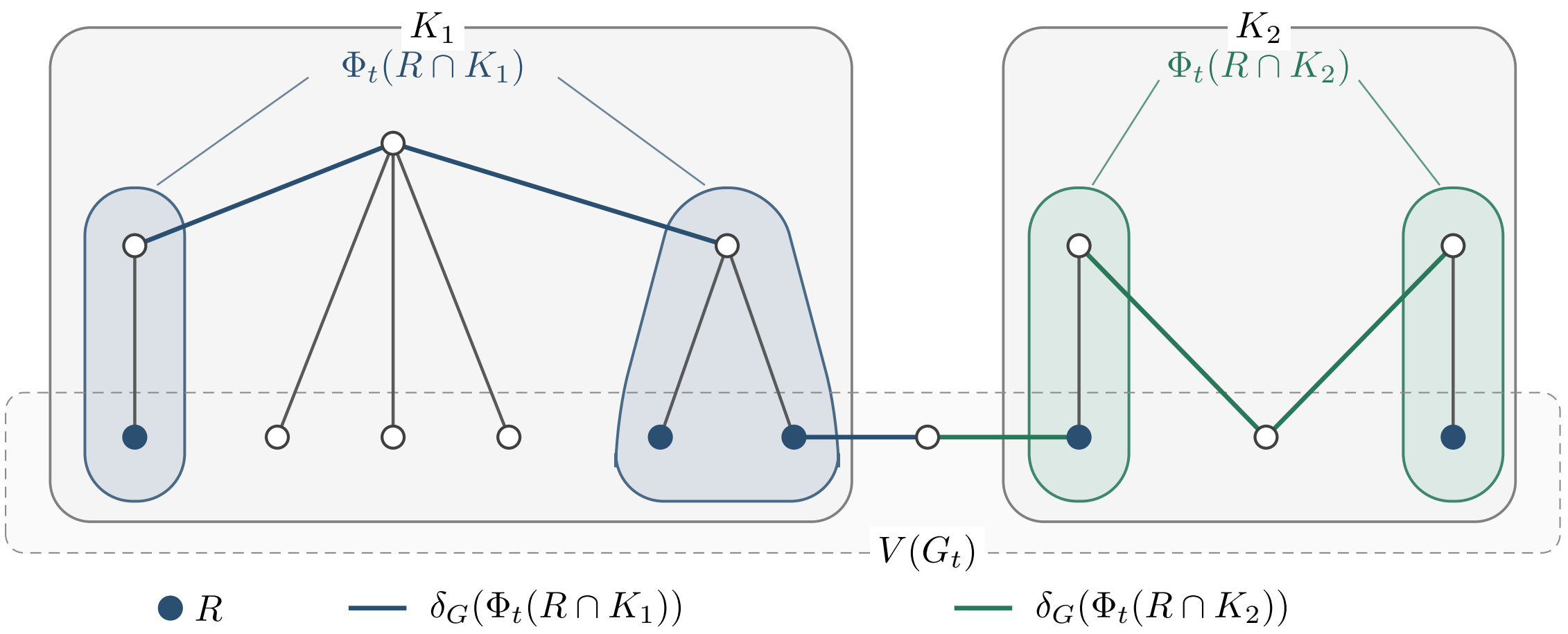

Conversely, fix $K\in\mathcal K_S$.  We show that
$\Phi_t(R\cap K)\subseteq K$ (\Cref{fig:component-localization}).  By
\cref{lem:canonical-minimizers},
$\Phi_t(R\cap K)\subseteq\Phi_t(S)$.  By
\cref{lem:component-intersection}, every connected component of
$G[\Phi_t(R\cap K)]$ contains a vertex of $V(G_t)$.  Since
$\Phi_t(R\cap K)\cap V(G_t)=R\cap K$, this vertex lies in $R\cap K\subseteq K$.
Since $K$ is a connected component of $G[\Phi_t(S)]$, every connected
component of $G[\Phi_t(R\cap K)]$ lies in $K$.  Hence
$\Phi_t(R\cap K)\subseteq K$.  Consequently,
$\bigcup_{K\in\mathcal K_S}\Phi_t(R\cap K)$ intersects $V(G_t)$ exactly in
$R$.  Moreover, since no edge joins distinct members of $\mathcal K_S$,
the edge sets $\delta_G(\Phi_t(R\cap K))$ for $K\in\mathcal K_S$ are
pairwise disjoint.  Therefore,
\[
  \mu_t(R)\le
  \left|\delta_G\left(\bigcup_{K\in\mathcal K_S}\Phi_t(R\cap K)\right)\right|
  =\sum_{K\in\mathcal K_S}|\delta_G(\Phi_t(R\cap K))|
   =\sum_{K\in\mathcal K_S}\mu_t(R\cap K),
\]
which proves \cref{eq:mu-additivity}.  This proves
\cref{eq:restricted-factorization}, and hence $\Phi_t(S)$ is connected.

Now we prove $|\delta_G(\Phi_t(S))|\le c$.
By the definitions of $Z_t$ and the formal logarithm,
\[
  \log Z_t(\boldsymbol y,w)
  =\sum_{j\ge1}\frac{(-1)^{j+1}}j
    \left(
      \sum_{\substack{\emptyset\ne A_t\subseteq V(G_t)\\\mu_t(A_t)\le k}}
        \left(w^{\mu_t(A_t)}\cdot\prod_{v\in A_t}y_v\right)
    \right)^j.
\]
Since the coefficient of $w^c\prod_{v\in V(G_t)}y_v^{a_v}$ is nonzero,
this monomial occurs in the expansion of one of the powers on the right-hand
side.  Hence, for some $j\ge1$, there are nonempty sets
$A_{t,1},\ldots,A_{t,j}\subseteq V(G_t)$ such that
\[
  w^c\prod_{v\in V(G_t)}y_v^{a_v}
  =\prod_{\ell=1}^j
    \left(w^{\mu_t(A_{t,\ell})}\cdot\prod_{v\in A_{t,\ell}}y_v\right)
  =w^{\sum_{\ell=1}^j\mu_t(A_{t,\ell})}
   \prod_{v\in V(G_t)}y_v^{|\{\ell:v\in A_{t,\ell}\}|}.
\]
Comparing the exponents of $w$ and the variables $y_v$ gives
\[
  \sum_{\ell=1}^j\mu_t(A_{t,\ell})=c,
  \qquad
  \bigcup_{\ell=1}^j A_{t,\ell}=S.
\]
Using $\Phi_t(A_{t,\ell})\cap V(G_t)=A_{t,\ell}$, we obtain
\[
  \left(\bigcup_{\ell=1}^j\Phi_t(A_{t,\ell})\right)\cap V(G_t)
  =\bigcup_{\ell=1}^j A_{t,\ell}=S.
\]
Moreover,
\[
  \delta_G\left(\bigcup_{\ell=1}^j\Phi_t(A_{t,\ell})\right)
  \subseteq\bigcup_{\ell=1}^j\delta_G(\Phi_t(A_{t,\ell})).
\]
Hence
\[
  \mu_t(S)
  \le\left|\delta_G\left(\bigcup_{\ell=1}^j\Phi_t(A_{t,\ell})\right)\right|
  \le\sum_{\ell=1}^j|\delta_G(\Phi_t(A_{t,\ell}))|
  =\sum_{\ell=1}^j\mu_t(A_{t,\ell})=c.
\]
By the definition of $\Phi_t$, we have
$|\delta_G(\Phi_t(S))|=\mu_t(S)\le c$.
\end{proof}

The next lemma supplies the second part toward Lemma~3.3.

\begin{lemma}\label{lem:important-support-family}
In $\Ostar(2^{\bigO(k)})$ time, one can enumerate a downward-closed family
$\cC_t\subseteq2^{B_t}$ of size $\Ostar(2^{\bigO(k)})$ such that every member
has size at most $2k$ and $A\cap B_t\in\cC_t$ for every connected
$A\subseteq V(G)$ with $|\delta_G(A)|\le k$ and $|A\cap B_t|\le2k$.
\end{lemma}

\begin{proof}
Let $H_t$ be obtained from $G$ by adding a new vertex $r_t$ adjacent to every
vertex of $B_t$.  For every $C\subseteq V(G)$, we have
$|\delta_{H_t}(C)|=|\delta_G(C)|+|C\cap B_t|$.
For each $v\in B_t$, use \cref{lem:important-cuts} to enumerate all
important $\{v\}$--$\{r_t\}$ cuts $D$ in $H_t$ satisfying
$|\delta_{H_t}(D)|\le3k$.  Let
$\cD_t$ be the union over $v\in B_t$ of the enumerated cut families and set
\begin{equation}\label{eq:CB}
  \cC_t:=\{\varnothing\}\cup
  \{T:T\subseteq D\cap B_t,\ |T|\le2k,
        \ D\in\cD_t\}.
\end{equation}
By definition, $\cC_t$ is downward closed.  Moreover,
\cref{lem:important-cuts} gives $|\cD_t|\le |B_t|4^{3k}$, and each
$D\in\cD_t$ contributes at most $2^{|D\cap B_t|}\le2^{3k}$ members to
$\cC_t$.  Hence
$|\cC_t|\le1+|B_t|\,4^{3k}2^{3k}=\Ostar(2^{\bigO(k)})$.

Let $A\subseteq V(G)$ satisfy the hypotheses of the lemma and set
$T:=A\cap B_t$.  If $T=\varnothing$, then $T\in\cC_t$ by \cref{eq:CB}.
Otherwise choose $v\in T$.  Then
$|\delta_{H_t}(A)|=|\delta_G(A)|+|T|\le3k$.  Hence $A$ is a connected
$\{v\}$--$\{r_t\}$ cut and is contained in an important such cut $D$ with
$|\delta_{H_t}(D)|\le|\delta_{H_t}(A)|$.  The enumeration contains $D$ by
\cref{lem:important-cuts}.  Since $T\subseteq D\cap B_t$ and $|T|\le2k$,
\cref{eq:CB} gives $T\in\cC_t$.
\end{proof}

Now we prove \cref{lem:log-support-family}.
\logsupportfamilylemma*

\begin{proof}
Let $\cC_t$ be the family produced by
\cref{lem:important-support-family}.  We show that it satisfies the statement
of \cref{lem:log-support-family}.  The running time, downward-closedness, the
size of $\cC_t$, and the bound on the size of its members follow immediately
from \cref{lem:important-support-family}.  It remains to show that the
$B_t$-support of every monomial satisfying the hypothesis belongs to $\cC_t$.

Choose nonnegative integers $b,\ell$ and $(a_v)_{v\in B_t}$ such that
\[
  \left[x^b w^\ell\prod_{v\in B_t}y_v^{a_v}\right]
  \log Z_t^{\le2k}
    (\boldsymbol\omega_t^{B_t}(\boldsymbol y,x),w)\ne0,
  \qquad
  \ell\le k,
  \qquad
  |\{v\in B_t:a_v>0\}|\le2k.
\]
Set $T:=\{v\in B_t:a_v>0\}$.
We claim that the exponents $(a_v)_{v\in B_t}$ can be extended to
$(a_v)_{v\in V(G_t)}$ such that
\begin{equation}\label{eq:monomial-extension}
  \left[w^\ell\prod_{v\in V(G_t)}y_v^{a_v}\right]
  \log Z_t(\boldsymbol y,w)\ne0.
\end{equation}
Given such an extension, set
$S:=\{v\in V(G_t):a_v>0\}$ and $A:=\Phi_t(S)$.
Then $A\cap B_t=S\cap B_t=T$, and \cref{lem:connected-support} shows that $A$
is connected and $|\delta_G(A)|\le\ell$.  Since $\ell\le k$ and $|T|\le2k$,
\cref{lem:important-support-family} gives $T=A\cap B_t\in\cC_t$, which proves
the lemma.

We now construct the extension in \cref{eq:monomial-extension}.
Since $\boldsymbol\omega_t^{B_t}(\boldsymbol y,x)$ replaces every $y_v$ with
$v\in V(G_t)\setminus B_t$ by $x$, we have
\[
  \left[x^b w^\ell\prod_{v\in B_t}y_v^{a_v}\right]
  \log Z_t^{\le2k}(\boldsymbol\omega_t^{B_t}(\boldsymbol y,x),w)
  =
  \sum
  \left[w^\ell\prod_{v\in V(G_t)}y_v^{a_v}\right]
  \log Z_t^{\le2k}(\boldsymbol y,w),
\]
where the sum is taken over all choices of $a_v\in\mathbb Z_{\ge0}$ for
$v\in V(G_t)\setminus B_t$ satisfying
$\sum_{v\in V(G_t)\setminus B_t}a_v=b$.
The left-hand side is nonzero by the hypothesis, so at least one summand on
the right-hand side is nonzero.  Fix an extension
$(a_v)_{v\in V(G_t)}$ indexing a nonzero summand.  Then,
\[
  \left[w^\ell\prod_{v\in V(G_t)}y_v^{a_v}\right]
  \log Z_t^{\le2k}(\boldsymbol y,w)\ne0.
\]

Set $S:=\{v\in V(G_t):a_v>0\}$.
Define the substitution
$\boldsymbol\pi_t^S(\boldsymbol y)
:=(\pi_{t,v}^S)_{v\in V(G_t)}$, where, for each $v\in V(G_t)$,
\[
  \pi_{t,v}^S:=
  \begin{cases}
    y_v,&v\in S,\\
    0,&v\in V(G_t)\setminus S.
  \end{cases}
\]
By the definition of $S$, the monomial in \cref{eq:monomial-extension} has
support $S$.  The substitution $\boldsymbol\pi_t^S$ fixes $y_v$ for every
$v\in S$ and sends only the variables outside $S$ to zero, so it preserves
the coefficient of this monomial in
$\log Z_t^{\le2k}(\boldsymbol y,w)$.  Thus,
\[
  \left[w^\ell\prod_{v\in V(G_t)}y_v^{a_v}\right]
  \log Z_t^{\le2k}(\boldsymbol\pi_t^S(\boldsymbol y),w)\ne0.
\]
Every term of $Z_t(\boldsymbol\pi_t^S(\boldsymbol y),w)$ is indexed by a set
$A_t\subseteq S$ and hence satisfies
$|A_t\cap B_t|\le|S\cap B_t|=|T|\le2k$.  Therefore,
\[
  Z_t^{\le2k}(\boldsymbol\pi_t^S(\boldsymbol y),w)
  =Z_t(\boldsymbol\pi_t^S(\boldsymbol y),w).
\]
Taking the formal logarithm of both sides gives
\[
  \log Z_t^{\le2k}(\boldsymbol\pi_t^S(\boldsymbol y),w)
  =\log Z_t(\boldsymbol\pi_t^S(\boldsymbol y),w).
\]
The substitution $\boldsymbol\pi_t^S$ likewise preserves the coefficient of
the monomial in \cref{eq:monomial-extension} in
$\log Z_t(\boldsymbol y,w)$.  Hence,
\begin{equation*}
  \left[w^\ell\prod_{v\in V(G_t)}y_v^{a_v}\right]
  \log Z_t(\boldsymbol y,w)
  =
  \left[w^\ell\prod_{v\in V(G_t)}y_v^{a_v}\right]
  \log Z_t^{\le2k}(\boldsymbol y,w)\ne0.\qedhere
\end{equation*}
\end{proof}

\section{Matching Cut}\label{sec:matching-cut}

In this section, we consider \textsc{Matching Cut}.  We use color coding to
reduce \textsc{Matching Cut} to an auxiliary problem called
\textsc{Colorful Cut}, and solve \textsc{Colorful Cut} using almost the
same algorithm as for \textsc{Minimum Bisection} in \cref{sec:bisection}.
Without loss of generality, assume throughout the section that $G$ is
connected.

For an integer $q\ge1$, a coloring $h:V(G)\to[2q]$, and a cut
$A\subseteq V(G)$, let
$\chi_h(A):=\{h(v):v\in V(G),\
\ \delta_G(A)\cap\delta_G(\{v\})\ne\varnothing\}$ be the set of colors
that appear at endpoints of edges in $\delta_G(A)$.

\boxproblem{Colorful Cut}
  {A simple undirected graph $G$, a positive integer $q$, and a coloring
   $h:V(G)\to[2q]$.}
  {Is there a set $\varnothing\subsetneq A\subsetneq V(G)$ such that
   $|\delta_G(A)|=q$ and $\chi_h(A)=[2q]$?}
  {$q$.}

The next lemma reduces \textsc{Matching Cut} to solving
$\Ostar(2^{\bigO(k)})$ instances of \textsc{Colorful Cut}.

\begin{lemma}\label{lem:matching-to-colorful}
A \textnormal{\textsc{Matching Cut}} instance $(G,k)$ with
$n\ge2$ and $k\ge1$ can be solved by solving
$\Ostar(2^{\bigO(k)})$ instances of \textnormal{\textsc{Colorful Cut}}.
\end{lemma}

\begin{proof}
For every $q\in[\min\{k,\lfloor n/2\rfloor\}]$, construct as above an
$(n,2q)$-perfect hash family $\mathcal H_q$ of colorings
$h:V(G)\to[2q]$, and add $(G,q,h)$ to $\mathcal I$ for every
$h\in\mathcal H_q$.  The construction time and the number of instances are
$\Ostar(2^{\bigO(k)})$.

Suppose that $G$ admits a matching cut $A$ with
$|\delta_G(A)|=q\le k$.  Since $G$ is connected, $q\ge1$.  Let $X$ be the
set of endpoints of $\delta_G(A)$.  Then $|X|=2q\le n$, so some
$h\in\mathcal H_q$ is injective on $X$.  The equality $h(X)=[2q]$ shows
that $A$ is feasible for $(G,q,h)$.

Conversely, take a \textsc{Colorful Cut} instance
$(G,q,h)\in\mathcal I$, and suppose that $A$ is a feasible solution for
$(G,q,h)$.  Let $X$ be the set of endpoints of $\delta_G(A)$.  Since
$|\delta_G(A)|=q$, we have
$|X|\le2q$.  Since $\chi_h(A)=h(X)=[2q]$, the set $X$ contains a vertex of
each of the $2q$ colors, so $|X|\ge2q$.  Therefore $|X|=2q$, and no two
edges of $\delta_G(A)$ share an endpoint.  Thus $\delta_G(A)$ is a matching
and $|\delta_G(A)|=q\le k$.
\end{proof}

We now solve \textsc{Colorful Cut} by slightly modifying the algorithm for
\textsc{Minimum Bisection} in \cref{sec:bisection}.  Intuitively, it suffices
to record in each dynamic-programming state the set of colors that occur at
the endpoints of cut edges.  To apply the algorithm of \cref{sec:bisection},
we express $\chi_h(A)=[2q]$ as a condition on the polynomial.  Fix an instance
$(G,q,h)$, and introduce one variable $u_i$ for each color $i\in[2q]$.  Write
$\boldsymbol u=(u_1,\ldots,u_{2q})$.  For every node $t$, define the polynomial
over $\mathbb Q$ by
\[
  Z_t(\boldsymbol y,w,\boldsymbol u)
  :=\sum_{\substack{A_t\subseteq V(G_t)\\\mu_t(A_t)\le q}}
      \left(\prod_{uv\in\delta_{G_t}(A_t)}u_{h(u)}u_{h(v)}\right)
      \cdot w^{\mu_t(A_t)}\cdot\prod_{v\in A_t}y_v
.\]
To compute the transition, the only nontrivial change from
\cref{sec:bisection} is the definition of $Q_{t,R}$.

Use the tree decomposition $\mathcal T$ and bottom-up computation of
\cref{sec:bisection} with $q$ in place of $k$.  Define the contribution of
the sets $A_t$ satisfying
$A_t\cap B_t=R$ by
\[
  Q_{t,R}(x,w,\boldsymbol u)
  :=\sum_{\substack{A_t\subseteq V(G_t)\\
      \mu_t(A_t)\le q,\ A_t\cap B_t=R}}
    \left(
      \left(\prod_{uv\in\delta_{G_t}(A_t)}u_{h(u)}u_{h(v)}\right)\cdot
      x^{|A_t\setminus B_t|}\cdot w^{\mu_t(A_t)}
    \right).
\]
By \cref{eq:completion}, the monomial of either
$Z_t(\boldsymbol y,w,\boldsymbol u)$ or
$Q_{t,R}(x,w,\boldsymbol u)$ indexed by $A_t$ has total
$\boldsymbol u$-degree $2|\delta_{G_t}(A_t)|$ and $w$-degree $\mu_t(A_t)$.
Consequently, every monomial $m$ in
$Z_t(\boldsymbol y,w,\boldsymbol u)$ or
$Q_{t,R}(x,w,\boldsymbol u)$ satisfies
$\sum_{i=1}^{2q}\deg_{u_i}(m)\le2\deg_w(m)$.  Hence the
$w$-degree truncation at $q$ also bounds the total $\boldsymbol u$-degree by
$2q$ throughout the algorithm, even after taking the formal logarithm or
exponential.  The number of monomials in
$\boldsymbol u$ of total degree at most $2q$ is
$\binom{4q}{2q}=2^{\bigO(q)}$.  Thus, managing the
variables $\boldsymbol u$ adds only a $2^{\bigO(q)}$ factor to the total
running time.

For $R=A_t\cap B_t$ and $A_s=A_t\cap V(G_s)$, we have
\[
  \delta_{G_t}(A_t)
  =\delta_{G_t[B_t]}(R)
   \mathbin{\dot\cup}
   \mathop{\dot\bigcup}_{s\in\operatorname{child}(t)}
     \delta_{G_s}(A_s).
\]
Therefore,
\cref{eq:fixed-pattern-recurrence} becomes
\begin{equation}\label{eq:colorful-fixed-pattern}
\begin{aligned}
  Q_{t,R}(x,w,\boldsymbol u)
  ={}&\mathopen{\left\langle
      \vphantom{\displaystyle\prod_{s\in\operatorname{child}(t)}}
      \right.}
      \left(\prod_{uv\in\delta_{G_t[B_t]}(R)}u_{h(u)}u_{h(v)}\right)
      \cdot w^{|\delta_{G_t[B_t]}(R)|+\eta_t(R\cap\partial G_t)}\\[-1mm]
    &{}\cdot
      \prod_{s\in\operatorname{child}(t)}
      \left(
        w^{-\eta_s(R\cap\partial G_s)}\cdot
        \left[\prod_{v\in R\cap\partial G_s}y_v\right]_{\boldsymbol y}
        Z_s\bigl(
          \boldsymbol\omega_s^{\partial G_s}(\boldsymbol y,x),w,\boldsymbol u
        \bigr)
      \right)
      \mathclose{\left.
      \vphantom{\displaystyle\prod_{s\in\operatorname{child}(t)}}
      \right\rangle}_{\deg_w\le q}.
\end{aligned}
\end{equation}
Apart from \cref{eq:colorful-fixed-pattern}, use the algorithm of
\cref{sec:bisection} without change, with
$Z_t(\boldsymbol y,w,\boldsymbol u)$ and
$Q_{t,R}(x,w,\boldsymbol u)$ in place of $Z_t(\boldsymbol y,w)$ and
$Q_{t,R}(x,w)$.  Accept exactly when
\begin{equation}\label{eq:colorful-root-test}
  \left[x^bw^q\prod_{i=1}^{2q}u_i\right]
  Z_{t_0}\bigl(
    \boldsymbol\omega_{t_0}^{\partial G_{t_0}}(\boldsymbol y,x),
    w,\boldsymbol u
  \bigr)>0
\end{equation}
for some $1\le b\le n-1$.

The next lemma proves the correctness and running time of the
\textsc{Colorful Cut} algorithm.

\begin{lemma}\label{lem:colorful-cut-algorithm}
\textnormal{\textsc{Colorful Cut}} can be decided deterministically in
$\Ostar(2^{\bigO(q)})$ time.
\end{lemma}

\begin{proof}
Let $Z_t^{\le2q}(\boldsymbol y,w,\boldsymbol u)$ denote the restriction of
$Z_t(\boldsymbol y,w,\boldsymbol u)$ to the terms indexed by sets $A_t$ with
$|A_t\cap B_t|\le2q$.  For $S\subseteq B_t$, let
$L_{t,S}(\boldsymbol y,w,\boldsymbol u)$ consist of the terms in
$\log Z_t^{\le2q}(\boldsymbol y,w,\boldsymbol u)$ whose $B_t$-support is $S$.

In \cref{lem:projected-series}, replace $Q_{t,R}(x,w)$ by
$Q_{t,R}(x,w,\boldsymbol u)$, adjoin $\boldsymbol u$ as the last argument of
every child polynomial $Z_s$, and replace \cref{eq:fixed-pattern-recurrence} by
\cref{eq:colorful-fixed-pattern}.  For $R=A_t\cap B_t$ and
$A_s=A_t\cap V(G_s)$, the edge partition used in
\cref{lem:projected-series} gives
\[
  \left(\prod_{uv\in\delta_{G_t[B_t]}(R)}u_{h(u)}u_{h(v)}\right)
  \prod_{s\in\operatorname{child}(t)}
    \left(\prod_{uv\in\delta_{G_s}(A_s)}u_{h(u)}u_{h(v)}\right)
  =\prod_{uv\in\delta_{G_t}(A_t)}u_{h(u)}u_{h(v)}.
\]
All factors independent of $\boldsymbol u$ are identical in
\cref{eq:fixed-pattern-recurrence,eq:colorful-fixed-pattern}.  Hence the
proof of \cref{lem:projected-series} requires no other change.

In \cref{eq:bag-to-boundary}, adjoin $\boldsymbol u$ as the last argument of
$Q_{t,R}$ and $Z_t$.  The derivation of \cref{eq:bag-to-boundary} partitions
the terms according to $A_t\cap B_t=R$, so
\cref{eq:bag-to-boundary} holds coefficientwise in $\boldsymbol u$.

In \cref{lem:large-bag-message}, replace $Z_t(\boldsymbol y,w)$ and
$Z_t^{\le2q}(\boldsymbol y,w)$ by $Z_t(\boldsymbol y,w,\boldsymbol u)$ and
$Z_t^{\le2q}(\boldsymbol y,w,\boldsymbol u)$, respectively.  Since
$\delta_{G_t}(V(G_t)\setminus A_t)=\delta_{G_t}(A_t)$, the proof of
\cref{lem:large-bag-message} requires no other change.

In \cref{lem:log-support-family}, replace
$Z_t^{\le2q}(\boldsymbol y,w)$ by
$Z_t^{\le2q}(\boldsymbol y,w,\boldsymbol u)$ and the coefficient ring
$\mathbb Q[x,w]$ by $\mathbb Q[x,w,\boldsymbol u]$.  The proof of
\cref{lem:log-support-family} is unchanged: \cref{eq:restricted-factorization}
holds over $\mathbb Q[x,w,\boldsymbol u]$, and the $B_t$-support depends only
on the exponents of the variables $y_v$.

In \cref{lem:local-log-support}, adjoin $\boldsymbol u$ as the last argument
of $L_{t,S}$, $Q_{t,R}$, and $Z_t^{\le2q}$.  The subset inversion in
\cref{eq:truncated-LS} holds coefficientwise for every monomial in
$\boldsymbol u$.
Therefore, \cref{lem:local-log-support} holds without modification.

In \cref{lem:reconstruct-small-side}, replace
$Z_t^{\le2q}(\boldsymbol y,w)$ by
$Z_t^{\le2q}(\boldsymbol y,w,\boldsymbol u)$.  The substitutions
$\boldsymbol\sigma_t(\boldsymbol y,x,x')$ and $x'\mapsto x$ fix
$\boldsymbol u$, and \cref{eq:fps-log-exp,eq:fps-exp} remain valid over
$\mathbb Q[\boldsymbol u]$.  Therefore,
\cref{lem:reconstruct-small-side} holds without modification.

Finally, in \cref{lem:bisection-message-correctness}, adjoin $\boldsymbol u$
as the last argument of every polynomial $Z_t$.  Applying
\crefrange{lem:projected-series}{lem:reconstruct-small-side} with
$Q_{t,R}(x,w,\boldsymbol u)$,
$Z_t(\boldsymbol y,w,\boldsymbol u)$, and
$Z_t^{\le2q}(\boldsymbol y,w,\boldsymbol u)$ proves the small-bag and
large-bag cases of \cref{lem:bisection-message-correctness}.  Thus, the
bottom-up computation obtains
$Z_{t_0}(\boldsymbol\omega_{t_0}^{\partial G_{t_0}}(\boldsymbol y,x),
w,\boldsymbol u)$.

At the root, a cut $A$ with $|\delta_G(A)|=q$ contributes a monomial of
total $\boldsymbol u$-degree $2q$.  The equality $\chi_h(A)=[2q]$ holds if
and only if every variable $u_i$ for $i\in[2q]$ occurs in the monomial.
Hence every such variable has exponent one.  Therefore,
\cref{eq:colorful-root-test} accepts exactly the feasible cuts of
\textsc{Colorful Cut}.

Throughout the computation, the total $\boldsymbol u$-degree is at most $2q$.
At most $\binom{4q}{2q}=2^{\bigO(q)}$ monomials in $\boldsymbol u$ occur.
Thus, each operation from \cref{sec:bisection} acquires only a
$2^{\bigO(q)}$ factor, and the total running time remains
$\Ostar(2^{\bigO(q)})$.
\end{proof}

Now we prove \cref{thm:matching-cut}.
\matchingcutthm*

\begin{proof}
Construct the family $\mathcal I$ from \cref{lem:matching-to-colorful}, and
apply \cref{lem:colorful-cut-algorithm} to every instance in $\mathcal I$.
By \cref{lem:matching-to-colorful}, some instance in $\mathcal I$ is a
yes-instance if and only if $(G,k)$ is a yes-instance of
\textsc{Matching Cut}.  For a yes-instance $(G,q,h)\in\mathcal I$, vertex
self-reduction using the decision algorithm of
\cref{lem:colorful-cut-algorithm} recovers a feasible cut with at most $n$
additional calls.  By \cref{lem:matching-to-colorful}, the feasible cut
defines the required matching cut.  The family $\mathcal I$ has
$\Ostar(2^{\bigO(k)})$ instances.  Since $q\le k$ for every
$(G,q,h)\in\mathcal I$, each instance takes $\Ostar(2^{\bigO(k)})$ time.
Therefore, the total running time is
$\Ostar(2^{\bigO(k)})$.
\end{proof}

\section{Unbreakable decomposition}\label{sec:structure}

In this section, we prove \cref{thm:decomp}, completing the decomposition
construction used by the algorithms in
\cref{sec:bisection,sec:matching-cut}.

For a graph $H$ and $U\subseteq V(H)$, let
$N_H(U):=\{v\in V(H)\setminus U:uv\in E(H)\text{ for some }u\in U\}$ denote
the open neighborhood of $U$ in $H$.
For $T\subseteq V(H)$, let $\operatorname{adh}_H(T)$ be the maximum of
$|N_H(C)|$ over the components $C$ of $H-T$, or zero if $H-T$ is empty.

We construct the tree decomposition from the root down.  If $V(G)$ is
$(2k,k)$-edge-unbreakable in $G$, we output the one-node tree decomposition
with bag $V(G)$.  Otherwise, we find a set $B\subseteq V(G)$ that is
$(2k,k)$-edge-unbreakable in $G$ and satisfies
$\operatorname{adh}_G(B)\le2k$, use $B$ as the root bag, and recurse for every
connected component of $G-B$.

Consider a recursive call on $H_C:=H[C\cup N_H(C)]$, where $C$ is a component
created by the parent call.  We terminate if $V(H_C)$ is
$(2k,k)$-edge-unbreakable in $H_C$.  Otherwise, let
$X_C:=N_H(C)\cup\{v_C\}$, where $v_C$ is an arbitrary vertex of
$C=V(H_C)\setminus N_H(C)$.  We construct a bag $B_C$ with
$X_C\subseteq B_C$ and recurse for every connected component of $H_C-B_C$.
We require $N_H(C)\subseteq B_C$ because every vertex of $N_H(C)$ occurs in
the parent bag, and the bags containing any fixed vertex must form a connected
subtree.  We
require $v_C\in B_C$ to rule out $B_C=N_H(C)$, which would repeat the same
recursive call.

To find such a bag $B_C$ in single-exponential time, we repeatedly use the
next lemma.

\begin{lemma}\label{lem:bag-shrink}
Let $X\subseteq B\subseteq V(H)$ satisfy $|X|\le2k+1$ and
$\operatorname{adh}_H(B)\le2k$.  In $\Ostar(2^{\bigO(k)})$ time one can
either find $B'\subseteq V(H)$ such that $X\subseteq B'$, $|B'|<|B|$, and
$\operatorname{adh}_H(B')\le2k$, or certify that $B$ is
$(2k,k)$-edge-unbreakable.
\end{lemma}

\begin{proof}
If $|B|\le4k+1$, every bipartition of $B$ has a side of size at most
$2k$, so $B$ is $(2k,k)$-edge-unbreakable.  Hence assume $|B|>4k+1$.
Since $|X|\le2k+1$, we have $B\setminus X\ne\varnothing$.

Guess a set $Z\subseteq X$ with $|X\setminus Z|\le k$ and a vertex
$r\in B\setminus X$.  There are $\Ostar(2^{\bigO(k)})$ choices for $(Z,r)$.
For each guess, first delete $X\setminus Z$ from $H$ and set
$H':=H-(X\setminus Z)$.  Then enumerate in $H'$ the important
$\{r\}$--$Z$ cuts $D$ with $|\delta_{H'}(D)|\le k$ using
\cref{lem:important-cuts}.

\begin{samepage}
The key point is that if $B$ is not $(2k,k)$-edge-unbreakable, then the
enumeration finds a cut $D$ containing more vertices of $B$ than the number of
edges leaving $D$ in $H'$, which helps us construct a smaller separator.

\begin{claim}\label{clm:large-enumerated-cut}
If $B$ is not $(2k,k)$-edge-unbreakable, then for some guess $(Z,r)$, the
enumeration finds a cut $D$ satisfying
$|D\cap B|>|\delta_{H'}(D)|$.
\end{claim}
\end{samepage}

\begin{proof}[Proof of \cref{clm:large-enumerated-cut}]
We construct a guess $(Z,r)$ for which the enumeration finds $D$.
Since $B$ is not $(2k,k)$-edge-unbreakable in $H$, there is a set
$A\subseteq V(H)$ such that $|\delta_H(A)|\le k$, $|A\cap B|>2k$, and
$|B\setminus A|>2k$.  Since $|X|\le2k+1$, we may assume without loss of
generality that $|A\cap X|\le k$.  Set $Z:=X\setminus A$.
See \cref{fig:cut-enumeration} for an illustration.

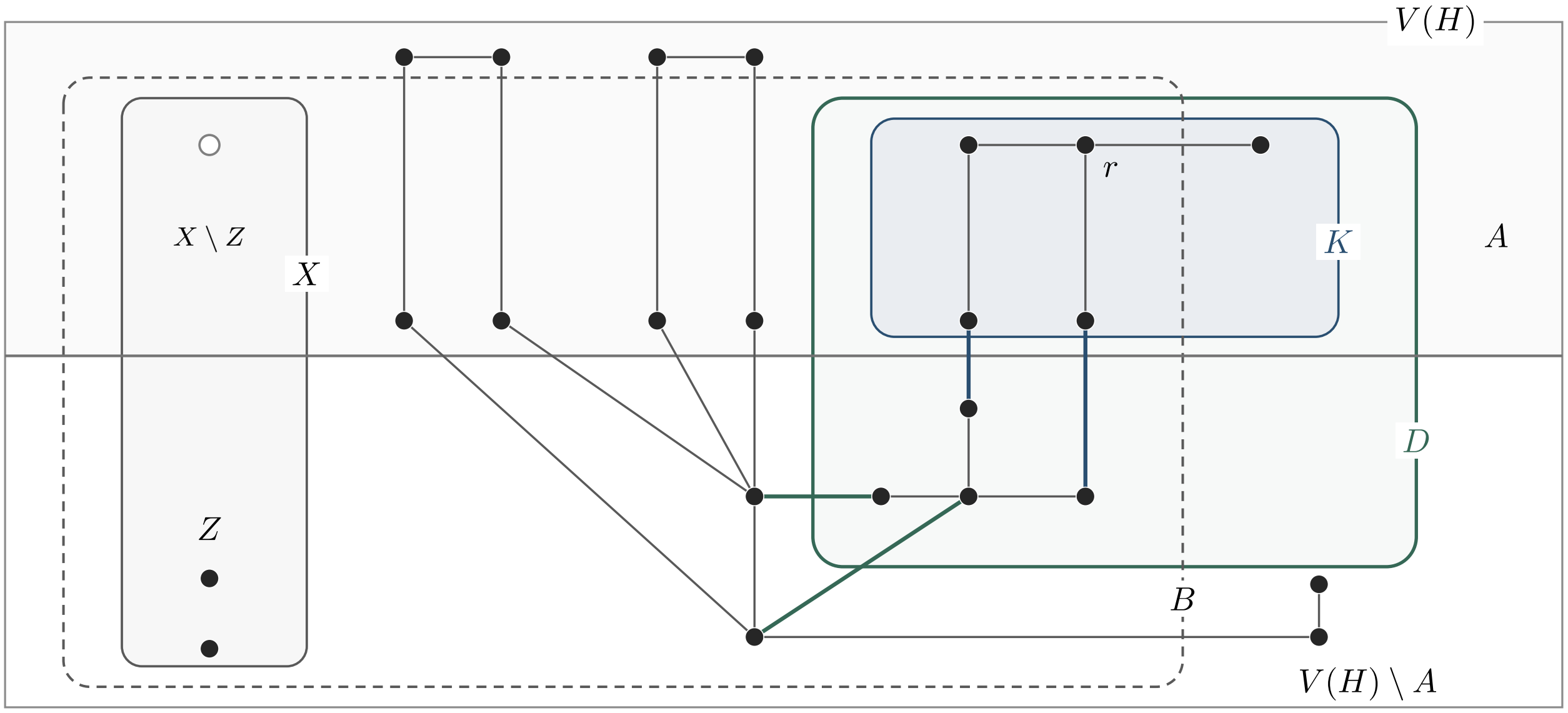
\begin{figure}[t]
\centering
\begingroup
\definecolor{enumcomponent}{RGB}{42,79,113}
\definecolor{enumshore}{RGB}{53,104,86}
\pgfmathsetlengthmacro{\enumHorizontalUnit}{(\linewidth-.8pt)/8}
\begin{tikzpicture}[
  x={(0cm,-.62cm)},y={(\enumHorizontalUnit,0cm)},font=\small,
  vertex/.style={circle,fill=black!85,draw=white,line width=.3pt,
    minimum size=5.6pt,inner sep=0pt},
  set label/.style={fill=white,inner sep=2pt},
  graph edge/.style={draw=black!65,line width=.65pt}]

  \path[fill=black!2] (-.6,0) rectangle (5.1,8);
  \draw[black!45,line width=.6pt] (-.6,0) rectangle (11.1,8);
  \path[fill=enumshore!4,draw=enumshore,line width=1pt,rounded corners=9pt]
    (0.7,4.15) rectangle (8.7,7.25);
  \path[fill=enumcomponent!10,draw=enumcomponent,line width=.7pt,rounded corners=7pt]
    (1.05,4.45) rectangle (4.775,6.85);
  \draw[black!65,line width=.75pt,dash pattern=on 3pt off 2.2pt,rounded corners=8pt]
    (0.35,0.3) rectangle (10.75,6.05);
  \path[fill=black!3,draw=black!65,line width=.7pt,rounded corners=6pt]
    (0.7,0.6) rectangle (10.4,1.55);
  \draw[black!55,line width=.8pt] (5.1,0)--(5.1,8);

  \coordinate (r) at (1.5,5.55);
  \coordinate (a) at (4.5,5.55);
  \coordinate (b) at (1.5,4.95);
  \coordinate (d) at (4.5,4.95);
  \coordinate (h) at (1.5,6.45);
  \coordinate (c1) at (0,3.85);
  \coordinate (c2) at (4.5,3.85);
  \coordinate (c3) at (0,3.35);
  \coordinate (c4) at (4.5,3.35);
  \coordinate (e1) at (0,2.55);
  \coordinate (e2) at (4.5,2.55);
  \coordinate (e3) at (0,2.05);
  \coordinate (e4) at (4.5,2.05);
  \coordinate (s) at (1.5,1.05);
  \coordinate (u1) at (7.5,5.55);
  \coordinate (u2) at (6,4.95);
  \coordinate (m) at (7.5,4.95);
  \coordinate (j1) at (7.5,3.85);
  \coordinate (jm) at (7.5,4.5);
  \coordinate (v) at (9.9,3.85);
  \coordinate (w1) at (9.9,6.75);
  \coordinate (w2) at (9.0,6.75);
  \coordinate (z1) at (8.9,1.05);
  \coordinate (z3) at (10.1,1.05);

  \draw[graph edge]
    (h)--(r)--(a) (r)--(b)--(d)
    (u1)--(m)--(u2)
    (c1)--(c2)--(j1) (c1)--(c3)--(c4)--(j1)
    (e1)--(e2)--(j1) (e1)--(e3)--(e4)--(v)
    (j1)--(v)--(w1)--(w2) (m)--(jm);
  \draw[enumcomponent,line width=1.2pt] (a)--(u1) (d)--(u2);
  \draw[enumshore,line width=1.2pt] (m)--(v) (jm)--(j1);

  \foreach \n in {r,a,b,d,h,c1,c2,c3,c4,e1,e2,e3,e4,u1,u2,m,j1,jm,v,w1,w2,z1,z3}
    \node[vertex] at (\n) {};
  \node[circle,fill=white,draw=black!50,line width=.7pt,
    minimum size=6pt,inner sep=0pt] at (s) {};

  \node[set label] at (-.6,7.35) {$V(H)$};
  \node at (3.05,7.66) {$A$};
  \node at (10.7,7) {$V(H)\setminus A$};
  \node[set label,text=enumcomponent] at (3.15,6.85) {$K$};
  \node[set label,text=enumshore] at (6.55,7.25) {$D$};
  \node[set label] at (9.25,6.05) {$B$};
  \node[set label] at (3.7,1.55) {$X$};
  \node at (8.05,1.05) {$Z$};
  \node[font=\scriptsize] at (3.1,1.05) {$X\setminus Z$};
  \node[below right=2pt] at (r) {$r$};
\end{tikzpicture}
\endgroup
\caption{The choice of $K$ and $D$.  Here $K$ is a component of
$H[A\setminus X]$, and $D\supseteq K$ is an important cut.}
\label{fig:cut-enumeration}
\end{figure}

Let $\mathcal K$ be the family of connected components of
$H[A\setminus X]$.  Observe that $A\setminus X$ is the part of $A$ that
remains in $H'$.  Since the sets in $\mathcal K$ partition $A\setminus X$,
\[
  \sum_{K\in\mathcal K}|K\cap B|
  =|(A\setminus X)\cap B|
  \ge |A\cap B|-|A\cap X|>k.
\]
For every $K\in\mathcal K$, we have
$\delta_{H'}(K)\subseteq\delta_H(A)$.  Moreover, the sets $\delta_{H'}(K)$ for
$K\in\mathcal K$ are pairwise disjoint.  Hence
\[
  \sum_{K\in\mathcal K}|\delta_{H'}(K)|
  \le |\delta_H(A)|\le k.
\]
Therefore some $K\in\mathcal K$ satisfies
$|K\cap B|>|\delta_{H'}(K)|$.

Fix such a component $K$ and choose $r\in K\cap B$.  Since
$K\subseteq A\setminus X$, we have $r\in B\setminus X$, so $(Z,r)$ is among
the guesses.  Also, $K\cap Z=\varnothing$ and $H'[K]$ is connected, so $K$ is
a connected $\{r\}$--$Z$ cut in $H'$.  Choose an important connected
$\{r\}$--$Z$ cut $D$ such that $K\subseteq D$ (possibly $D=K$).  Then,
$|\delta_{H'}(D)|\le|\delta_{H'}(K)|\le k$.  By
\cref{lem:important-cuts}, the enumeration contains $D$.
Finally,
$|D\cap B|\ge|K\cap B|>|\delta_{H'}(K)|\ge|\delta_{H'}(D)|$.
\end{proof}

For a cut $D$ guaranteed by \cref{clm:large-enumerated-cut}, construct
$B'$ as follows.  Let $Y$ be the set of vertices of $D$ incident with an edge of
$\delta_{H'}(D)$, and let $P\subseteq D$ be a minimum
$Y$--$(D\cap B)$ vertex separator in
$H[D]$.  Set $B':=(B\setminus D)\cup P$.  The sets $B$, $D$, $Y$, $P$,
and $B'$ are illustrated in
\cref{fig:bag-shrinking}.

\begin{figure}[htbp]
\centering
\begingroup
\definecolor{shrinksep}{RGB}{53,104,86}
\definecolor{shrinknewbag}{RGB}{137,87,43}
\begin{tikzpicture}[
  x=1.43cm,y=1.08cm,font=\small,
  vertex/.style={circle,fill=black!85,draw=white,line width=.3pt,
    minimum size=5.6pt,inner sep=0pt},
  separator vertex/.style={vertex,fill=shrinksep,minimum size=6.5pt},
  set label/.style={fill=white,inner sep=2pt},
  graph edge/.style={draw=black!65,line width=.65pt}]

  \path[fill=black!5,draw=black!50,line width=.6pt,rounded corners=5pt]
    (2.35,1.15) rectangle (2.85,3.60);
  \path[fill=shrinksep!9,draw=shrinksep,line width=.8pt,rounded corners=5pt]
    (5.15,1.25) rectangle (5.65,4.40);
  \draw[black!60,line width=.7pt,rounded corners=9pt]
    (2.15,.30) rectangle (9.80,5.15);
  \draw[black!45,line width=.6pt,rounded corners=5pt]
    (.48,1.43) rectangle (1.10,3.58);
  \draw[black!45,line width=.6pt,rounded corners=5pt]
    (.48,.60) rectangle (1.10,1.23);

  \draw[black!70,line width=.75pt,dash pattern=on 3pt off 2.2pt,rounded corners=7pt]
    (.20,.20)--(1.40,.20)--(1.40,3.85)--(7.50,3.85)
    --(7.50,1.05)--(9.25,1.05)--(9.25,4.60)--(.20,4.60)--cycle;
  \draw[shrinknewbag,line width=1.05pt,rounded corners=7pt]
    (.35,-.55)--(5.90,-.55)--(5.90,4.85)--(4.90,4.85)
    --(4.90,-.10)--(1.20,-.10)--(1.20,3.98)--(.35,3.98)--cycle;

  \coordinate (z1) at (.82,3.30);
  \coordinate (z2) at (.82,2.475);
  \coordinate (z3) at (.82,1.65);
  \coordinate (deleted1) at (.64,1.04);
  \coordinate (deleted2) at (.95,1.04);
  \coordinate (deleted3) at (.80,.80);
  \coordinate (a1) at (1.65,3.30);
  \coordinate (a2) at (1.65,2.475);
  \coordinate (a3) at (1.65,1.65);
  \coordinate (y1) at (2.60,3.30);
  \coordinate (y2) at (2.60,2.475);
  \coordinate (y3) at (2.60,1.65);
  \coordinate (w1) at (3.85,3.30);
  \coordinate (w2) at (3.85,2.475);
  \coordinate (w3) at (3.85,1.65);
  \coordinate (p1) at (5.40,4.15);
  \coordinate (p2) at (5.40,1.65);
  \coordinate (u1) at (6.62,4.15);
  \coordinate (u2) at (6.62,1.65);
  \coordinate (b1) at (7.88,4.15);
  \coordinate (b2) at (8.88,4.15);
  \coordinate (b3) at (7.88,1.65);
  \coordinate (b4) at (8.88,1.65);
  \coordinate (n1) at (6.62,.65);
  \coordinate (n2) at (7.88,.65);

  \draw[graph edge,line width=1pt]
    (a1)--(y1) (a2)--(y2) (a3)--(y3);
  \draw[graph edge]
    (z1)--(a1) (z2)--(a2) (z3)--(a3)
    (a1)--(a2)--(a3) (z1)--(a2)
    (y1)--(w1)--(p1)--(u1)--(b1)--(b2)
    (y2)--(w2)--(p1) (w2)--(p2)
    (y3)--(w3)--(p2)--(u2)--(b3)--(b4)
    (p2)--(n1)--(n2);

  \node[set label] at (.82,3.58) {$Z$};
  \node[set label,font=\scriptsize,text=black!70,inner sep=1pt] at (.79,.60) {$X\setminus Z$};
  \node[set label] at (3.65,5.15) {$D$};
  \node[set label] at (2.60,3.60) {$Y$};
  \node[set label,text=shrinksep] at (5.40,4.40) {$P$};
  \node[set label] at (8.38,4.60) {$B$};
  \node[set label,text=shrinknewbag] at (5.40,4.85) {$B'$};

  \foreach \v in {z1,z2,z3,a1,a2,a3,y1,y2,y3,w1,w2,w3,u1,u2,b1,b2,b3,b4,n1,n2}
    \node[vertex] at (\v) {};
  \foreach \v in {p1,p2}
    \node[separator vertex] at (\v) {};
  \foreach \v in {deleted1,deleted2,deleted3}
    \node[circle,fill=white,draw=black!50,line width=.7pt,
      minimum size=5.8pt,inner sep=0pt] at (\v) {};
\end{tikzpicture}
\endgroup
\caption{The construction of $B'=(B\setminus D)\cup P$, where $P$ is a
minimum $Y$--$(D\cap B)$ vertex separator in $H[D]$.}
\label{fig:bag-shrinking}
\end{figure}
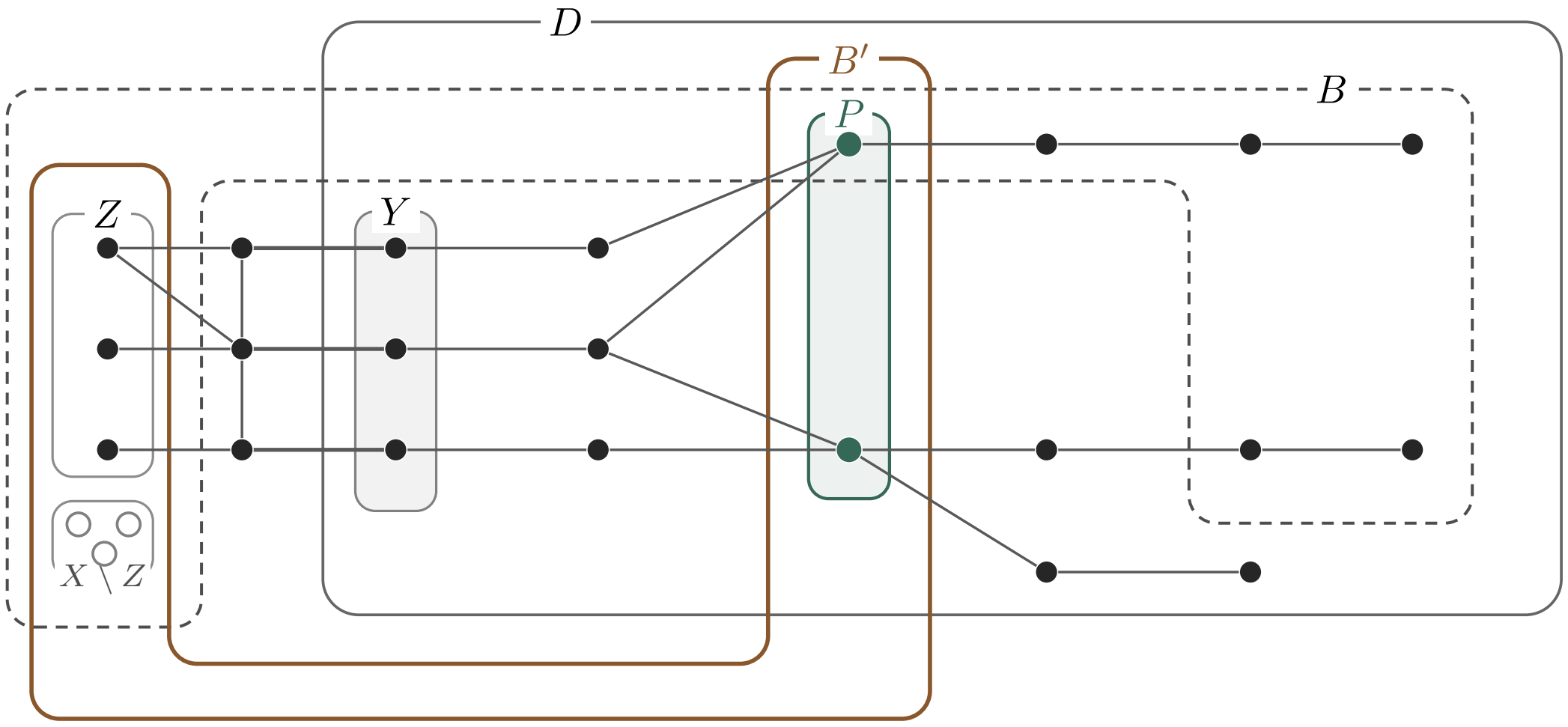

The set $D$ is disjoint from $X$, so
$X\subseteq B\setminus D\subseteq B'$.
Both $Y$ and $P$ are $Y$--$(D\cap B)$ vertex separators in $H[D]$,
so the minimality of $P$ gives $|P|\le|Y|$.  Since
$|Y|\le|\delta_{H'}(D)|$, \cref{clm:large-enumerated-cut} gives
$|Y|\le|\delta_{H'}(D)|<|D\cap B|$.  Hence $|P|<|D\cap B|$.  Moreover,
$P\subseteq D$, so
\[
  |B'|=|B\setminus D|+|P|
  =|B|-|D\cap B|+|P|<|B|.
\]

We now prove that $\operatorname{adh}_H(B')\le2k$.
Let $K$ be a component of $H-B'$.  It suffices to show that
$|N_H(K)|\le2k$.  Suppose first that $K\cap B\ne\varnothing$.
We first show that $K\subseteq D$.  Since $B\setminus D\subseteq B'$ and
$K\cap B'=\varnothing$, every vertex of $K\cap B$ belongs to $D$.
Moreover, $P\subseteq B'$ gives $K\cap P=\varnothing$.  Suppose that
$K\not\subseteq D$.
Choose a path $Q$ in $H[K]$ from $K\cap B$ to $K\setminus D$, and let $uv$
be the first edge of $Q$ with $u\in D$ and $v\notin D$.  Since
$X\setminus Z\subseteq B'$, we have
$v\notin X\setminus Z$.  Thus $uv\in\delta_{H'}(D)$ and $u\in Y$.
The initial subpath of $Q$ ending at $u$ lies in $H[D]-P$ and joins
$D\cap B$ to $Y$, contradicting that $P$ separates $Y$ from $D\cap B$.
Hence $K\subseteq D$.

Since $H[K]$ is connected and $K\cap P=\varnothing$, the separator $P$
also gives $K\cap Y=\varnothing$.  Every neighbor of $K$ in $D$ belongs to
$B'\cap D=P$.  Every neighbor $v\notin D$ of $K$ belongs to $X\setminus Z$,
since any edge $uv\in E(H)$ with $u\in K$ and
$v\notin D\cup(X\setminus Z)$ belongs to $\delta_{H'}(D)$, contrary to
$K\cap Y=\varnothing$.  Hence
$N_H(K)\subseteq P\cup(X\setminus Z)$, and
$|N_H(K)|\le |P|+|X\setminus Z|\le2k$.

Suppose now that $K\cap B=\varnothing$, and let $K_0$ be the component of
$H-B$ that contains $K$.  We show that
$|N_H(K)|\le |N_H(K_0)|\le\operatorname{adh}_H(B)$ by constructing an
injection from $N_H(K)$ into $N_H(K_0)$.
\Cref{fig:bag-adhesion} illustrates an injection from $N_H(K)$ into
$N_H(K_0)$ when $K\cap B=\varnothing$.

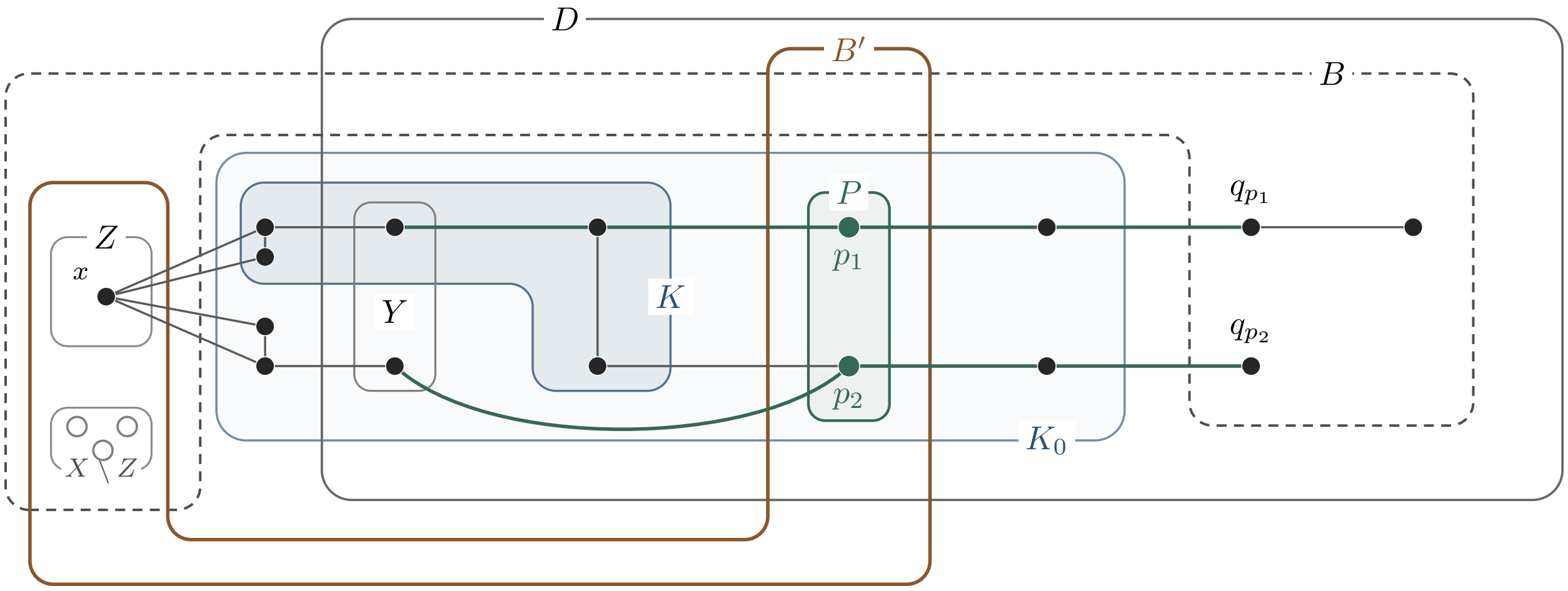
\begin{figure}[htbp]
\centering
\begingroup
\definecolor{adhcomponent}{RGB}{42,79,113}
\definecolor{adhseparator}{RGB}{53,104,86}
\definecolor{adhnewbag}{RGB}{137,87,43}
\pgfmathsetlengthmacro{\adhHorizontalUnit}{(\linewidth-1.1pt)/9.6}
\begin{tikzpicture}[
  x=\adhHorizontalUnit,y=1.05cm,font=\small,
  vertex/.style={circle,fill=black!85,draw=white,line width=.3pt,
    minimum size=5.6pt,inner sep=0pt},
  pvertex/.style={vertex,fill=adhseparator,minimum size=6.5pt},
  set label/.style={fill=white,inner sep=2pt},
  edge/.style={draw=black!65,line width=.65pt}]

  \path[draw=adhcomponent!65,fill=adhcomponent!3,line width=.65pt,rounded corners=9pt]
    (1.50,.90) rectangle (7.10,3.80);
  \path[draw=adhcomponent!80,fill=adhcomponent!12,line width=.65pt,rounded corners=7pt]
    (1.65,2.48)--(3.45,2.48)--(3.45,1.40)--(4.30,1.40)
    --(4.30,3.50)--(1.65,3.50)--cycle;
  \draw[black!60,line width=.7pt,rounded corners=9pt]
    (2.15,.30) rectangle (9.80,5.15);
  \draw[black!50,line width=.6pt,rounded corners=5pt]
    (2.35,1.40) rectangle (2.85,3.30);
  \path[fill=adhseparator!9,draw=adhseparator,line width=.8pt,rounded corners=5pt]
    (5.15,1.10) rectangle (5.65,3.40);
  \draw[black!45,line width=.6pt,rounded corners=5pt]
    (.48,1.85) rectangle (1.10,2.95);
  \draw[black!45,line width=.6pt,rounded corners=5pt]
    (.48,.60) rectangle (1.10,1.23);

  \draw[black!70,line width=.75pt,dash pattern=on 3pt off 2.2pt,rounded corners=7pt]
    (.20,.20)--(1.40,.20)--(1.40,3.98)--(7.50,3.98)
    --(7.50,1.05)--(9.25,1.05)--(9.25,4.60)--(.20,4.60)--cycle;
  \draw[adhnewbag,line width=1.05pt,rounded corners=7pt]
    (.35,-.55)--(5.90,-.55)--(5.90,4.85)--(4.90,4.85)
    --(4.90,-.10)--(1.20,-.10)--(1.20,3.50)--(.35,3.50)--cycle;

  \coordinate (x) at (.82,2.35);
  \coordinate (deleted1) at (.64,1.04);
  \coordinate (deleted2) at (.95,1.04);
  \coordinate (deleted3) at (.80,.80);
  \coordinate (a1) at (1.80,3.05);
  \coordinate (a2) at (1.80,2.75);
  \coordinate (a3) at (1.80,1.65);
  \coordinate (a4) at (1.80,2.05);
  \coordinate (y1) at (2.60,3.05);
  \coordinate (y2) at (2.60,1.65);
  \coordinate (w1) at (3.85,3.05);
  \coordinate (w2) at (3.85,1.65);
  \coordinate (p1) at (5.40,3.05);
  \coordinate (p2) at (5.40,1.65);
  \coordinate (r1) at (6.62,3.05);
  \coordinate (r2) at (6.62,1.65);
  \coordinate (q1) at (7.88,3.05);
  \coordinate (q2) at (7.88,1.65);
  \coordinate (q3) at (8.88,3.05);

  \draw[edge]
    (x)--(a1) (x)--(a2) (x)--(a3) (x)--(a4)
    (a1)--(a2) (a3)--(a4)
    (a1)--(y1) (a3)--(y2) (w2)--(p2)
    (w1)--(w2) (q1)--(q3);
  \draw[adhseparator,line width=1.05pt]
    (y1)--(w1)--(p1)--(r1)--(q1)
    (y2) .. controls (3.15,.80) and (4.85,.80) .. (p2)
    --(r2)--(q2);

  \node[set label] at (.82,2.95) {$Z$};
  \node[set label,font=\scriptsize,text=black!70,inner sep=1pt] at (.79,.60) {$X\setminus Z$};
  \node[set label] at (3.65,5.15) {$D$};
  \node[set label] at (2.60,2.20) {$Y$};
  \node[set label,text=adhseparator] at (5.40,3.40) {$P$};
  \node[set label] at (8.38,4.60) {$B$};
  \node[set label,text=adhnewbag] at (5.40,4.85) {$B'$};
  \node[set label,text=adhcomponent] at (4.30,2.35) {$K$};
  \node[set label,text=adhcomponent] at (6.62,.90) {$K_0$};

  \foreach \v in {x,a1,a2,a3,a4,y1,y2,w1,w2,r1,r2,q1,q2,q3}
    \node[vertex] at (\v) {};
  \foreach \v in {p1,p2}
    \node[pvertex] at (\v) {};
  \foreach \v in {deleted1,deleted2,deleted3}
    \node[circle,fill=white,draw=black!50,line width=.7pt,
      minimum size=5.8pt,inner sep=0pt] at (\v) {};
  \node[above left=2pt,font=\scriptsize] at (x) {$x$};
  \node[below=3pt,text=adhseparator] at (p1) {$p_1$};
  \node[below=3pt,text=adhseparator] at (p2) {$p_2$};
  \node[above=3pt] at (q1) {$q_{p_1}$};
  \node[above=3pt] at (q2) {$q_{p_2}$};
\end{tikzpicture}
\endgroup
\caption{The injection from $N_H(K)$ into $N_H(K_0)$ when
$K\cap B=\varnothing$.
The green $Y$--$(D\cap B)$ paths contain subpaths connecting the vertices of
$N_H(K)\cap(P\setminus B)$ to distinct vertices of
$N_H(K_0)\setminus B'$.}
\label{fig:bag-adhesion}
\end{figure}

By Menger's theorem, there are $|P|$
vertex-disjoint
$Y$--$(D\cap B)$ paths in $H[D]$.  Truncate each path at its first vertex in
$D\cap B$.  Since the paths are vertex-disjoint and every path intersects $P$,
each path contains exactly one vertex of $P$, and every vertex of $P$ lies on
one path.  For each $p\in P\setminus B$, let
$q_p\in(D\cap B)\setminus P$ be the endpoint of the path through $p$.
The vertices $q_p$ are distinct.

Since $K$ is a component of $H-B'$, every vertex of $N_H(K)$ belongs
to $B'$.  Together with $B'=(B\setminus D)\cup P$, this shows that every
vertex of $N_H(K)$ belongs either to $N_H(K)\cap B$ or to
$N_H(K)\cap(P\setminus B)$, and these two events are mutually exclusive.
For $x\in N_H(K)\cap B$, the vertex $x$ has a neighbor in
$K\subseteq K_0$, so $x\in N_H(K_0)$.

Now fix $p\in N_H(K)\cap(P\setminus B)$.  Both $p$ and $K$ avoid $B$,
and $p$ has a neighbor in $K$, so $p$ belongs to $K_0$.  On the path
through $p$, the subpath from $p$ to $q_p$ contains no vertex of $B$
except $q_p$, because $q_p$ is the first vertex of $D\cap B$ on the
path.  Hence all vertices of this subpath other than $q_p$ belong to
$K_0$, and therefore $q_p\in N_H(K_0)$.  Moreover,
$q_p\notin B'$ because $q_p\in(D\cap B)\setminus P$ and
$B'=(B\setminus D)\cup P$.

Define $f\colon N_H(K)\to N_H(K_0)$ by $f(x)=x$ for
$x\in N_H(K)\cap B$ and $f(p)=q_p$ for
$p\in N_H(K)\cap(P\setminus B)$.  The vertices $q_p$ are distinct
because the paths are vertex-disjoint.  Moreover,
$f(N_H(K)\cap B)\subseteq B'$, whereas
$f(N_H(K)\cap(P\setminus B))\subseteq N_H(K_0)\setminus B'$.
Hence no image of a vertex in $N_H(K)\cap B$ equals an image of a vertex
in $N_H(K)\cap(P\setminus B)$.  Thus $f$ is injective, and hence
$|N_H(K)|\le |N_H(K_0)|\le\operatorname{adh}_H(B)$.

Since $|P|\le|\delta_{H'}(D)|\le k$ and $|X\setminus Z|\le k$, the bounds for
$K\cap B\ne\varnothing$ and $K\cap B=\varnothing$ give
\[
  \operatorname{adh}_H(B')
  \le\max\{\operatorname{adh}_H(B),|P\cup(X\setminus Z)|\}
  \le2k.
\]
The minimum separator $P$ and the set $B'$ are computable in polynomial time,
so the total running time is
$\Ostar(2^{\bigO(k)})$.
\end{proof}

Now we prove \cref{thm:decomp}.
\decompositionthm*

\begin{proof}
To construct the root bag, choose an arbitrary vertex $v\in V(G)$, set
$X:=\{v\}$ and $B:=V(G)$, and apply \cref{lem:bag-shrink} to $(X,B)$.
Whenever the lemma returns $B'$, replace $B$ by $B'$ and apply the lemma to
$(X,B)$ again.  Each replacement satisfies $X\subseteq B$,
$\operatorname{adh}_G(B)\le2k$, and $|B|$ strictly decreases.  Hence, after
at most $|V(G)|$ applications, the lemma certifies that $B$ is
$(2k,k)$-edge-unbreakable in $G$.

We recurse on the components as follows.  At the root, set $H:=G$ and let
$B$ be the bag obtained by the repeated applications of
\cref{lem:bag-shrink}.  After constructing a bag $B$ in an induced subgraph
$H$ of $G$, create a node with bag $B$.  For every component $C$ of $H-B$,
choose an arbitrary vertex $v_C\in C$, define
$H_C:=H[C\cup N_H(C)]$ and $X_C:=N_H(C)\cup\{v_C\}$.  Since
$|N_H(C)|\le\operatorname{adh}_H(B)\le2k$, we have $|X_C|\le2k+1$.
Starting with $B_C:=V(H_C)$, the same repeated application of
\cref{lem:bag-shrink} as for the root produces a set $B_C$ such that
$X_C\subseteq B_C$, $\operatorname{adh}_{H_C}(B_C)\le2k$, and $B_C$ is
$(2k,k)$-edge-unbreakable in $H_C$.  Create a child node with bag $B_C$,
attach it below the node with bag $B$, and repeat the construction for every
component of $H_C-B_C$.  Since $v_C\in X_C\subseteq B_C$, the vertex $v_C$
belongs to no descendant component; thus the recursion terminates after at
most $|V(G)|$ recursive steps.

We verify that the procedure actually yields a tree decomposition of $G$.
Every vertex outside $B$ belongs to some component $C$ of $H-B$ and hence to
$V(H_C)$, so the recursion covers all vertices.  For an edge $xy\in E(H)$, either
$x,y\in B$, or both endpoints lie in $C\cup N_H(C)=V(H_C)$ for some component
$C$ of $H-B$; thus some bag contains both endpoints.

Fix $x\in V(H)$.  If $x\notin B$, it occurs only in the child instance for the
component containing $x$.  If $x\in B$, every child instance containing $x$
has $x\in N_H(C)\subseteq X_C\subseteq B_C$.  Induction therefore shows that
the bags containing $x$ form a connected subtree.

For the parent--child edge associated with $C$, since
$N_H(C)\subseteq X_C\subseteq B_C$ and
$B\cap V(H_C)=N_H(C)$, we have $B\cap B_C=N_H(C)$.  Thus the adhesion between
$B$ and $B_C$ has size $|N_H(C)|\le\operatorname{adh}_H(B)\le2k$.

It remains to verify that every bag is $(2k,k)$-edge-unbreakable in $G$.
Consider a bag $B$ created in a recursive step on $H=G[W]$.  For every
$A\subseteq V(G)$ with $|\delta_G(A)|\le k$, we have
$\delta_H(A\cap W)\subseteq\delta_G(A)$.  The $(2k,k)$-edge-unbreakability of
$B$ in $H$ therefore gives
$\min\{|A\cap B|,|B\setminus A|\}\le2k$.  Hence every bag is
$(2k,k)$-edge-unbreakable in $G$.  There are at most $|V(G)|$
recursive steps, and each step applies \cref{lem:bag-shrink} at most
$|V(H)|$ times,
so the total running time is $\Ostar(2^{\bigO(k)})$.
\end{proof}

\section{Reduction from 3-SAT}\label{sec:lower-bound}

We prove \cref{thm:bisection-lower-bound} by combining a bounded-degree
reduction from \textsc{3-SAT} to \textsc{Max Cut} with a reduction from
\textsc{Max Cut} to \textsc{Minimum Bisection}.

\boxproblemwithoutparameter{Max Cut}
  {A simple undirected graph $H$ and an integer $t\ge0$.}
  {Is there a set $X\subseteq V(H)$ such that $|\delta_H(X)|\ge t$?}

Garey, Johnson, and Stockmeyer~\cite{GareyJohnsonStockmeyer1976} proved the
NP-completeness of \textsc{Max Cut} via a reduction from \textsc{3-SAT}.
Here we need a reduction whose constructed \textsc{Max Cut} instance has
linear size and bounded degree.  We therefore give a reduction satisfying both
requirements.

\begin{lemma}
\label{lem:3sat-to-bounded-degree-max-cut}
There is a polynomial-time reduction from \textnormal{\textsc{3-SAT}} to
\textnormal{\textsc{Max Cut}} that maps a $3$-CNF formula $\varphi$ with a
total of $s$ variables and clauses to an instance $(H,t)$ such that $\varphi$
is satisfiable if and only if $(H,t)$ is a yes-instance, $H$ has maximum
degree $\bigO(1)$, and $|V(H)|+|E(H)|=\bigO(s)$.
\end{lemma}

\begin{proof}
Introduce a variable $z$ and, for every clause $C=(a\lor b\lor c)$ of
$\varphi$, a variable $w_C$.  Replace $C$ by the two constraints
$\operatorname{NAE}(a,b,w_C)$ and $\operatorname{NAE}(\neg w_C,c,z)$, where
$\operatorname{NAE}$ requires its three arguments not to be all equal.  Let
$\psi$ be the conjunction of these constraints.  Complementing every variable
preserves all constraints of $\psi$.  Hence, if $\psi$ is satisfiable, it has a
satisfying assignment with $z=0$.  If $c=1$, one can choose $w_C$ so that
$\operatorname{NAE}(a,b,w_C)$ holds, while
$\operatorname{NAE}(\neg w_C,c,z)$ holds for either value of $w_C$.  If $c=0$
and $a\lor b=1$, setting $w_C=0$ satisfies both constraints.  If $a=b=c=0$,
$\operatorname{NAE}(\neg w_C,c,z)$ forces $w_C=0$, which violates
$\operatorname{NAE}(a,b,w_C)$.  Hence, $\varphi$ is satisfiable if and only if
$\psi$ is satisfiable.

Let $r$ be the number of constraints of $\psi$.  Construct a simple undirected
graph $H$ with one vertex for every literal occurrence in $\psi$.  For each
NAE constraint, add the three edges joining the vertices representing its three
literal occurrences.

We next add a chain for every variable $x$ of $\psi$.  List the occurrence
vertices of literals $x$ and $\neg x$ in an arbitrary order
$u_1,\ldots,u_d$.  For every $1\le h<d$, add a path from $u_h$ to $u_{h+1}$
whose internal vertices are new.  The path has length two if the two
occurrences have the same sign and length three otherwise.  Let $L$ be the
total length of all these paths and set $t:=2r+L$.

The three edges added for any NAE constraint contribute at most two edges to
any cut, and the paths between consecutive occurrences contain $L$ edges in
total.  Hence no cut has more than $t$ edges.

Assume first that $\varphi$ is satisfiable.  Then $\psi$ has a satisfying
assignment.  For every literal occurrence, place its representing vertex in
$X$ exactly when the literal is true under this assignment.  For each
path joining two consecutive occurrences of a variable, its endpoints are on
the same side of $X$ if the path has length two and on opposite sides if the
path has length three.  We can therefore add its internal vertices to $X$ so
that every path edge crosses the cut.  For each NAE constraint, its three
literals are not all equal, so exactly two of the three edges among their
vertices cross the cut.  Hence $|\delta_H(X)|=2r+L=t$.

Conversely, assume that $X\subseteq V(H)$ satisfies
$|\delta_H(X)|\ge t$.  The preceding upper bound implies that equality holds,
every path edge crosses the cut, and exactly two of the three edges added for
each NAE constraint cross the cut.  It follows along the chain for each
variable $x$ that any two occurrences with the same sign lie on the same side
of $X$, whereas any two occurrences with opposite signs lie on opposite sides.
Hence there is a truth value for $x$ such that each occurrence of $x$ or
$\neg x$ is true exactly when its vertex belongs to $X$.  Finally, the three
vertices corresponding to any NAE constraint
are not all on the same side of $X$, because exactly two of the three edges
among them cross the cut.  Thus every constraint of $\psi$ is satisfied, and
hence $\varphi$ is satisfiable.

Each vertex representing a literal occurrence has degree at most four, and
each internal path vertex has degree two.  The number of constraints, literal
occurrences, internal path vertices, and edges is $\bigO(s)$.  Therefore, $H$
has maximum degree $\bigO(1)$ and $|V(H)|+|E(H)|=\bigO(s)$.
\end{proof}

For an instance $(H,t)$, write $n:=|V(H)|$ and $m:=|E(H)|$.
We transform a \textsc{Max Cut} instance $(H,t)$ with
$0\le t\le m$ into a \textsc{Minimum Bisection} instance.  Let $q\ge2$ be
arbitrary.
Partition $V(H)$ into nonempty sets
$V_0,\ldots,V_{\lceil n/q\rceil-1}$ of size at most $q$.  For every
$0\le i<\lceil n/q\rceil$, define
$D_i:=1+\max_{v\in V_i}\deg_H(v)$.  Set
\[
  K:=\sum_{i=0}^{\lceil n/q\rceil-1}
  \bigl(D_i\cdot |V_i|\cdot(|V_i|-1)\bigr)+2(m-t).
\]

For every $v\in V_i$ and $a\in\{0,1\}$, create a clique $P(v,a)$ of size
$(K+n+2)(q+1)^i$.  Add edges between these cliques as follows.

\begin{itemize}
  \item For every $i$, every unordered pair of distinct vertices
  $u,v\in V_i$, and every $a,b\in\{0,1\}$, add $D_i$ edges with distinct
  endpoint pairs between $P(u,a)$ and $P(v,b)$.

  \item For every $uv\in E(H)$, add an arbitrary edge between $P(u,0)$ and
  $P(v,1)$ and an arbitrary edge between $P(u,1)$ and $P(v,0)$, distinct
  from all edges already added.
\end{itemize}

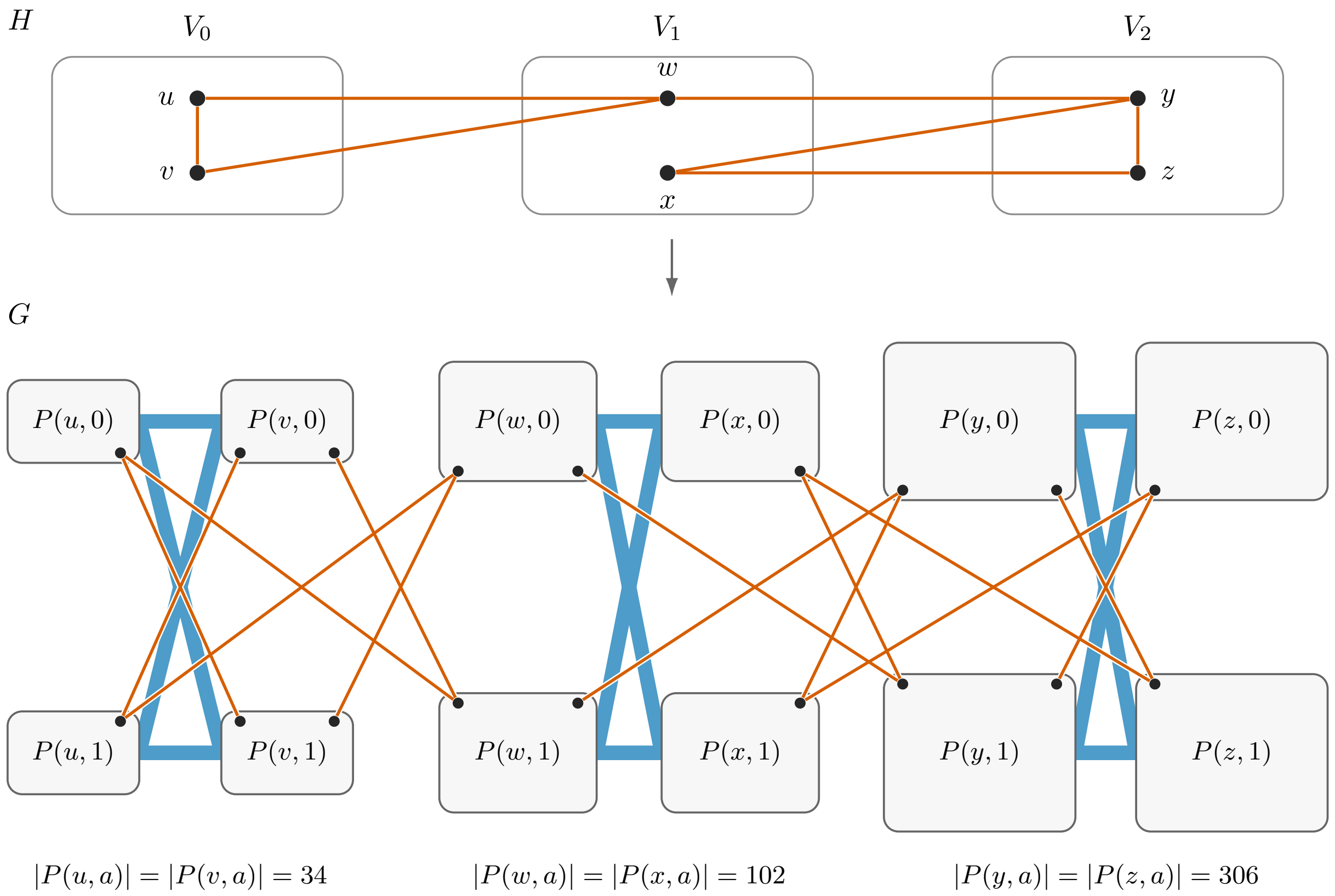
\begin{figure}[tb]
\centering
\begingroup
\definecolor{reductiongroup}{RGB}{77,156,201}
\definecolor{reductionedge}{RGB}{213,94,0}
\pgfmathsetlengthmacro{\reductionunit}{\linewidth/16}
\begin{tikzpicture}[
  x=\reductionunit,y=1cm,font=\small,
  vertex/.style={circle,fill=black!85,draw=white,line width=.22pt,
    minimum size=4pt,inner sep=0pt},
  input vertex/.style={vertex,minimum size=5.6pt},
  group edge/.style={draw=reductiongroup,line width=5pt},
  original edge/.style={draw=reductionedge,line width=1.05pt},
  original halo/.style={draw=white,line width=2pt},
  clique/.style={draw=black!60,fill=black!3,line width=.7pt,rounded corners=5pt},
  transition/.style={-{Latex[length=2.3mm,width=1.5mm]},black!60,line width=.8pt}]
  \path[use as bounding box] (0,-.20) rectangle (16,10.75);

  \node[anchor=west] at (.10,10.50) {$H$};
  \foreach \xx/\ii in {2.45/0,7.95/1,13.45/2}{
    \draw[black!45,line width=.6pt,rounded corners=7pt]
      ({\xx-1.70},8.15) rectangle ({\xx+1.70},10.05);
    \node[above=2pt] at (\xx,10.05) {$V_{\ii}$};
  }
  \coordinate (hu) at (2.45,9.55);
  \coordinate (hv) at (2.45,8.65);
  \coordinate (hw) at (7.95,9.55);
  \coordinate (hx) at (7.95,8.65);
  \coordinate (hy) at (13.45,9.55);
  \coordinate (hz) at (13.45,8.65);
  \foreach \s/\t in {u/v,u/w,v/w,w/y,x/y,x/z,y/z}
    \draw[original edge] (h\s)--(h\t);
  \foreach \p/\t/\side in {hu/u/left,hv/v/left,hw/w/above,hx/x/below,hy/y/right,hz/z/right}{
    \node[input vertex] at (\p) {};
    \node[\side=4pt] at (\p) {$\t$};
  }
  \draw[transition] (8,7.85)--(8,7.15);

  \node[anchor=west] at (.10,6.95) {$G$};
  \def\reductioncliques{
    u0/1.00/5.65/.77/.50,u1/1.00/1.65/.77/.50,
    v0/3.50/5.65/.77/.50,v1/3.50/1.65/.77/.50,
    w0/6.20/5.65/.92/.72,w1/6.20/1.65/.92/.72,
    x0/8.80/5.65/.92/.72,x1/8.80/1.65/.92/.72,
    y0/11.60/5.65/1.12/.95,y1/11.60/1.65/1.12/.95,
    z0/14.55/5.65/1.12/.95,z1/14.55/1.65/1.12/.95}
  \foreach \p/\xx/\yy/\ww/\hh in \reductioncliques{
    \coordinate (\p) at (\xx,\yy);
    \coordinate (\p-group-east) at ({\xx+\ww},\yy);
    \coordinate (\p-group-west) at ({\xx-\ww},\yy);
  }

  \foreach \s/\t in {
    u0/v0,u1/v1,u0/v1,u1/v0,
    w0/x0,w1/x1,w0/x1,w1/x0,
    y0/z0,y1/z1,y0/z1,y1/z0}
    \draw[group edge] (\s-group-east)--(\t-group-west);

  \foreach \p/\xx/\yy/\ww/\hh in \reductioncliques
    \path[clique] ({\xx-\ww},{\yy-\hh}) rectangle ({\xx+\ww},{\yy+\hh});

  \foreach \v/\xx/\ww/\hh in {
    u/1.00/.77/.50,v/3.50/.77/.50,
    w/6.20/.92/.72,x/8.80/.92/.72,
    y/11.60/1.12/.95,z/14.55/1.12/.95}{
    \foreach \side/\sgn in {left/-1,right/1}{
      \coordinate (\v0-\side) at ({\xx+\sgn*(\ww-.22)},{5.65-\hh+.12});
      \coordinate (\v1-\side) at ({\xx+\sgn*(\ww-.22)},{1.65+\hh-.12});
    }
  }

  \foreach \edgestyle in {original halo,original edge}
    \foreach \s/\t in {u/v,u/w,v/w,w/y,x/y,x/z,y/z}{
      \draw[\edgestyle] (\s0-right)--(\t1-left);
      \draw[\edgestyle] (\s1-right)--(\t0-left);
    }

  \foreach \v/\side in {
    u/right,v/left,v/right,w/left,w/right,
    x/right,y/left,y/right,z/left}
    \foreach \a in {0,1}
      \node[vertex] at (\v\a-\side) {};
  \foreach \p/\vv/\aa in {
    u0/u/0,u1/u/1,v0/v/0,v1/v/1,w0/w/0,w1/w/1,
    x0/x/0,x1/x/1,y0/y/0,y1/y/1,z0/z/0,z1/z/1}
    \node[font=\footnotesize] at (\p) {$P(\vv,\aa)$};

  \node[font=\footnotesize] at (2.25,.15) {$|P(u,a)|=|P(v,a)|=34$};
  \node[font=\footnotesize] at (7.50,.15) {$|P(w,a)|=|P(x,a)|=102$};
  \node[font=\footnotesize] at (13.08,.15) {$|P(y,a)|=|P(z,a)|=306$};
\end{tikzpicture}
\endgroup
\caption{The construction for the six-vertex graph $H$ with $q=2$ and $t=5$.
The blocks are $V_0=\{u,v\}$, $V_1=\{w,x\}$, and $V_2=\{y,z\}$.
Here $D_0=3$, $D_1=D_2=4$, and $K=26$.
A thick blue line between $P(u,a)$ and $P(v,b)$ represents $D_i$ distinct
edges, where $u,v\in V_i$. Internal clique edges are omitted.}
\label{fig:max-cut-to-bisection}
\end{figure}

These choices are always possible: it adds at most
$n+1<K+n+2$ edges between any pair of cliques.  The resulting graph $G$ is
simple, and the \textsc{Minimum Bisection} instance is $(G,K)$.  The graph
$G$ has an even number of vertices because every vertex of $H$ contributes
two cliques of the same size.
\Cref{fig:max-cut-to-bisection} illustrates the construction of $G$.

\begin{lemma}\label{lem:max-cut-to-bisection}
There is a set $X\subseteq V(H)$ with $|\delta_H(X)|\ge t$ if and only if
there is a set $Y\subseteq V(G)$ with $|Y|=|V(G)\setminus Y|$ and
$|\delta_G(Y)|\le K$.
\end{lemma}

\begin{proof}
Suppose first that $X\subseteq V(H)$ satisfies $|\delta_H(X)|\ge t$, and set
\[
  Y:=\bigcup_{v\in X}P(v,0)\ \cup\!
     \bigcup_{v\in V(H)\setminus X}P(v,1).
\]
For each $v\in V(H)$, the set $Y$ contains exactly one of the two equally
sized cliques $P(v,0)$ and $P(v,1)$.  Hence
$|Y|=|V(G)\setminus Y|$.  We evaluate $|\delta_G(Y)|$.  The edges introduced
inside group $V_i$ contribute to $|\delta_G(Y)|$ by
$D_i|V_i|(|V_i|-1)$.  The two edges representing
$uv\in E(H)$ contribute zero when $uv\in\delta_H(X)$ and two otherwise.
Therefore,
\[
  |\delta_G(Y)|=\sum_{i=0}^{\lceil n/q\rceil-1}
  \bigl(D_i\cdot |V_i|\cdot(|V_i|-1)\bigr)
  +2\bigl(m-|\delta_H(X)|\bigr)\le K.
\]

Conversely, let $Y\subseteq V(G)$ minimize $|\delta_G(Y)|$ among all sets
satisfying $|Y|=|V(G)\setminus Y|$.  Assume $|\delta_G(Y)|\le K$.
Every clique $P(v,a)$ lies entirely in $Y$ or in $V(G)\setminus Y$: a
nontrivial partition of a clique of size at least $K+n+2$ cuts more than $K$
edges.  Let $a_i$ be the number of cliques from group $V_i$ contained in $Y$.
Since each such clique has size $(K+n+2)(q+1)^i$, we have
\[
  0=|Y|-|V(G)\setminus Y|
   =2(K+n+2)\sum_{i=0}^{\lceil n/q\rceil-1}
      \bigl((a_i-|V_i|)\cdot(q+1)^i\bigr).
\]
We have $-q\le a_i-|V_i|\le q$, and the only multiple of $q+1$ in this
interval is zero.  Reducing the balance equation modulo $q+1$ therefore gives
$a_0=|V_0|$.  Divide the equation by $q+1$ and repeat the argument to obtain
$a_i=|V_i|$ for every $i$.

We claim that, for every $v\in V(H)$, exactly one of $P(v,0)$ and $P(v,1)$
is contained in $Y$.  Suppose otherwise, and fix a group $V_i$ in which this
fails.  Because exactly $|V_i|$ of the $2|V_i|$ cliques
$P(x,a)$ are contained in $Y$, there exist
$u,v\in V_i$ such that $P(u,0)\cup P(u,1)\subseteq Y$ and
$P(v,0)\cup P(v,1)\subseteq V(G)\setminus Y$.  Define
$Y':=(Y\setminus P(u,0))\cup P(v,0)$.  Since
$|P(u,0)|=|P(v,0)|$, we have $|Y'|=|Y|$.

Under the first rule, for every $w\in V_i\setminus\{u,v\}$ and
$a\in\{0,1\}$, the number of cut edges between $P(w,a)$ and
$P(u,0)\cup P(v,0)$ is the same for $Y$ and $Y'$, while the number of cut
edges between $P(u,a)$ and $P(v,b)$ over $a,b\in\{0,1\}$ decreases from
$4D_i$ to $2D_i$.  Every clique other than $P(u,0)$ and $P(v,0)$ lies on the same side
of the cut for $Y$ and $Y'$.  Therefore, a second-rule edge can change its
membership in the cut only if it has an endpoint in
$P(u,0)\cup P(v,0)$.  There are at most $\deg_H(u)+\deg_H(v)$ such edges, so
\[
  |\delta_G(Y)|-|\delta_G(Y')|
  \ge 2D_i-\deg_H(u)-\deg_H(v)
  \ge 2.
\]
This contradicts the choice of $Y$ and proves the claim.

We now recover a cut of $H$ from this bisection.  Define
$X:=\{v\in V(H):P(v,0)\subseteq Y\}$.  For each group $V_i$, exactly
$D_i|V_i|(|V_i|-1)$ edges added by the first rule belong to $\delta_G(Y)$.
For each edge of $E(H)\setminus\delta_H(X)$, both edges added by the second
rule belong to $\delta_G(Y)$.  Thus,
\[
  |\delta_G(Y)|=\sum_{i=0}^{\lceil n/q\rceil-1}
  \bigl(D_i\cdot |V_i|\cdot(|V_i|-1)\bigr)
  +2\bigl(m-|\delta_H(X)|\bigr)\le K.
\]
The definition of $K$ therefore gives $|\delta_H(X)|\ge t$, so $X$ is the
required cut of $H$.
\end{proof}

We have
\begin{align*}
  |V(G)|
  &=2(K+n+2)\sum_{i=0}^{\lceil n/q\rceil-1}
    \bigl(|V_i|\cdot(q+1)^i\bigr)\\
  &\le2(K+n+2)\sum_{i=0}^{\lceil n/q\rceil-1}
    \bigl(q\cdot(q+1)^i\bigr)\\
  &\le8q^2(n+m)(q+1)^{n/q+1}
    =2^{(n/q+1)\log_2(q+1)+\log_2(8q^2(n+m))}.
\end{align*}
Since $G$ is simple, it has fewer than $|V(G)|^2$ edges.  Thus, the construction
takes time polynomial in $|V(G)|$.

\bisectionlowerboundthm*

\begin{proof}
Suppose that \textsc{Minimum Bisection} can be solved in
$\Ostar(2^{h(k)})$ time, where $h(k)=o(g(k))$.  We may replace $h(x)$ by
$\max_{0\le j\le x}h(j)$ and assume that $h$ is nondecreasing.  The redefined
$h$ is still $o(g)$ because $g$ is nondecreasing and unbounded.  Let $\varphi$
be a $3$-CNF formula of size $s$,
and fix an integer
$q$ with $2\le q\le s$.  Apply
\cref{lem:3sat-to-bounded-degree-max-cut} to obtain $(H,t)$, and then apply the
construction above to obtain $(G,K)$.  Write $n:=|V(H)|$ and $m:=|E(H)|$.
The two reductions are correct by
\cref{lem:3sat-to-bounded-degree-max-cut,lem:max-cut-to-bisection}.

The maximum-degree bound in \cref{lem:3sat-to-bounded-degree-max-cut} gives
$D_i=\bigO(1)$.  Since $|V_i|\le q$ and $\sum_i|V_i|=n$, the definition of $K$
gives $K=\bigO(qn+m)=\bigO(qs)$.  Hence there is an absolute constant $C$ such
that $K\le Cqs$.  Since $g$ is nondecreasing
and $g(Cx)=\bigO(g(x))$, we have
$h(K)\le h(Cqs)=o(g(Cqs))=o(g(qs))$.  Moreover,
\[
  |V(G)|
  \le 8q^2(n+m)(q+1)^{n/q+1}
  =2^{\bigO(s\log q/q+\log s)}.
\]
The hypothetical algorithm applied to $(G,K)$ thus decides whether $\varphi$ is
satisfiable in
$2^{o(g(qs))}\cdot 2^{\bigO(s\log q/q+\log s)}$ time, as claimed.

We finally verify the three consequences in the statement.  We choose $q$
separately for each consequence.  Let $r(s)$ tend to infinity sufficiently
slowly.  For the first consequence, put $\beta:=\alpha/(2-\alpha)$, use
$g(k):=(k/\log k)^\beta$, and choose
$q:=\lceil s^{1-\alpha}\log s\cdot r(s)\rceil$.  For the second consequence,
use $g(k):=k/((\log k)^{2c}\log\log k)$ and choose
$q:=\lceil(\log s)^c\log\log s\cdot r(s)\rceil$.  For the third consequence,
use $g(k):=k$ and choose $q:=\lceil r(s)\rceil$.  In the three cases, $r$ can
be chosen so slowly that the exponent
$o(g(qs))+\bigO(s\log q/q+\log s)$ is, respectively,
$o(s^\alpha)$, $o(s/(\log s)^c)$, and $o(s)$.  These three bounds contradict
the three hypotheses on \textsc{3-SAT} in the theorem statement.
\end{proof}

\section*{Acknowledgments}
\addcontentsline{toc}{section}{Acknowledgments}
The author thanks Tomohiro Koana for useful discussions and for improving the
presentation of the paper.

\section*{Declaration of generative AI use}
\addcontentsline{toc}{section}{Declaration of generative AI use}
ChatGPT 6 Pro generated the proofs of
\crefrange{thm:decomp}{thm:matching-cut}.  GPT-6 Astra generated the
gadget construction for \cref{thm:bisection-lower-bound} as part of the
ETH-tightness proof, before either OpenAI or Anthropic publicly announced a
refutation of ETH.  GPT-5.6 Sol and GPT-6 Astra were
used in drafting and editing the manuscript.
The author verified and revised the proofs and the manuscript and takes full
responsibility for the paper.

\bibliographystyle{alphaurl}
\begingroup
\raggedright
\phantomsection
\addcontentsline{toc}{section}{References}
\bibliography{references}
\endgroup

\end{document}